\documentclass[11pt]{article}
\usepackage{amssymb, amsthm, amsmath}
\usepackage{booktabs}
\usepackage{multirow}
\usepackage{graphicx}
\usepackage{natbib}
\usepackage[left=25.4mm,right=25.4mm,top=30mm,bottom=25.4mm,a4paper]{geometry}
\usepackage{subfigure}
\newcommand{\Var}{\mathrm{Var}}

\newcommand{\Corr}{\mathrm{Corr}}
\newcommand{\D}{\mathrm{d}}

\theoremstyle{theorem}
\newtheorem{theorem}{Theorem}[section]
\newtheorem{lemma}[theorem]{Lemma}

\theoremstyle{definition}
\newtheorem{assumption}[theorem]{Assumption}
\newtheorem{definition}[theorem]{Definition}

\begin{document}

\title{Conditional correlation in asset return and GARCH intensity model}
\author{Geon Ho Choe\footnote{Department of Mathematical Sciences, KAIST, Daejeon 305-701, Korea}
 and Kyungsub Lee\footnote{Corresponding author, klee@euclid.kaist.ac.kr, Department of Mathematical Sciences, KAIST, Daejeon 305-701, Korea}
}
\date{}
\maketitle

\begin{abstract}
In an asset return series there is a conditional asymmetric dependence between current return and past volatility depending on the current return's sign.
To take into account the conditional asymmetry, we introduce new models for asset return dynamics in which frequencies of the up and down movements of asset price
have conditionally independent Poisson distributions with stochastic intensities.
The intensities are assumed to be stochastic recurrence equations of the GARCH type in order to capture the volatility clustering and the leverage effect.
We provide an important linkage between our model and existing GARCH,
explain how to apply maximum likelihood estimation to determine the parameters in the intensity model and show empirical results with the S\&P 500 index return series.
\end{abstract}
\section*{Acknowledgement}
This research was supported by Basic Science Research Program through the National Research Foundation of Korea(NRF) funded by the Ministry of Education(NRF-2011-0012073).
\section{Introduction}\label{Sect:intro}
Serial dependence in a financial return series is one of the most important topics in empirical finance.
Although the modeling of asset price movements using geometric Brownian motion has offered a great insight
into option pricing theory,
it is unable to incorporate the serial dependence of return such as volatility clustering.
Volatility clustering refers to the observation that large volatility tends to be followed by large volatility and small volatility tends to be followed by small volatility in financial return series.
One of the successful models to take into account volatility clustering is the ARCH\citep{Engle} (extended by the GARCH\citep{Bollerslev}) model
where the key idea is that the conditional volatility is a function of past information of squared innovations.

Another well-known property of financial return series is the leverage effect.
The leverage effect refers to the fact that today's volatility is negatively correlated with past returns.
More specifically, if the current volatility is large, then the past returns are more likely to be negative than to be positive,
and if the current volatility is small, then the past returns are more likely to be positive.
On the other hand, there are no significant correlations between today's return and past volatilities, implying that it is hard to predict today's return based on the information of past returns.

One can successfully incorporate the leverage effect with modifications to volatility functions in the original models of ARCH and GARCH.
Numerous studies have been devoted to take into account the leverage effect: e,g., \cite{Black}, \cite{Pagan&Schwert}, \cite{Nelson},  \cite{EngleNg}, \cite{GJR}, \cite{Zakoian}, \cite{Hentschel}, \cite{ChrisJ}, \cite{Bollerslev2006} and \cite{Dufour2012}.
In some literature, the leverage effect is also known as dynamic asymmetry.

We show there is another asymmetric property between return and past information, called conditional asymmetry.
This concept of asymmetry was first introduced by \cite{Babsiri} as a contemporaneous asymmetry which states that the volatility processes for up and down price moves are different from each other.
The conditional asymmetry is an asymmetric correlation between current return and past volatility
on whether the current return and the past return are positive or negative.
For example, when today's return is negative, the correlation between return and past volatility is less than the correlation between the return and past volatility when today's return is positive.
(Recall that the unconditional correlation between current return and past information , including past volatility, is almost zero.)
This phenomenon is different from the leverage effect since the leverage effect is an asymmetric relation between today's volatility and past information.

Similar approaches are found in several literatures such as in \cite{Pelagatti}, the skewness in return dynamics are examined with two different dynamics in positive and negative returns.
\cite{Palandri} investigates whether positive and negative returns share the same dynamic volatility process using a bivariate generalization of the standard EGARCH model.

We provide a new approach to explain volatility clustering, the leverage effect and conditional asymmetry in terms of conditional correlation.
We examine conditional serial correlations on the conditions of current and past returns' sign
and explain how to take volatility clustering, the leverage effect and conditional asymmetry into account in a unified way.

To incorporate conditional asymmetry in modeling asset price dynamic,
we employ a new approach, different from existing GARCH models based on volatility modeling.
A natural approach to deal with the asymmetry is that we separate the up and down movements of price dynamic and model them differently.
Therefore, we introduce intensity modeling with two pure jump processes, where the one jump process is for up movement and the other is for down movement.
In this way, the jump rates of up movement and down movement are different functions of past information, thus providing flexibility to deal with conditional dependency.
We suggest GARCH type modeling for the jump frequency (intensities) to incorporate volatility clustering and the leverage effect.

Conditional distribution of the number of jumps within a given fixed short period of time is assumed to be Poisson distributed and the jump sizes are assumed to be a small constant.
Then the conditional distribution of return is a Skellam distribution with a closed form density function and we easily employ the maximum likelihood method to estimate the parameters.
We exclude drift and diffusion terms for parsimony,
which is different from the general modeling of price dynamics of \cite{DaiSingleton}, \cite{Eraker} or \cite{Broadie}.
Intensity itself already contains the drift term; for example, greater intensity for up movement than down movement implies an upward drift.
In addition, small jumps play a role in diffusion in traditional price dynamic modeling,
as we remark on the differences between the diffusion model and our model later.
Recently, a discrete-valued small jump model has been studied to describe the tick structure of high frequency data by \cite{Barndorff}.
In this model, a Skellam process is considered to show price changes with a pure mid-price technique.
We also provide important linkages between our model and the GARCH model such that
if we constrain a parameter condition in our model, then our model and the GARCH are almost equivalent except for the difference in conditional distributions.

Related models are time-varying jump intensity models studied by \cite{Chan} and \cite{Maheu}, where the conditional jump intensity follows Poisson distribution and the jump size distribution is Gaussian.
To model microstructure noise, \cite{Bacry2013} introduced two dimensional mutually excited Hawkes processes where two intensity processes are used to model up and down price movements in microscopic level.
Our approach is different from the above literature since in our model, positive and negative jumps processes have different parameters.

The remainder of paper is organized as follows:
In Section~\ref{Sect:asymmetry}, we discuss conditional asymmetry in asset return.
In Section~\ref{Sect:model}, we introduce the Poisson intensity model to take into account the dependence structure of return series, explain the properties of our model and the relationship between our model and existing GARCH model.
In Section~\ref{Sect:Empirical}, empirical studies are employed with S\&P 500 index return series.
Section~\ref{Sect:Conclusion} concludes the paper.
In Appendix, we have interesting plots.

\section{Conditional asymmetry in asset return}\label{Sect:asymmetry}

In this section, we explain the conditional asymmetry in financial return series.
We examine the conditional serial dependencies of return series where the condition is whether the signs of returns are positive or negative.
From this point of view, we not only explain the conditional asymmetry but also discuss a unified approach to explain the other dependence structures in return series including the volatility clustering and the leverage effect.
We use the data of the S\&P 500 daily log-returns (1990.01.03--2009.12.31) with a sample size of 5,027.

\begin{figure}
\begin{center}
\subfigure[$\Corr(X_t, X_{t-\ell} | X_t > 0, X_{t-\ell} < 0)$]{
\includegraphics[width=5cm]{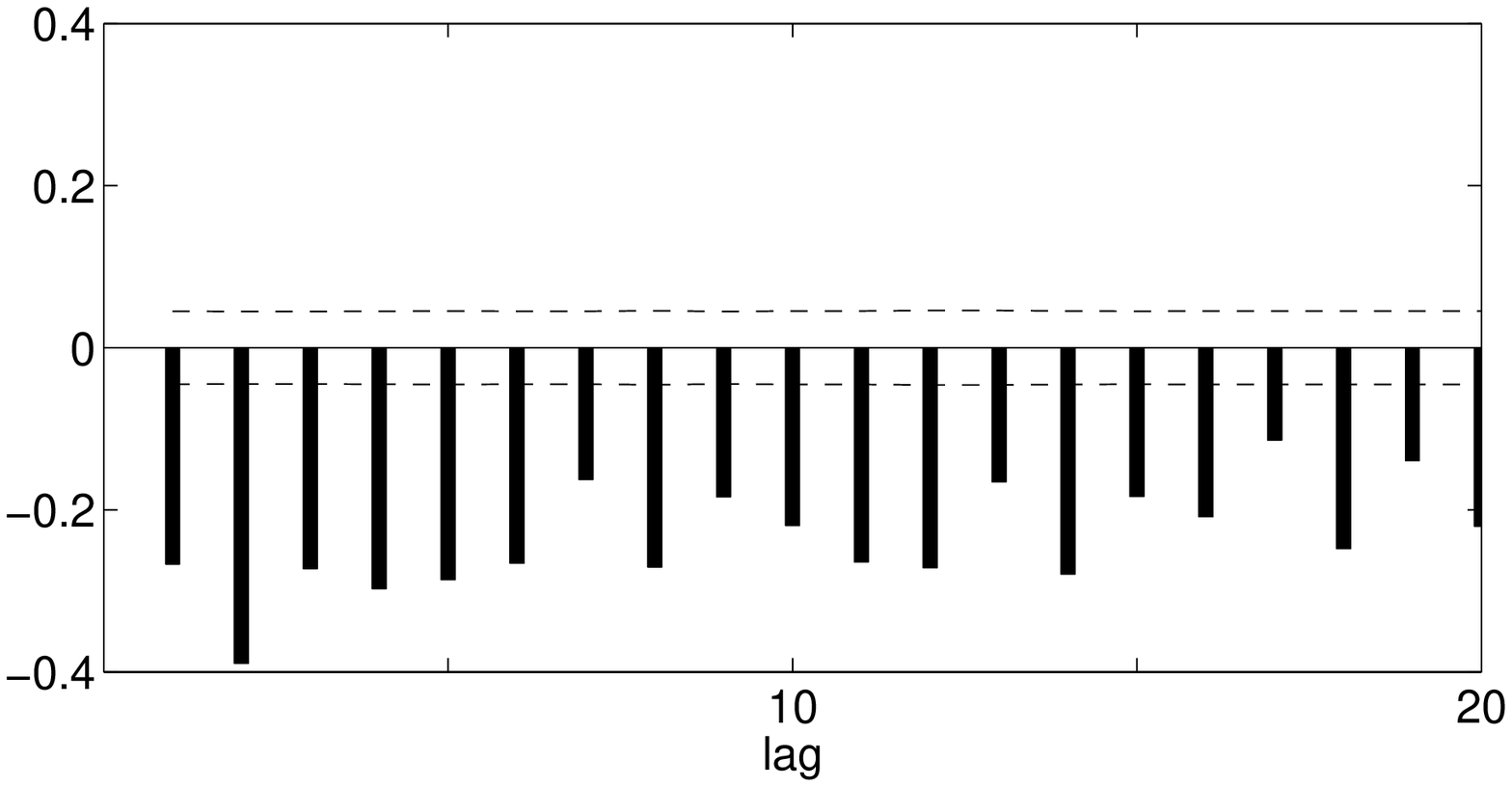}
}
\quad
\subfigure[$\Corr(X_t, X_{t-\ell} | X_t > 0, X_{t-\ell} > 0)$]{
\includegraphics[width=5cm]{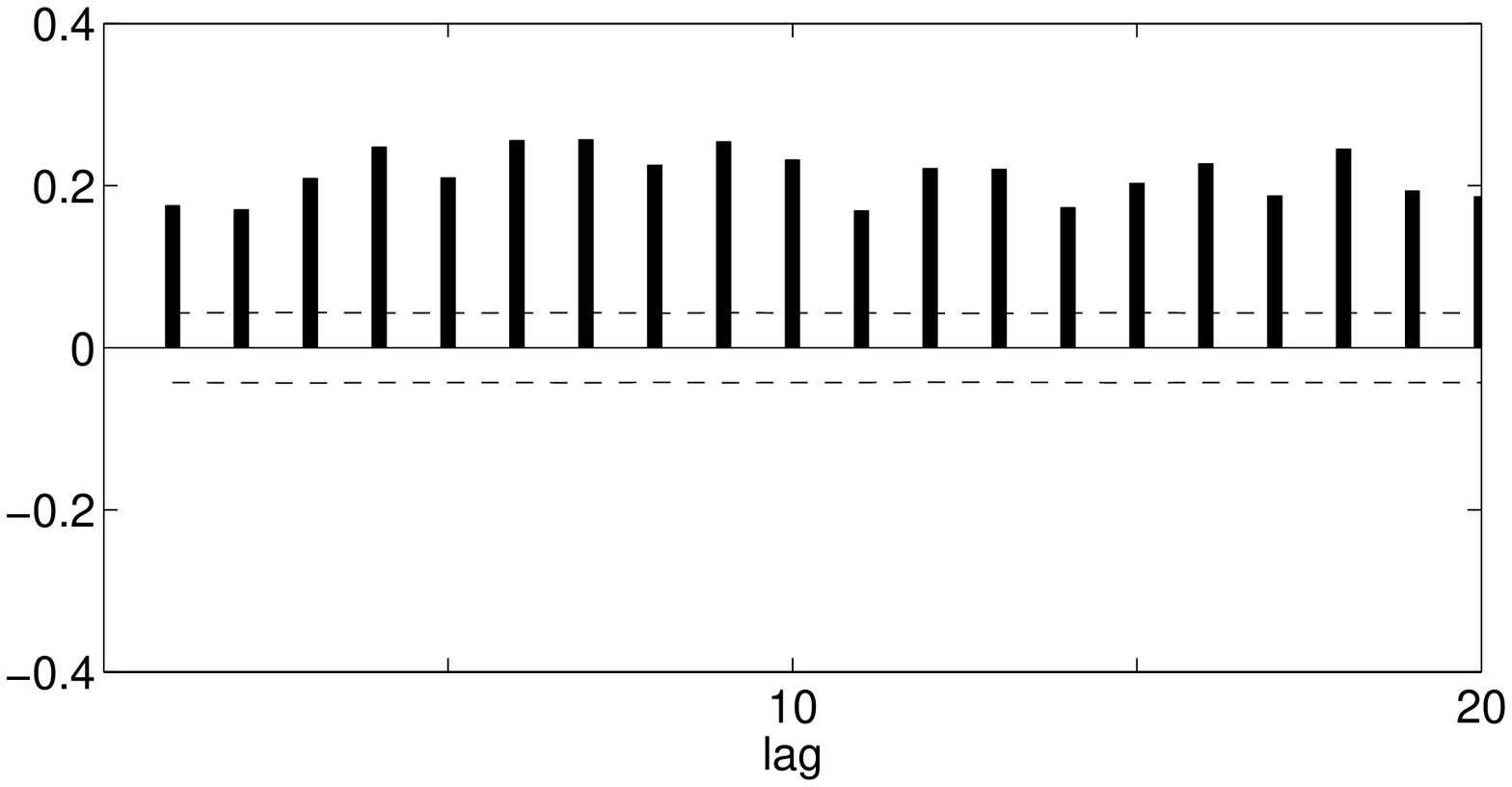}
}
\vspace{0.5cm}
\subfigure[$\Corr(X_t, X_{t-\ell} | X_t < 0, X_{t-\ell} < 0)$]{
\includegraphics[width=5cm]{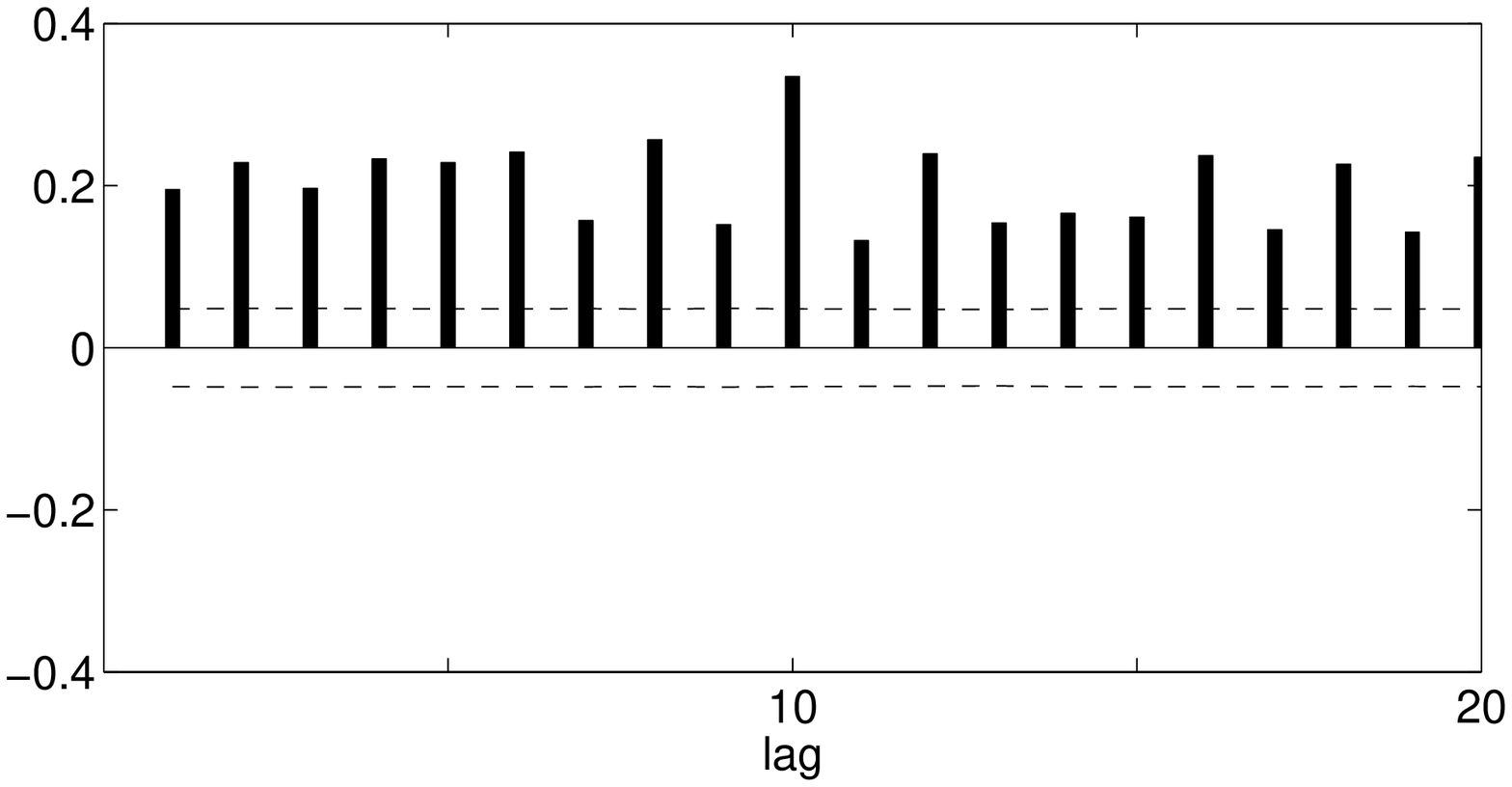}
}
\quad
\subfigure[$\Corr(X_t, X_{t-\ell} | X_t < 0, X_{t-\ell} > 0)$]{
\includegraphics[width=5cm]{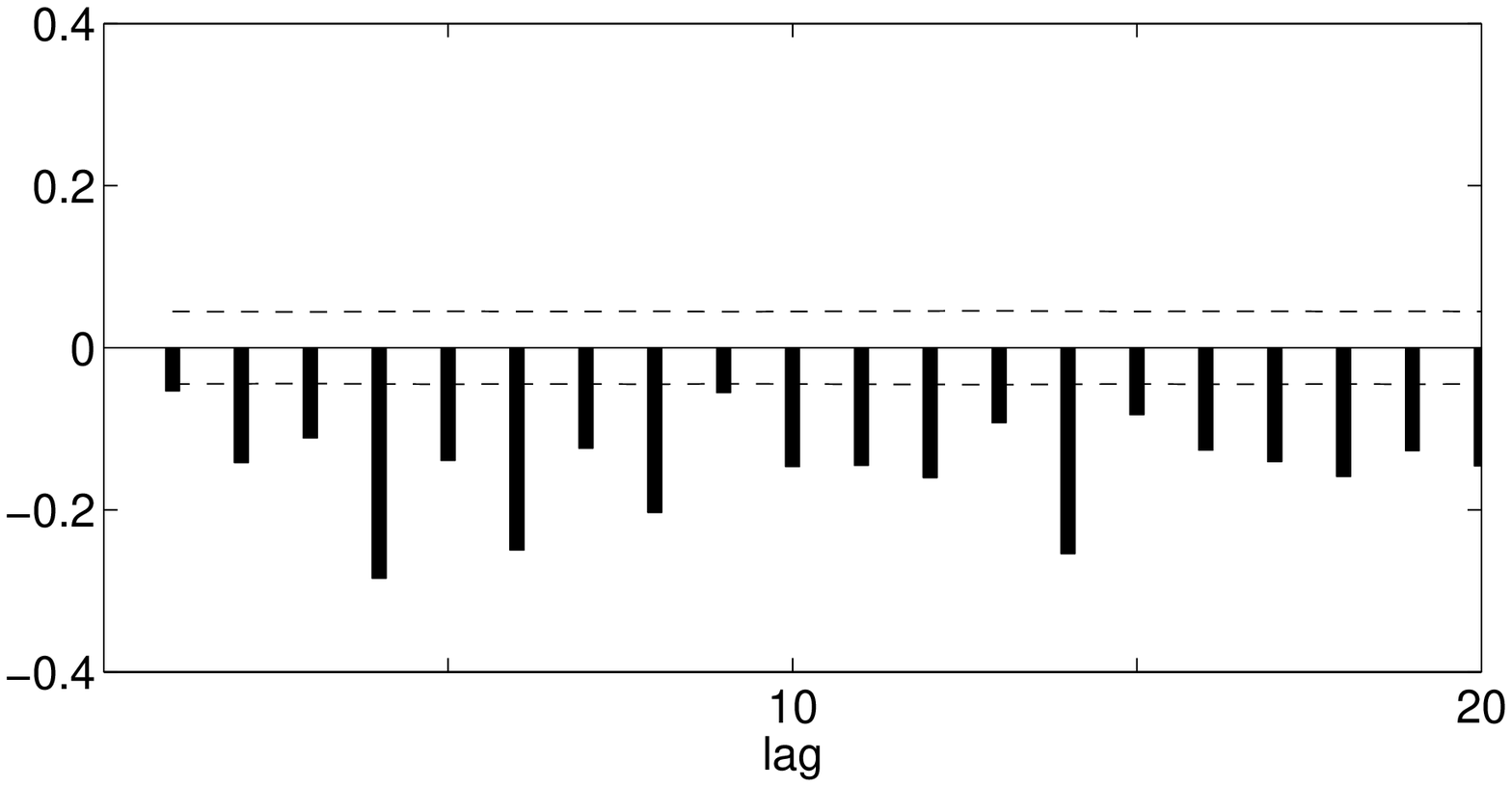}
}
\end{center}
\caption{S\&P 500: conditional correlations depending on current and past returns' signs}\label{Fig:asymmetry}
\end{figure}

Let $X_{t}$ denote the log return at time $t$, whose absolute value is regarded as a measurement of volatility.
In Figure~\ref{Fig:asymmetry}, we plot four kinds of conditional serial correlations depending on the signs of returns:

(a)$\Corr(X_t, X_{t-\ell} | X_t > 0, X_{t-\ell} > 0)$

(b)$\Corr(X_t, X_{t-\ell} | X_t > 0, X_{t-\ell} < 0)$

(c)$\Corr(X_t, X_{t-\ell} | X_t < 0, X_{t-\ell} < 0)$

(d)$\Corr(X_t, X_{t-\ell} | X_t < 0, X_{t-\ell} > 0)$\\
with lag $=\ell>0$.
The dashed lines in the figures are $1.96/\sqrt{n}$ with $n=$ sample size under the corresponding condition, which is the $2.5\%$ and $97.5\%$ quantiles of the normal distribution.
We use it as usual approximations of the standard errors of correlations in the absence of appropriate measure of the errors.
Note that all kinds of conditional correlations have significant values despite the fact that the unconditional serial return correlations, $\Corr(X_t, X_{t-\ell})$, are generally negligible.

First, the absolute values of correlations of the top panels are larger than those of the bottom panels,
and this represents that the serial correlations between returns are stronger when the current return, $X_t$, is positive.
The right bottom panel has the smallest magnitudes.
We call this phenomenon ``conditional asymmetry'', which refers to the difference between the dependence structures of up and down movements on past information.
This is different from the leverage effect, another kind of asymmetric relationship between return and volatility,
which we will explain later.
With these correlograms, we also provide the stylized facts of the stock return process such as the volatility clustering and the leverage effect.

\begin{figure}
\begin{center}
\includegraphics[width=5cm]{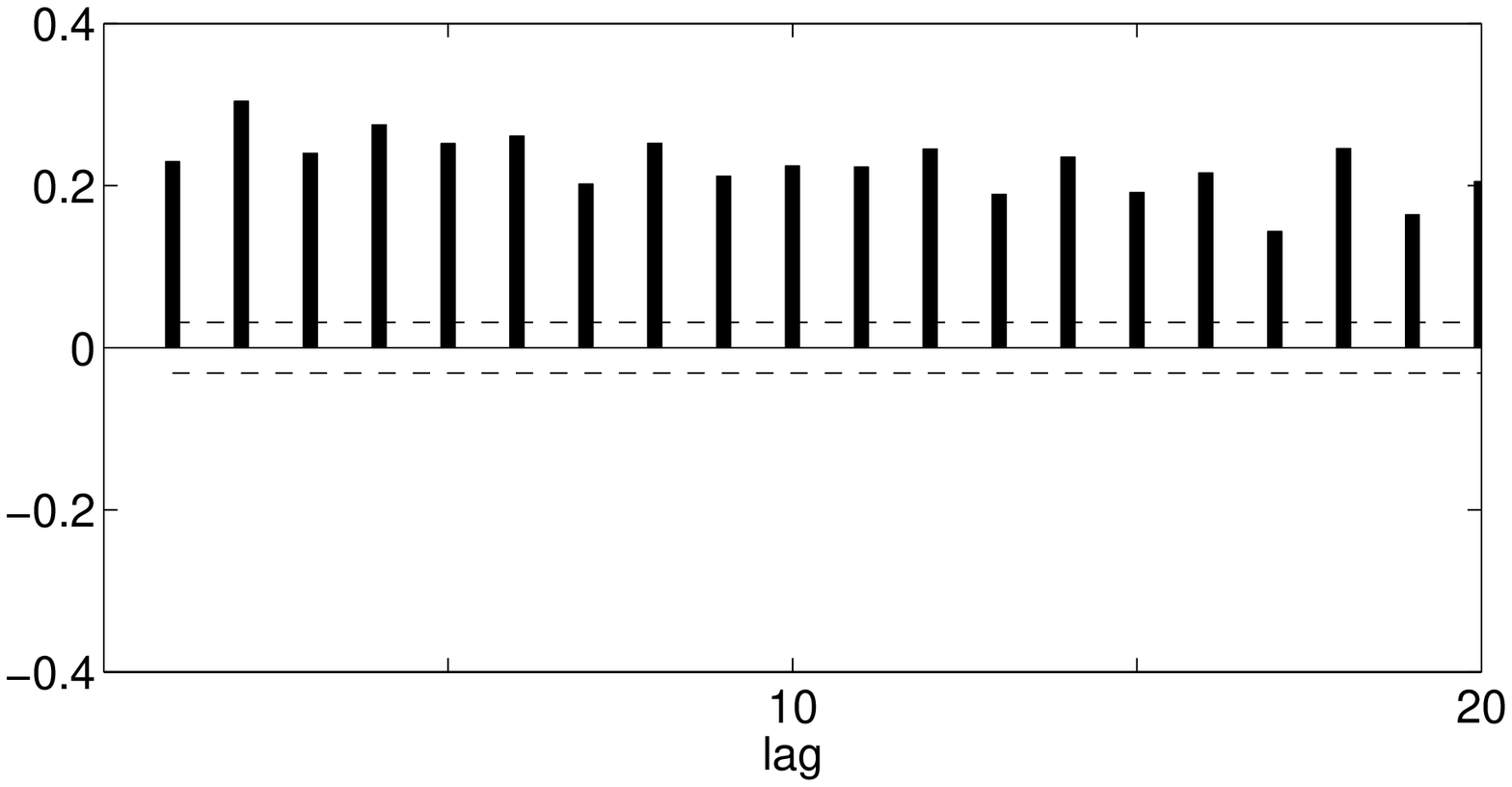}
\quad
\includegraphics[width=5cm]{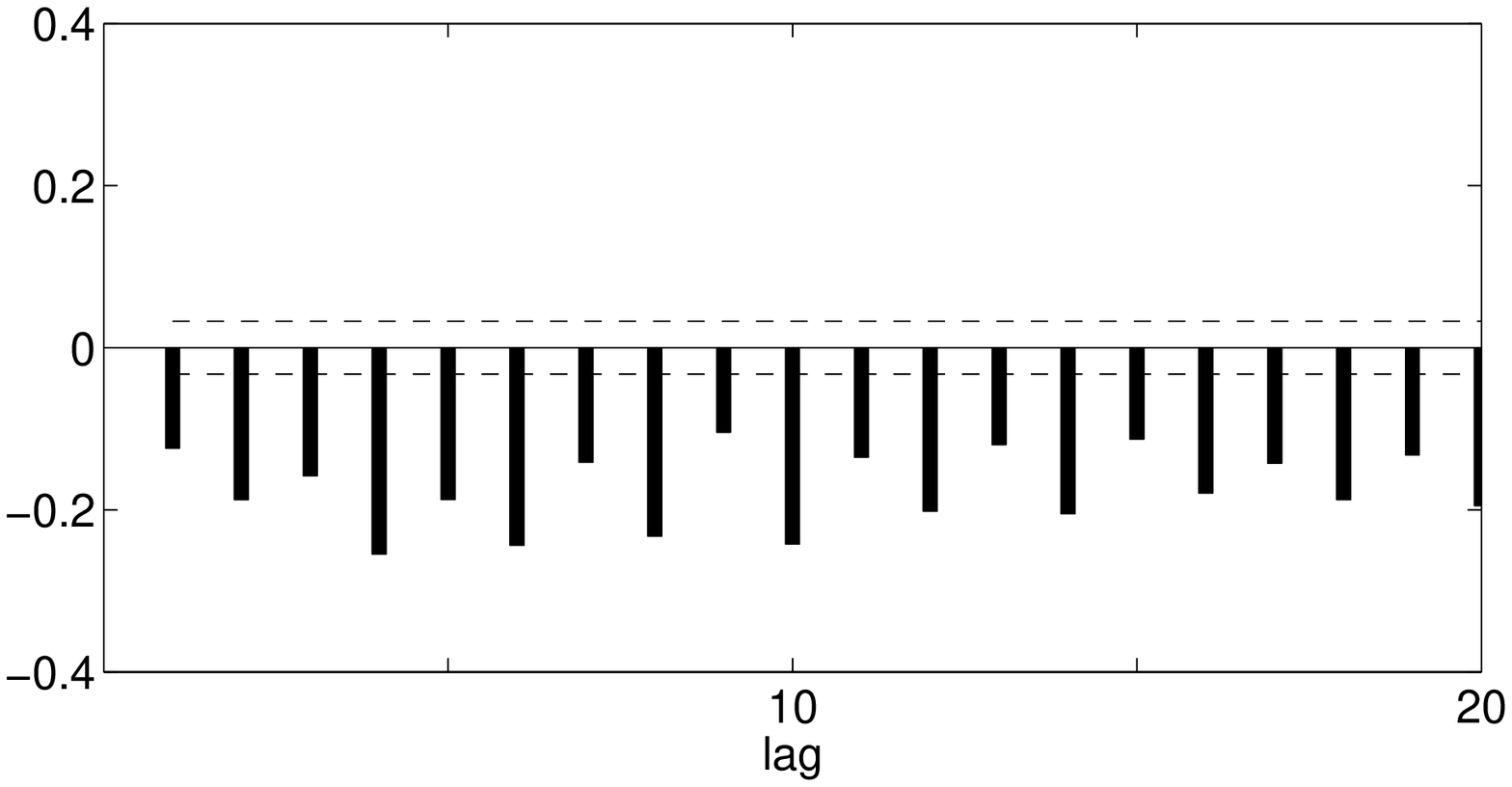}
\end{center}
\caption{S\&P 500: $\Corr(X_t, |X_{t-\ell}| \, | X_t>0)$  and $\Corr(X_t, |X_{t-\ell}| \, | X_t< 0 )$ for $\ell\geq 1$
(from left to right)}
\label{Fig:leverage22}
\end{figure}

Combining correlations in the top panels and the bottom panels, respectively, we show more evidence of conditional asymmetry.
In Figure~\ref{Fig:leverage22}, we plot the conditional correlations of the current return and the past volatilities on the conditions of the current return's sign.
Differences in magnitudes between the bars in the left and the right represent the conditional asymmetry.
Note that the magnitude of $\Corr(X_t, |X_{t-\ell}| \,| X_t>0 )$ is larger than that of $\Corr(X_t, |X_{t-\ell}| \,| X_t<0)$;
that is, today's price movement is less affected by the previous information $|X_{t-\ell}|$ when the price falls than the case when the price rises.

Also note that the positive values on the left are expected since
$$\Corr(X_t, |X_{t-\ell}| \, | X_t>0)= \Corr(|X_t|, |X_{t-\ell}| \, | X_t>0)$$
and $\Corr(|X_t|, |X_{t-\ell}|)>0$ due to the volatility clustering.
Similarly, the negative values on the right are expected since
$$\Corr(X_t, |X_{t-\ell}| \, | X_t< 0 )=-\Corr(|X_t|, |X_{t-\ell}| \, | X_t< 0 ).$$

\begin{figure}
\begin{center}
\includegraphics[width=5cm]{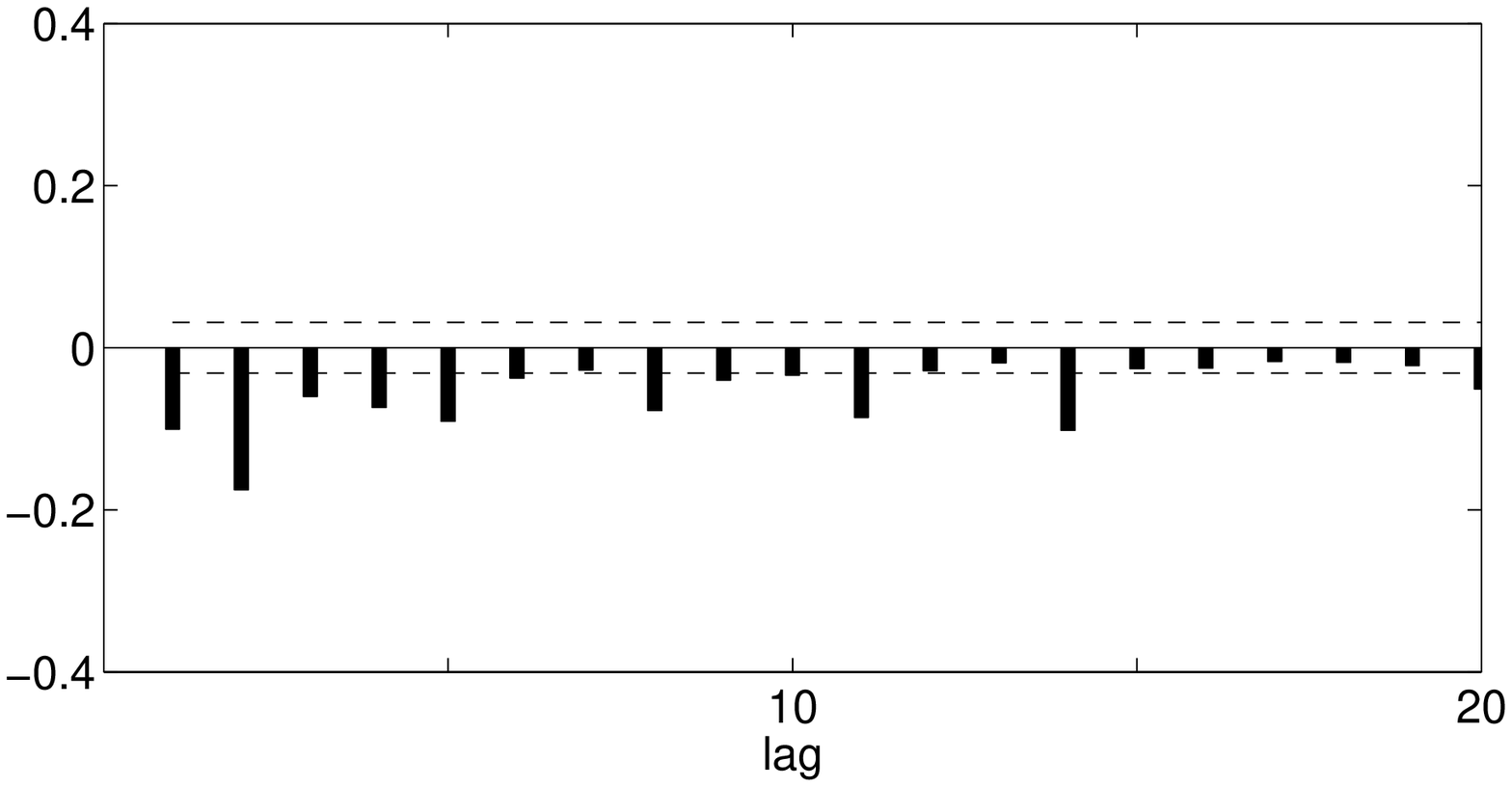}
\quad
\includegraphics[width=5cm]{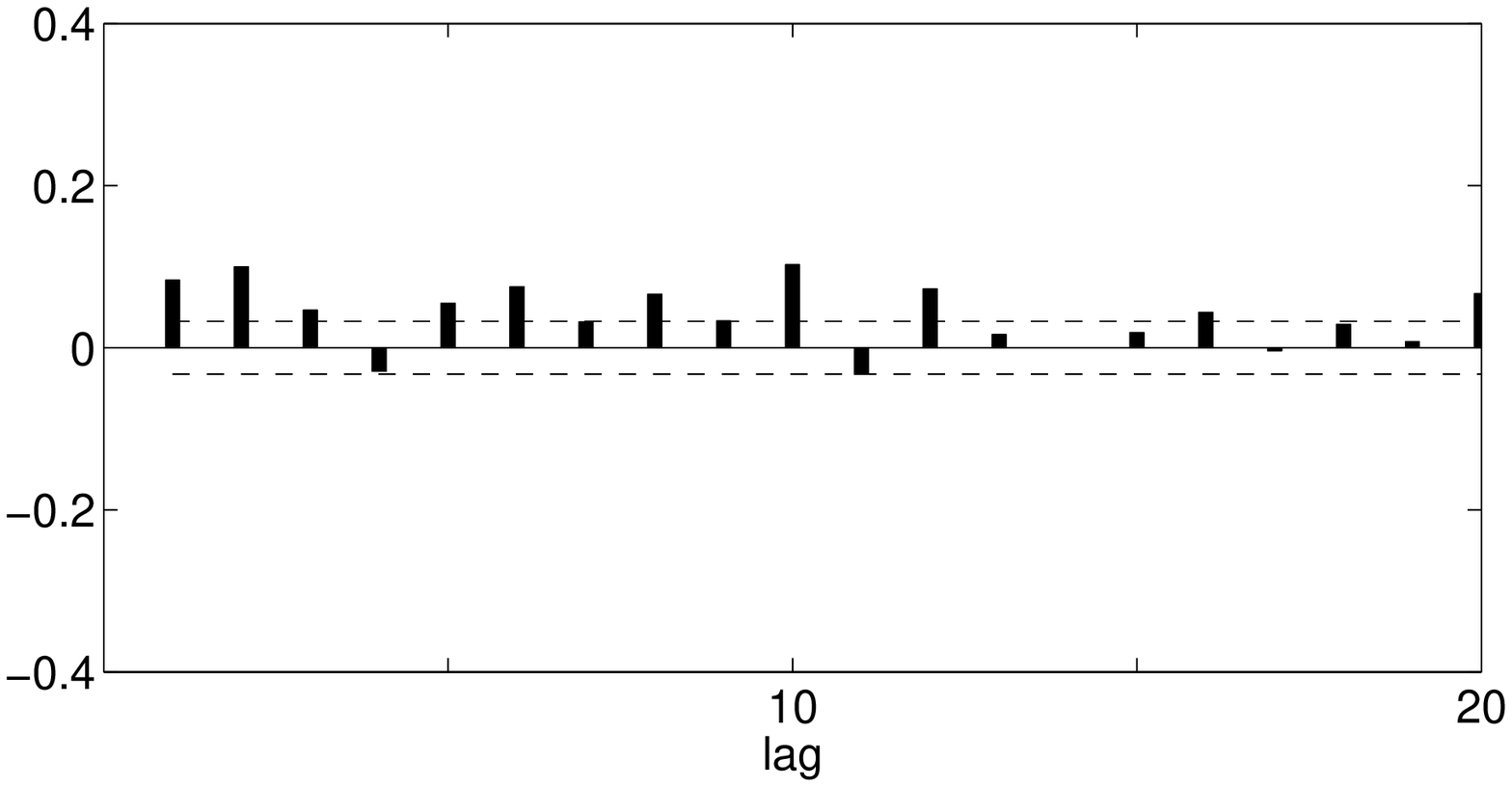}
\end{center}
\caption{S\&P 500: $\Corr(X_t, X_{t-\ell} | X_t>0)$ and $\Corr(X_t, X_{t-\ell} | X_t<0)$ for $\ell\geq 1$ (from left to right)}
\label{Fig:leverage}
\end{figure}

Second, the differences between the magnitudes of the bars on left and right of Figure~\ref{Fig:asymmetry} represent the leverage effect.
Recall that the differences between the top and bottom represents the conditional asymmetry.
The magnitudes of the bars in the left panels are generally larger than the bars in the right.
Combining correlations in the left and right panels, respectively, we have Figure~\ref{Fig:leverage}.
The left is all negative, and the right is more or less positive.
This interpretation of the leverage effect is slightly different from the traditional argument of the leverage effect, the relationship between past return and current volatility.
Note that the traditional argument of the leverage effect implies $\Corr(|X_t|, X_{t-\ell})<0$ and this is the combined result of the left and right of the figure.

\begin{figure}
\begin{center}
\includegraphics[width=5cm]{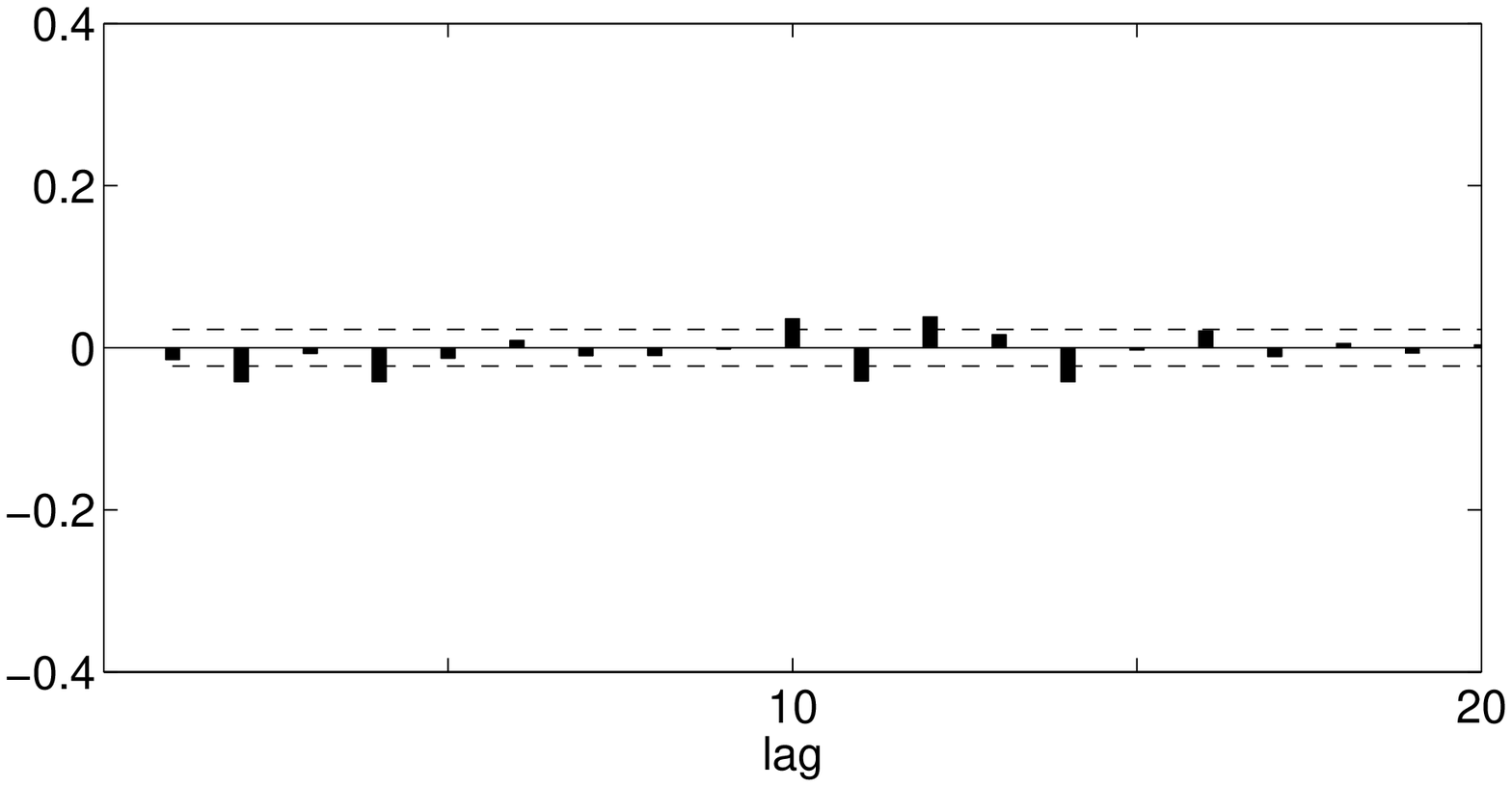}
\quad
\includegraphics[width=5cm]{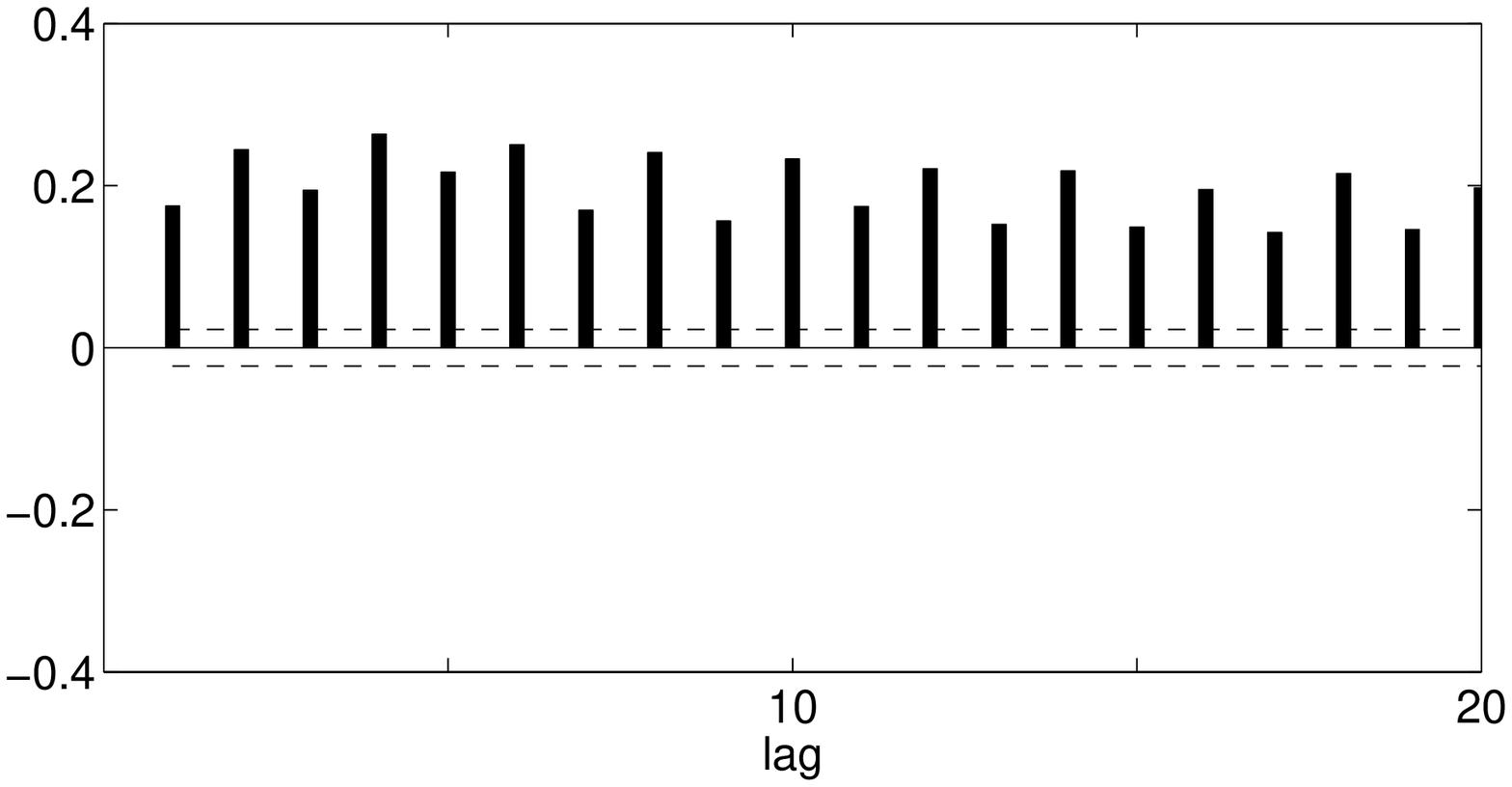}
\end{center}
\caption{S\&P 500: $\Corr(X_t, X_{t-\ell})$ and $\Corr(|X_t|, |X_{t-\ell}|)$ for $\ell \geq 1$ (from left to right).
Volatility clustering is observed in the right.}
\label{Fig:ACF}
\end{figure}

Finally, the fact that all conditional correlations are significant in Figure~\ref{Fig:asymmetry} implies volatility clustering.
Combining all conditional correlations, we have Figure~\ref{Fig:ACF}, where the traditional representation of the volatility clustering is in the right,
in contrast with the absence of unconditional autocorrelation of return series.
It is interesting to note that, with the conditional correlograms in Figure~\ref{Fig:asymmetry},
we observe the existence of volatility clustering, the leverage effect and the conditional asymmetry in a unified way.

Table~\ref{Table:SP_Corr} summarizes the various conditional correlations depending on the current sign of S\&P 500 returns with lags $\ell=1,2,3,5,10,20$.
All reported correlations are significant when we use $1/\sqrt{n}$ with $n=$ sample size satisfying the corresponding condition as approximations to the standard deviations of the conditional correlations.
We also show the result of a modified Ljung-Box test in the table.
The modified Ljung-Box statistics of time series $Y$ and $Z$ is defined by
$$ Q_{N} = \bar T (\bar T +2) \sum_{\ell=1}^{N} \left( \frac{\Corr(Y_t, Z_{t-\ell} | A)^2}{T_\ell-\ell} \right)$$
where $A$ denotes corresponding conditions, $T_{\ell}$ is the number of samples with lag $\ell$ and $\bar T$ is the average of $\{ T_\ell \}_{\ell=1}^{N}$.
Ljung-Box statistics is used to check whether the autocorrelation of given time series is different from zero \citep{Ljung}.
Though the test is known to be valid under a strong white noise assumption, we use the test statistics to quantify the serial conditional correlations.
We observe the conditional asymmetry as the statistics $Q_{20}$ are larger when $X_t>0$ than $X_t<0$.
We also observe the leverage effect as the statistics $Q_{20}$ are larger when $X_{t-\ell}<0$ than $X_{t-\ell}>0$.

\begin{table}
\centering
\caption{Sample conditional correlations of S\&P 500}\label{Table:SP_Corr}
\begin{tabular}{lccccccc}
\toprule
$\ell$ & 1 & 2 & 3 & 5 & 10 & 20 & $Q_{20}$ \\
\midrule
$\Corr(X_t, |X_{t-\ell}| \; | X_t >0)$ & 0.230 & 0.304 & 0.240 & 0.252 & 0.224 & 0.205 & 4139.6\\
$\Corr(-X_t, |X_{t-\ell}| \; | X_t <0)$ & 0.124 & 0.188 & 0.158 & 0.187 & 0.242 & 0.195 & 2329.7\\
\midrule
$\Corr(X_t, X_{t-\ell} | X_t>0, X_{t-\ell} >0)$ & 0.175 & 0.170 & 0.209 & 0.209 & 0.232 & 0.186 & 1993.0 \\
$\Corr(X_t, -X_{t-\ell} | X_t > 0, X_{t-\ell}<0)$ & 0.267 & 0.389 & 0.272 & 0.286 & 0.219 & 0.220 & 2239.3\\
$\Corr(-X_t, -X_{t-\ell} | X_t<0, X_{t-\ell} <0)$ & 0.195 & 0.228 & 0.196 & 0.228 & 0.334 & 0.235 & 1470.5\\
$\Corr(-X_t, X_{t-\ell} | X_t<0, X_{t-\ell}>0)$ & 0.053 & 0.141 & 0.111 & 0.139 & 0.147 & 0.146 & 968.0 \\
\bottomrule
\end{tabular}
\end{table}

\section{Modeling asymmetry}\label{Sect:model}
In the previous section, we showed that the dependence structure of return series differs depending on the current return's sign.
When today's price movement is up, then the dependency with past volatility is strong
and when today's price movement is down, then the dependency with past volatility is weak.
One natural way to incorporate the conditional asymmetry is modeling up and down movements separately.

\subsection{Intensity modeling}
In our model, an asset price is assumed to move as the result of economic shock, such as the changes in an asset's expected payoff, investor's preference or other economic state variables, arriving at the financial market.
More frequent arrivals of shocks imply larger volatility and less frequent occurrences of shocks imply smaller volatility.
With this point of view, we develop an intensity-based asset price model to describe the time-varying volatility and the asymmetric relations between return and volatility.

The key idea is that the asset price dynamic consists of two pure jump processes and each jump process describes positive and negative shocks, respectively.
The jump size is small enough to capture small movements of the price, which are commonly captured by a diffusion term. (Thus, a diffusion term is absent in our model.)
Stochastic jump size would be possible under a similar framework but for simplicity and parsimoniousness, we deal with only constant jump size in this paper.
For a constant jump size, we may consider a tick structure of asset price movements or a discretization of a continuous price path.
The intensities of the jump arrivals are modeled separately
and hence we provide more flexibility to our model to incorporate not only volatility clustering and the leverage effect but also the conditional asymmetry.
Furthermore, extra modeling of drift is also absent since the intensity modeling itself contains the movements of the mean of return, in contrast with the conditional variance modeling in the GARCH.

More formally, we are given a probability space $(\Omega, \mathcal F = \mathcal F(T), \mathbb P)$ with a filtration $\mathcal F(t)$, $0\leq t\leq T$,
where the $\sigma$-algebra $\mathcal F(t)$ is an information set available to the investors at time $t$.
Every stochastic process and random variable introduced in this paper is defined on $(\Omega, \mathcal F, \mathbb P)$.
We assume that good news and bad news arrive at the stock market independently of one another (at least over some short time period)
and they cause changes to the asset price immediately.

The frequencies of good news and bad news up to time $t$ are modeled by Poisson processes $N_{+} (t)$ and $N_{-}(t)$
with stochastic intensities $\lambda_+(t)$ and $\lambda_-(t)$, respectively.
Additionally, the intensity processes are assumed to be step processes,
i.e., $\lambda_{+}(t)$ and $\lambda_{-}(t)$ are constant over the time intervals $t_{i-1} \leq t < t_i$
where $\Delta t = T/N$ for some integer $N$ and $t_i = i\Delta t$, $0 \leq i \leq N$.
Finally, we assume that the size of stock price change caused by news is constant $\delta$.
These assumptions are summarized in Assumption~\ref{Assumption}.

Now we postulate the axioms for the intensity model.
\begin{assumption}\label{Assumption}
We are given $\mathcal F(t)$-adapted r.c.l.l. processes $N_{+}(t)$, $N_{-}(t)$
and positive $\mathcal F(t)$-adapted r.c.l.l. processes $\lambda_{+}(t)$, $\lambda_{-}(t)$
for $0 \leq t \leq T$ satisfying the following conditions:\\
(i) (Discrete observation time) $\Delta t = T/N$ and $t_i = i\Delta t$, $0 \leq i \leq N$.\\
(ii) (Conditional distribution)
$\left( N_{\pm}(t) - N_{\pm}(t_{i-1}) \right) | \mathcal F(t_{i-1})$ has Poisson distribution
with intensity $\lambda_{\pm}(t_{i-1})( t- t_{i-1})$,
 $t_{i-1} \leq t \leq t_{i}$.
Hence
$$\mathbb P( N_{\pm}(t_i) -  N_{\pm}(t_{i-1})= k | \mathcal F(t_{i-1}))
=  \frac{(\lambda_{\pm}(t_{i-1}) \Delta t)^k}{ k!}\exp{(-\lambda_{\pm}(t_{i-1}) \Delta t )}.$$
(iii) (Conditional independence)
$N_+(t) - N_+(t_{i-1})$ and $N_-(t) - N_-(t_{i-1})$ are conditionally independent given
$\mathcal F(t_{i-1})$, $t_{i-1} \leq t \leq t_{i}$.\\
(iv) (Step process)
$\lambda_{+}(t) = \lambda_{+}(t_{i-1})$ and $\lambda_{-}(t) = \lambda_{-}(t_{i-1})$, $t_{i-1} \leq t < t_{i}$.\\
(v) (Predictability) $\lambda_+(t_i)$  depends on $N_{\pm}(t_{i-k+1})$ and $\lambda_{\pm}(t_{i-k})$, $1\leq k \leq i-1$,
and similarly for $\lambda_-(t_i)$.\\
(vi) (Asset price) There exists a constant $\delta>0$ such that
$$S(t)= S(0)\exp\left( \delta ( N_{+}(t) - N_{-}(t))\right).$$
\end{assumption}

The assumption of constant jump size seems to be controversial, as it appears to be assumed that sizes of the impacts of news affecting asset price are the same.
However, we need a different point of view to understand that the choice of $\delta$ is rather an indifferent matter.
In our framework, the intensities are considered to be the expected time of the arrival of economic news that change the underlying return with a size of $\delta$.
For example, if $\delta = 0.002$, then the intensities imply the expected times of occurrence of economic shock that cumulatively changes $0.2\%$ of underling asset return.
If $\delta = 0.005$, then the intensities are for the occurrence of economic shock that changes $0.5\%$ of return and in this case, the values of intensity processes are smaller than the intensities in the previous case.
It turns out that when we compute the inferred conditional variance, the variances are consistent with various choices of $\delta$, see Section~\ref{Subsect:comparison}.
From this perspective, $\delta$ is rather a size of measurement to be counted than a predetermined jump size.

\begin{figure}
\caption{A tick structure of Samsung electronics at 2009.6.15.}\label{Fig:tick}
\centering
\includegraphics[width=7cm]{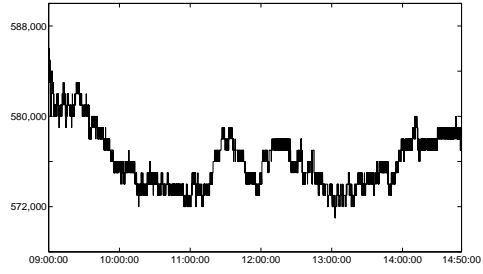}
\end{figure}

An interesting example is the tick structure of an equity (Figure~\ref{Fig:tick}), where the transactions are based on a predetermined fixed size of price change and hence every price change is constant.
As our model is based on counting processes, it may be possible to estimate the intensity processes (and related parameters) by directly counting every change of the equity price.
However, problems may arise.
Because of the existence of market structure noise\citep{Hansen}, one cannot distinguish the ``meaningful'' price changes from simple market noise.
Trades generally occur in milliseconds, so, first, it is difficult to count the number of every price changes and second it is also hard to adequately distinguish market microstructure noise from the valid movements of price.
In addition, although tick-by-tick data are available on individual stock, it is hard to observe the tick structure of an index, typically defined as a weighted average of more than one hundred stock prices.
Therefore, instead of counting every asset price movement, we rather use the daily return distribution to estimate the latent intensity processes based on the maximum likelihood estimation which will be explained later.

\begin{definition}[Decomposition of Log-Return]\label{Def:components}
Define $\mu(t_i)$, $\gamma(t_i)$, $\varepsilon(t_i)$ by
\begin{eqnarray*}
\mu(t_i) &=& \{ (e^\delta - 1) \lambda_{+} (t_{i-1}) + (e^{-\delta} - 1)\lambda_{-} (t_{i-1}) \} \Delta t \\
\gamma(t_i) &=& \{ (e^\delta - 1 -  \delta)  \lambda_+ (t_{i-1}) + (e^{-\delta} - 1 +  \delta) \lambda_- (t_{i-1}) \} \Delta t \\
\varepsilon(t_i) &=& X(t_i) - \mathbb E [X(t_i) | \mathcal F(t_{i-1})].
\end{eqnarray*}
Note that $\mu(t_i)$ and $\gamma(t_i)$ are $\mathcal F(t_{i-1})$-measurable, and $\varepsilon(t_i)$ is $\mathcal F(t_{i})$-measurable.
\end{definition}

\begin{lemma}\label{Lemma:Xdecomp}
Under Assumption~\ref{Assumption} we have
\begin{eqnarray*}
\mathbb E[X(t_i)|\mathcal F(t_{i-1})]  &=& \delta(\lambda_{+}(t_{i-1})- \lambda_{-}(t_{i-1})) \Delta t\\
\Var(X(t_i)|\mathcal F(t_{i-1}))&=& \delta^{2}(\lambda_{+}(t_{i-1})+\lambda_{-}(t_{i-1})) \Delta t\\
\mathbb E[ \exp(X(t_i))|\mathcal F(t_{i-1}))&=& \exp(\mu(t_i)).
\end{eqnarray*}
\end{lemma}

Lemma~\ref{Lemma:Xdecomp} implies
$$X(t_i) = \mu(t_i) - \gamma(t_i) + \varepsilon(t_i),$$
where $\mu(t_i)$ is regarded as a drift term, $\gamma(t_i)$ is an It\^{o} correction factor,
and $\varepsilon(t_i)$ is a shock, innovation, or residual during time interval $[t_{i-1}, t_i]$.

We compare our model with the Black-Scholes-Merton\citep{BlackScholes} framework based on the geometric Brownian motion
$\D S(t)=\alpha S(t) \D t + \sigma S(t) \D W(t)$
where $\alpha$ is drift coefficient and $\sigma$ is volatility.
Let
$$Y(t_i)  = \log \frac{S(t_i)}{S(t_{i-1})} = (\alpha-\frac{1}{2}\sigma^2) \Delta t + \sigma \Delta W .$$
In Table~\ref{Table:Comparison1},
various concepts in each model are compared.
In Table~\ref{Table:Comparison2},
the conditional expectations $\mathbb E[X(t_i)\, | \mathcal F(t_{i-1}) ]$,
$\mathbb E[\exp(X(t_i)) | \mathcal F(t_{i-1})]$
and the conditional variance $\Var[X(t_i)\, | \mathcal F(t_{i-1})]$ for $X(t_i)$
and the corresponding values for $Y(t_i)$ are compared.
In Table~\ref{Table:Comparison3},
conditional expectations and variances are compared for $\varepsilon(t_i)$ and $\sigma \Delta W$.

\begin{table}
\caption{Comparison of intensity models and Black-Scholes-Merton model}
\label{Table:Comparison1}
$$
\begin{array}{ccc}
\hline
\textrm{ }  & \textrm{Intensity}  & \textrm{BSM}\\
\hline
\textrm{Drift coefficient}  & \mu(t_i)  & \alpha \Delta t \\
\textrm{Ito correction}  & \gamma(t_i)  & \frac{1}{2} \sigma^2 \Delta t \\
\textrm{Randomness}  & \varepsilon(t_i)  &  \sigma \Delta W  \\
\textrm{Asset price}  &  S_0\exp\left( \delta ( N_{+}(t) - N_{-}(t))\right)  & S_0\exp((\alpha-\frac{1}{2}\sigma^2)t+\sigma W(t)) \\
\hline
\end{array}
$$
\end{table}

\begin{table}
\caption{Comparison of intensity models and Black-Scholes-Merton model}
\label{Table:Comparison2}
$$
\begin{array}{ccc}
\hline
\textrm{ }  & X(t_i)  & Y(t_i)\\
\hline
\mathbb E[\,\cdot\,| \mathcal F(t_{i-1})] & \mu(t_i)-\gamma(t_i)  & (\alpha-\frac{1}{2}\sigma^2) \Delta t \\
\Var[\,\cdot\,| \mathcal F(t_{i-1})] & \delta^2 (\lambda_{+}(t_{i-1})+ \lambda_{-}(t_{i-1}))\Delta t & \sigma^2 \Delta t \\
\mathbb E[\exp(\,\cdot\,) | \mathcal F(t_{i-1})] & \exp(\mu(t_i))  & \exp(\alpha \Delta t) \\
\hline
\end{array}
$$
\end{table}

\begin{table}
\caption{Comparison of intensity models and Black-Scholes-Merton model}
\label{Table:Comparison3}
$$
\begin{array}{ccc}
\hline
\textrm{ } & \varepsilon(t_i)  & \sigma \Delta W \\
\hline
\mathbb E[\,\cdot\, | \mathcal F(t_{i-1}) ]  & 0  & 0 \\
\Var[\,\cdot\, | \mathcal F(t_{i-1})] & \delta^2 (\lambda_{+}(t_{i-1})+ \lambda_{-}(t_{i-1}))\Delta t & \sigma^2\Delta t \\
\mathbb E[\exp(\,\cdot\,) | \mathcal F(t_{i-1}) ]  & \exp(\gamma(t_i))  & \exp(\frac{1}{2}\sigma^2 \Delta t) \\
\hline
\end{array}
$$
\end{table}

Note that if there is a jump in the price at $t$, then $S(t)=e^{\delta}S(t-)$ or $S(t)=e^{-\delta}S(t-)$ depending the direction of the change.
The asset price satisfies the stochastic differential equation given by
$$\D S(t)= (e^\delta -1)S(t-)\D N_{+} (t) + (e^{-\delta}-1)S(t-)\D N_{-} (t).$$

Let
$$X(t_i)= \log  \frac{S(t_{i})}{S(t_{i-1})}$$
be the log-return over the period $[t_{i-1},t_i]$.
Then the random variable $M_i$ defined by
$$M_i=\frac{X(t_i)}{\delta}=N_{+}(t_i)-N_{-}(t_i)-(N_{+}(t_{i-1})-N_{-}(t_{i-1}))$$ has integer values $m$,
and its conditional distribution given the information $\mathcal F(t_{i-1})$ is called Skellam distribution with its probability density function (p.d.f.)
\begin{eqnarray*}
& & f(m | \lambda_{+}(t_{i-1}),\lambda_{-}(t_{i-1})) \\
&=& \exp \{ -\lambda_{+}(t_{i-1}) - \lambda_{-}(t_{i-1}) \}
\left( \frac{\lambda_{+}(t_{i-1})}{\lambda_{-}(t_{i-1})} \right)^{m/2} I_{|m|}( 2\sqrt{ \lambda_{+}(t_{i-1})\lambda_{-}(t_{i-1})} )
\end{eqnarray*}
where $I_{a}$ is the modified Bessel function of the first kind defined by
$$ I_{a}(x) = \sum_{k=0}^{\infty} \frac{1}{ k! \,\Gamma(k+a+1)} \left( \frac{x}{2} \right)^{2k+a}.$$
See \citet{Haight} for more information.
Since the closed form of the conditional probability density exists, we are able to employ maximum likelihood estimation (MLE).

Let $X_1, \ldots, X_n$ denote the log-returns of asset at times $t_1< \cdots < t_n $ with their joint probability density function
$f_{\theta}(x_1, \ldots, x_n)$ for a parameter set $\theta$.
(We will explain the parameter set $\theta$ in the next section.)
For a given data, $x_{1}, \ldots, x_{n}$, we find $\theta$ which maximizes
the likelihood $f_{\theta}(x_{1}, \ldots,x_{n} )$.
The joint density function is represented by
$$f_{\theta}(x_1, \ldots, x_n | \lambda_{\pm}(t_0))
= f_{\theta}(x_1 | \lambda_{\pm}(t_0)) f_{\theta}(x_2 |\lambda_{\pm}(t_1 ))
\cdots  f_{\theta}(x_n | \lambda_{\pm}(t_{n-1}) ) $$
where
\begin{eqnarray*}
& & f_\theta(x_{i} | \lambda_{\pm}(t_{i-1})) \\
&=& \exp \{ -\lambda_{+}(t_{i-1}) - \lambda_{-}(t_{i-1}) \}
\left( \frac{\lambda_{+}(t_{i-1})}{\lambda_{-}(t_{i-1})} \right)^{x_{i}/2\delta}
I_{|x_{i}/\delta|}( 2\sqrt{ \lambda_{+}(t_{i-1})\lambda_{-}(t_{i-1})} ).
\end{eqnarray*}

\subsection{GARCH type intensity}
In order to capture the volatility clustering, we introduce GARCH type stochastic Poisson intensities,
where the innovation is a sum of two compensated Poisson distributions and the conditional intensity describes the heteroscedasticity.
Since the development of the ARCH model by \citet{Engle},
various conditional heteroscedastic models, including the GARCH models by \citet{Bollerslev}, have been proposed.
For example, if we take the basic GARCH(1,1) model for intensity, i.e.,
$$ \lambda_{\pm}(t_i) = \omega_{\pm} + \beta_{\pm} \lambda_{\pm}(t_{i-1}) + \alpha_{\pm} \varepsilon^2(t_{i-1}),$$
we have an infinite summation formula for the intensities
$$ \lambda_{\pm}(t_i) = \frac{\omega_{\pm}}{1-\beta_{\pm}} + \alpha_{\pm} \sum_{n=0}^{\infty} \beta_{\pm}^{n} \varepsilon^2(t_{i-n}).$$

If we set
\begin{equation}\label{Eq:condition}
\beta_{+} = \beta_{-}=:\beta
\end{equation}
in our model, then we obtain the basic GARCH model of \citet{Bollerslev}.
For, if  we let $h(t_i)$ be the normalized one-step-ahead conditional variance of return at $t_i$, then
$$h(t_i) = \Var(X(t_i)|\mathcal F(t_{i-1}))/\Delta t = \delta^2(\lambda_{+}(t_{i-1}) + \lambda_{-}(t_{i-1})),$$
and
\begin{equation}\label{Eq:variance-intensity}
h(t_i)
= \delta^{2}(\omega_{+} + \omega_{-}) + \beta h(t_{i-1}) + {\delta^2}(\alpha_{+} + \alpha_{-} )\varepsilon^2(t_{i-1})
\end{equation}
This is consistent with the conditional variance modeling in the GARCH, suggesting that the intensity model is an extension of the existing GARCH models.
Similarly, the one-step-ahead conditional variance of return is represented by
\begin{equation}\label{Eq:mean-intensity}
m(t_i) \equiv \mathbb E [X(t_i) | \mathcal F(t_{i-1})]/\Delta t = \delta(\omega_{+} - \omega_{-}) + \beta m(t_{i-1}) + {\delta}(\alpha_{+} - \alpha_{-} )\varepsilon^2(t_{i-1}).
\end{equation}

Furthermore,
$$
h(t_i) = \frac{\delta^{2}(\omega_{+} + \omega_{-})}{1-\beta} + \delta^2 (\alpha_+ + \alpha_-) \sum_{n=0}^{\infty}\beta^{n}\varepsilon^2(t_{i-n})
$$
and
$$
\mathbb E[ h(t_i)] = \frac{\delta^{2}(\omega_{+} + \omega_{-})}{1-\beta} + \delta^2 (\alpha_+ + \alpha_-) \sum_{n=0}^{\infty}\beta^{n}\mathbb E[ \varepsilon^2(t_{i-n})].
$$
Since
$$\mathbb E[\varepsilon^2(t_{i})] = \mathbb E[h(t_i)]\Delta t,$$
and by assuming weakly stationarity, we have
$$\mathbb E[ h(t_i)] = \frac{\delta^{2}(\omega_{+} + \omega_{-})}{1-\beta} + \delta^2 (\alpha_+ + \alpha_-) \sum_{n=0}^{\infty}\beta^{n}\mathbb E[ h(t_i)]\Delta t$$
and
\begin{equation}\label{Eq:variace}
\mathbb E[ h(t_i)] = \frac{\delta^2(\omega_+ + \omega_-)}{1-\beta-\delta^2(\alpha_+ + \alpha_-)\Delta t}.
\end{equation}
The above formula for unconditional variance is similar to the formula of the original GARCH of \cite{Bollerslev}.
If we define normalized parameters by
\begin{equation}\label{Eq:normalized}
\omega_{\pm}^* = \omega_{\pm}\delta^2, \quad \alpha_{\pm}^* = \alpha_{\pm}\delta^2,
\end{equation}
then the unconditional variance is
$$
\mathbb E[ h(t_i)] = \frac{\omega^*_+ + \omega^*_-}{1-\beta-(\alpha^*_+ + \alpha^*_-)\Delta t}.
$$
In the next section, we will show that the estimates of normalized parameters are similar across various jump sizes $\delta$.

\cite{Bollerslev} showed that in the usual GARCH(1,1) model, when $\alpha + \beta < 1$, the process is weakly stationary.
Similarly, in the intensity model, when Eq.~\eqref{Eq:condition} is satisfied, if $\alpha^*_+ +\alpha^*_- + \beta <1$, then the process is weakly stationary, where we set $\Delta t =1$ for simplicity.
The proof for the weakly stationarity is almost the same with the typical proof in the usual GARCH
except the fact that $\eta(t_i) := \varepsilon(t_i) / \sqrt{h(t_i)} $ is not identically distributed
but $\mathbb E [\eta^2(t_i) ] = 1$.

More precisely, we have
$$ h(t_i) = (\omega^*_+ + \omega^*_-) \sum_{k=0}^{\infty} M(t_i, k)$$
where
\begin{align*}
M(t_i, 0) = 1, \quad M(t_i, 1) = (\alpha_+^* + \alpha_-^*)\eta^2(t_{i-1}) + \beta
\end{align*}
and
$$ M(t_i, k+1) = (\alpha_+^* + \alpha_-^*) \eta^2(t_{i-1})M(t_{i-1},k) + \beta M(t_{i-1},k).$$
Note that
$$\mathbb E [\eta^2(t_i) ] = \mathbb E \left[ \frac{\varepsilon^2(t_i)}{h(t_i)} \right] =
\mathbb E \left[\frac{1}{h(t_i)} \mathbb E [ \varepsilon^2(t_i) | \mathcal F(t_{i-1}) ] \right] =1.$$
Furthermore, $M(t_i, k)$ involves all the terms of the form
$$ \prod_i (\alpha_+^* + \alpha_-^*)^{a_i} \prod_j \beta^{b_j} \prod_\ell \eta(t_{i-S_\ell}),$$
and the expectation of $M(t_i, k)$ does not depend on $t_i$.
By the recursive formula for $M$, we have
$ \mathbb E[M(t_i, k)] = (\alpha_+^* + \alpha_-^* + \beta)^k$
and $ \mathbb E[h(t_i)] = \mathbb E[\varepsilon^2(t_i)]$ does not depend on $t_i$.

\begin{table}
\caption{GARCH intensity models}\label{Table:intensity}
$$
\begin{array}{ll}
\hline
\textrm{Model}&
\lambda_\pm (t_i) \\
\hline
\textrm{Basic GARCH}&
 \omega_\pm + \beta_\pm\lambda_\pm(t_{i-1}) + \alpha_\pm \varepsilon^2(t_{i})\\
\textrm{Asymmetric GARCH}&
 \omega_\pm + \beta_\pm\lambda_\pm(t_{i-1}) + \alpha_\pm (\varepsilon(t_{i}) - \gamma_\pm)^2\\
\textrm{Nonlinear asymmetric}&
\omega_\pm +\beta_\pm\lambda_\pm(t_{i-1}) +\alpha_\pm (\varepsilon(t_{i}) -\gamma_\pm \sqrt{h(t_{i}}))^2\\
\textrm{GJR}&
 \omega_\pm + \beta_\pm\lambda_\pm(t_{i-1}) + (\alpha_\pm + \gamma_\pm I(t_{i}))\varepsilon^2(t_{i}) \\
\textrm{News type}&
\omega_\pm+\beta_\pm\lambda_\pm(t_{i-1}) +\alpha_\pm(\varepsilon(t_{i})+\gamma_\pm |\varepsilon(t_{i})|)^2\\
\textrm{QGARCH}&
 \omega_\pm +\beta_\pm\lambda_\pm(t_{i-1}) +\alpha_\pm\varepsilon^2(t_{i}) +\gamma_\pm \varepsilon(t_{i})\\
\textrm{Heston and Nandi}&
 \omega_\pm + \beta_\pm\lambda_\pm(t_{i-1}) + \alpha_\pm (z(t_{i}) - \gamma_\pm \sqrt{h(t_{i})})^2  \\
\textrm{VGARCH}&
 \omega_\pm + \beta_\pm\lambda_\pm(t_{i-1}) + \alpha_\pm (z(t_{i}) - \gamma_\pm )^2  \\
\hline
\end{array}
$$
\end{table}

For the various intensity models in Table~\ref{Table:intensity},
the current intensities for the occurrences of economic events are functions of past innovations and past intensities.
The models in the table correspond to various GARCH models
such as the GARCH \citep{Bollerslev}, asymmetric GARCH \citep{EngleNg},
nonlinear asymmetric GARCH \citep{EngleNg}, GJR \citep{GJR}, news type \citep{ChrisJ}, QGARCH \citep{Sentana},
VGARCH\citep{EngleNg} and \citet{Heston2000}.
In the models of Heston and Nandi, and VGARCH, we set
$$z(t_i)=\frac{\varepsilon(t_{i})}{\sqrt{h(t_{i})}},$$
and in GJR define
$$I(t_{i})=\left\{
  \begin{array}{ll}
    1, & \hbox{$\varepsilon(t_{i})<0$} \\
    0, & \hbox{$\varepsilon(t_{i})\geq 0$}.
  \end{array}
\right.
$$
If $\gamma_{\pm}=0$, then the GJR is reduced to the basic GARCH model.

It is not feasible to derive conditional correlations theoretically
due to the complicated form of the conditional probability distribution function in our model.
Thus, we show an example of simulation to show how an intensity model behaves with respect to conditional correlation under the condition of today return's sign.
Note that, for example, in the basic GARCH model the parameters $\alpha_{\pm}$ in front of $\varepsilon^2(t_{i-1})$ play important roles with respect to conditional correlations with past information $\varepsilon^2(t_{i-1})$.
For the simulation study, let
\begin{align*}
&\lambda_+(t_i) = 0.0210 + 0.9369 \lambda_+(t_{i-1}) + (86.99 + 1899 I(t_{i-1})) \varepsilon^2(t_{i-1}),\\
&\lambda_-(t_i) = 0.0167 + 0.9425 \lambda_-(t_{i-1}) + (38.23 + 1702 I(t_{i-1})) \varepsilon^2(t_{i-1})
\end{align*}
and we simulate 100 paths with 5000 daily returns.
(The parameter values come from the results of the maximum likelihood estimation that will be explained in the next subsection.)
We also presume $\lambda_+(t_0) = \lambda_-(t_0) = 5$ and $\delta = 0.005$.
The differences in $\alpha_{\pm}, \gamma_{\pm}$ in front of $\varepsilon^2(t_{i-1})$ causes different conditional correlations as reported in Table~\ref{Table:Simul_corr}.
As $\alpha_{+} > \alpha_{-}$ and $\gamma_{+} > \gamma_{-}$ , we have $\Corr(X_t, |X_{t-\ell} | ) | X_t > 0) > \Corr(-X_t, |X_{t-\ell}|) | X_t < 0)$.
Similarly, we observe $\Corr(X_t, X_{t-\ell} | X_t>0, X_{t-\ell} >0) > \Corr(-X_t, X_{t-\ell} | X_t<0, X_{t-\ell}>0)$.
In the table, the numbers in parenthesis imply corresponding standard errors.

Roughly speaking, positive return $X_t$ implies that the realization of $N_{-}(t)$ is small and close to zero so that $X_t \approx \delta N_{+}(t)$, where
the intensity of $N_{+}(t)$ is a linear function of the past squared shock with parameter $\alpha_{+}$, which is larger than $\alpha_{-}$.
Therefore, when today's return is positive, the return has a stronger correlation with the past square of shock than when today's return is negative.

\begin{table}
\centering
\caption{Simulation results of conditional correlations}\label{Table:Simul_corr}
\begin{tabular}{lcccccc}
\toprule
$\ell$ & 1 & 2 & 3 & 5 & 10 & 20 \\
\midrule
$\Corr(X_t, |X_{t-\ell}| \; | X_t >0)$ & 0.190 & 0.180 & 0.175 & 0.169 & 0.151 & 0.133 \\
& (0.061) & (0.055) & (0.058) & (0.059) & (0.050) & (0.054) \\
$\Corr(-X_t, |X_{t-\ell}| \; | X_t <0)$ & 0.153 & 0.149 & 0.148 & 0.143 & 0.137 & 0.1221 \\
& (0.049) & (0.050) & (0.050) & (0.050) & (0.047) & (0.047) \\
\midrule
$\Corr(X_t, X_{t-\ell} | X_t>0, X_{t-\ell} >0)$ & 0.171 & 0.156 & 0.152 & 0.151 & 0.139 & 0.118  \\
& (0.075) & (0.060) & (0.072) & (0.071) & (0.056) & (0.068) \\
$\Corr(X_t, -X_{t-\ell} | X_t > 0, X_{t-\ell}<0)$ & 0.210 & 0.204 & 0.198 & 0.190 & 0.166 & 0.149 \\
& (0.063) & (0.064) & (0.062) & (0.062) & (0.058) & (0.057) \\
$\Corr(-X_t, -X_{t-\ell} | X_t<0, X_{t-\ell} <0)$ & 0.189 & 0.181 & 0.179 & 0.178 & 0.166 & 0.150 \\
& (0.061) & (0.060) & (0.066) & (0.064) & (0.056) & (0.059) \\
$\Corr(-X_t, X_{t-\ell} | X_t<0, X_{t-\ell}>0)$ & 0.120 & 0.120 & 0.120 & 0.111 & 0.110 & 0.100  \\
& (0.050) & (0.052) & (0.047) & (0.047) & (0.047) & (0.048) \\
\bottomrule
\end{tabular}
\end{table}

\section{Empirical Study}\label{Sect:Empirical}
\subsection{Estimation for basic GARCH-type intensity model}\label{Sect:Calibration1}
In this section, the maximum likelihood method is employed to estimate the parameters in the GARCH intensity models
with the S\&P 500 daily log-return series from January 1990 through December 2009.
The basic statistics of the data are reported in Table~\ref{Table:statistics}, where the mean and standard deviation are annualized.
For the estimation, we assume strong stationarity of the intensity models.

\begin{table}
\centering
\caption{The statistics of S\&P 500 daily log-return series from 1990 to 2009}\label{Table:statistics}
\begin{tabular}{cccc}
\hline
mean & std. dev. & skewness & kurtosis \\
\hline
0.056 & 0.184 & -0.209 & 12.424\\
\hline
\end{tabular}
\end{table}

First, we consider the basic GARCH-type intensity with a constraint that $\beta = \beta_+ = \beta_-$,
and we have a parameter set $\theta = \{\omega_{\pm}, \beta, \alpha_{\pm}\}$.
The reason for $\beta$ constraint is that, under the constraint we are able to calculate the theoretic value of the unconditional standard deviation of the daily return distribution generated by the GARCH intensity model using Eq.~\eqref{Eq:variace}.
The estimates from maximum likelihood methods are presented in Table~\ref{Table:estimation0} for various jump sizes $\delta$.
The standard errors of the estimates calculated by a bootstrap method are reported in the corresponding parenthesis.

We observe that the minus log-likelihoods monotonically decrease as the jump sizes increase.
This implies we cannot apply the MLE to determine the size of $\delta$ since if we maximize the likelihood, then $\delta$ diverges.
This is because of the nature of the Skellam distribution.
The p.d.f. of the Skellam distribution is a typical bell shaped function and has the maximum value near the center.
The larger the presumed value of $\delta$, the closer to zero the realized aggregate jump numbers.
For simplicity, consider a Skellam distribution with $\lambda_+ = \lambda_-$.
Suppose that observed the one-day return is 0.01.
For the first case, if $\delta$ is assumed to be 0.01, the aggregated number of jumps over one day is positive one.
For the second case, if $\delta$ is assumed to be 0.005, the aggregated number of jumps over one day is positive two.
Since the value of the bell-shaped Skellam p.d.f. is larger as the jump number is nearer to zero,
we have larger likelihood for the first case where the jump size is assumed to be larger.
This is why we have larger likelihood as jump size $\delta$ increases.

\begin{table}
\caption{Parameter sets for basic GARCH type intensity model when $\beta = \beta_+ = \beta_-$}\label{Table:estimation0}
$$
\begin{array}{cccccc}
\hline
\delta & 0.05 & 0.01 & 0.005 & 0.002 & 0.001\\
\hline
\omega_{+} & 0.0057 & 0.0111 & 0.0140 & 0.0461 & 5.2428\\
           & (0.0014) & (0.0028) & (0.0038) & (0.0135) & (0.7535) \\
\beta      & 0.9040 & 0.9358 & 0.9402 & 0.9440 & 0.8200\\
           & (0.0061) & (0.0062) & (0.0065) & (0.0074) & (0.0224)\\
\alpha_{+} & 17.77 & 275.1  & 1095.3 & 6568.4 & 29364\\
& (1.44) & (27.5) & (130.1) & (836.8) & (3646) \\
\omega_{-} & 0.0053 & 0.0093 & 0.0107 & 0.0399 & 5.1899\\
& (0.0013) & (0.0027) &(0.0037) & (0.0137) & (0.7474) \\
\alpha_{-} & 16.17 & 262.8  & 1069.3 & 6524.6 & 29226\\
& (1.28) & (27.5)& (130.2)& (839.8) & (3601) \\
\hline
-\textrm{loglikelihood} & 2411 & 7037 & 10262 & 14910 & 19007\\
\textrm{std. dev.} & 0.785 & 0.222 & 0.165 & 0.153 & 0.147\\
\hline
\end{array}
$$
\end{table}

No matter which value of $\delta$ is chosen, we observe consistent properties for the estimates.
The fact that $\omega_{+} > \omega_{-}$ shows the upward drift of the asset price process in general.
The fact that $\alpha_{+} > \alpha_{-}$ for any presumed jump size implies that downward movements are less affected by the previous information
$\varepsilon^2$ than upward movements, which is the conditional asymmetry explained in Section~\ref{Sect:asymmetry}.

In the table, the theoretic unconditional standard deviations of daily return distribution based on the estimated parameters are presented.
The unconditional standard deviation varies over $\delta$.
When $\delta = 0.05$, the theoretical unconditional standard deviations are too large compared with the empirical standard deviation the sample in Table~\ref{Table:statistics}.
We observe that, as the jump size decreases, the estimated standard deviation decreases.
We may suppose the reasonable jump size is less than 0.01 (around 0.005) by comparing the sample standard deviation of standard deviation in Table~\ref{Table:statistics} and the theoretical values in Table~\ref{Table:statistics}.

The normalized values of parameters defined by Eq.~\eqref{Eq:normalized}
are reported in Table~\ref{Table:normalized_estimation0}.
In this way, we show that $\omega_{\pm}^*, \alpha_{\pm}^*$ and $\beta$ have similar values for all jump sizes.
(In fact, the normalized parameters $\omega_{\pm}^*, \alpha_{\pm}^*$ tend to increase slightly as the jump sizes increase, except for the case of $\delta=0.001$.
The parameter $\beta$ slightly decreases as the jump sizes increase.)
The consistency shows that the jump size $\delta$ is rather irrelevant to the dependence structure and the conditional asymmetry of return series.
From now on, we will show the estimation results with various jump sizes.

\begin{table}
\caption{Normalized parameter sets for basic GARCH type intensity model when $\beta = \beta_+ = \beta_-$}\label{Table:normalized_estimation0}
$$
\begin{array}{cccccc}
\hline
\delta & 0.05 & 0.01 & 0.005 & 0.002 & 0.001\\
\hline
\omega_{+}^{*} & 1.43\times10^{-5} & 1.11\times10^{-6} & 3.49\times10^{-7} & 1.84\times10^{-7} & 5.24\times10^{-6}\\
&(3.52\times10^{-6}) & (2.81\times10^{-7}) & (9.51\times10^{-8}) & (5.38\times10^{-8}) & (7.54\times10^{-7})\\
\beta & 0.9040 & 0.9358 & 0.9402 & 0.9440 & 0.8200\\
& (0.0061)& (0.0062) & (0.0065) & (0.0074) & (0.0224)\\
\alpha_{+}^{*} & 0.0442 & 0.0275 & 0.0274 & 0.0263 & 0.0294 \\
& (0.0036) & (0.0036) & (0.0033)& (0.0033) & (0.0036)\\
\omega_{-}^{*} & 1.32\times10^{-5} & 8.68\times10^{-7} & 2.68\times10^{-7} & 1.59\times10^{-7} & 5.19\times10^{-6}\\
& (3.37\times10^{-6})& (2.67\times10^{-7}) & (9.18\times10^{-8}) & (5.50\times10^{-8}) & (7.47\times10^{-7})\\
\alpha_{-}^{*} & 0.0404 & 0.0263 & 0.0267 & 0.0261 & 0.0292\\
& (0.0032) & (0.0027) & (0.0033) & (0.0034) & (0.0036)\\
\hline
\end{array}
$$
\end{table}

The parameters $\alpha_{+}$ and $\alpha_{-}$ play crucial roles to capture the conditional asymmetry.
Even if we observe that $\alpha_{+} > \alpha_{-}$ for all jump sizes,
we cannot be sure that the estimates are significantly different from each other since the differences are relatively small compared with the corresponding standard errors.
Thus, we report the p-values for the null hypothesis that $\alpha_{+}=\alpha_{-}$ in Table~\ref{Table:alpha} where p-values are calculated by a bootstrap method.
As results, at the $5\%$ level we reject the conditional symmetry that $\alpha_{+}=\alpha_{-}$ for all jump sizes.
At the $1\%$ level, we reject the conditional symmetry for all jump sizes, except for the case of $\delta = 0.001$.

\begin{table}
\caption{The p-values for the null hypothesis that $\alpha_{+}=\alpha_{-}$}\label{Table:alpha}
\centering
\begin{tabular}{cccccc}
\hline
$\delta$ & 0.05 & 0.01 & 0.005 & 0.002 & 0.001\\
\hline
p-value & 0.0000 & 0.0000 & 0.0003 & 0.0004 & 0.0233\\
\hline
\end{tabular}
\end{table}

Next, we consider the basic GARCH intensity model with
the parameter set  $\theta = \{\omega_{\pm}, \beta_{\pm}, \alpha_{\pm}\}$, i.e., without the constraint for $\beta_\pm$.
The estimates from the maximum likelihood methods are presented in Table~\ref{Table:estimation1} for various jump sizes $\delta$.
Similarly with the previous case, we observe the conditional asymmetry in the estimates
as we have $\alpha_{+} > \alpha_{-}$ for all jump sizes.

\begin{figure}
\begin{center}
\subfigure[$\Corr(X_t, X_{t-\ell} | X_t > 0, X_{t-\ell} < 0)$]{
\includegraphics[width=5cm]{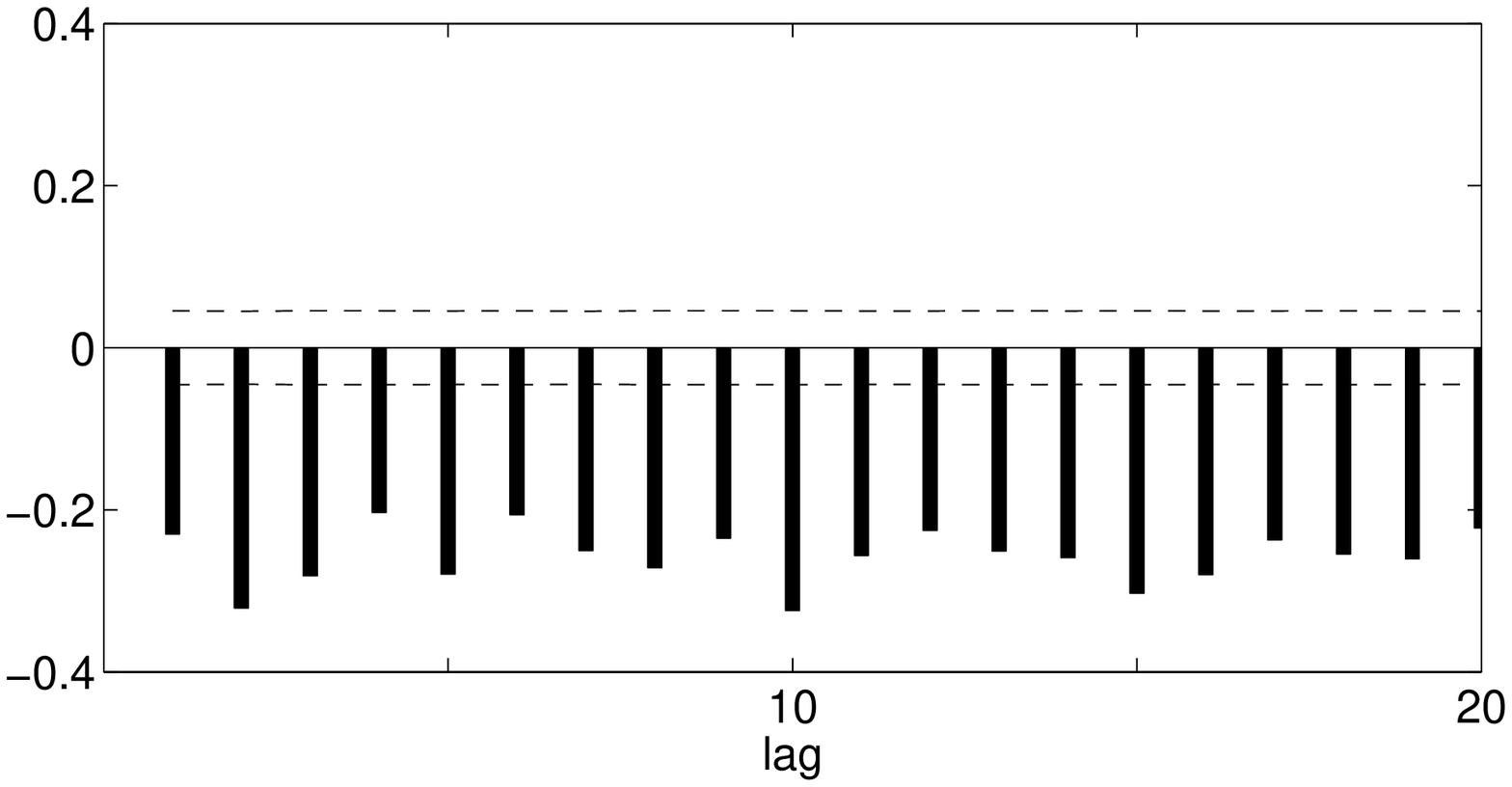}
}
\quad
\subfigure[$\Corr(X_t, X_{t-\ell} | X_t > 0, X_{t-\ell} > 0)$]{
\includegraphics[width=5cm]{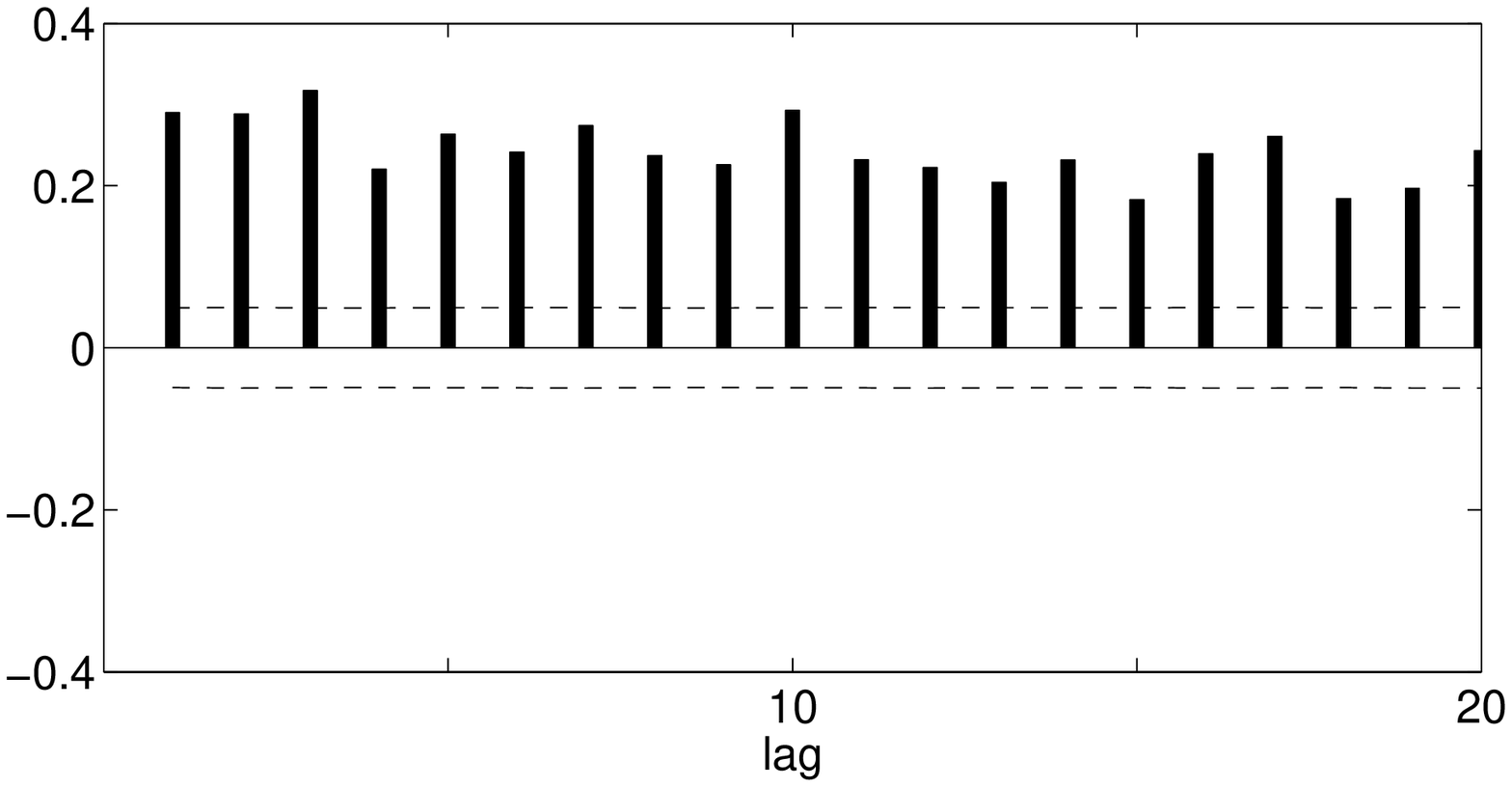}
}
\vspace{0.5cm}
\subfigure[$\Corr(X_t, X_{t-\ell} | X_t < 0, X_{t-\ell} < 0)$]{
\includegraphics[width=5cm]{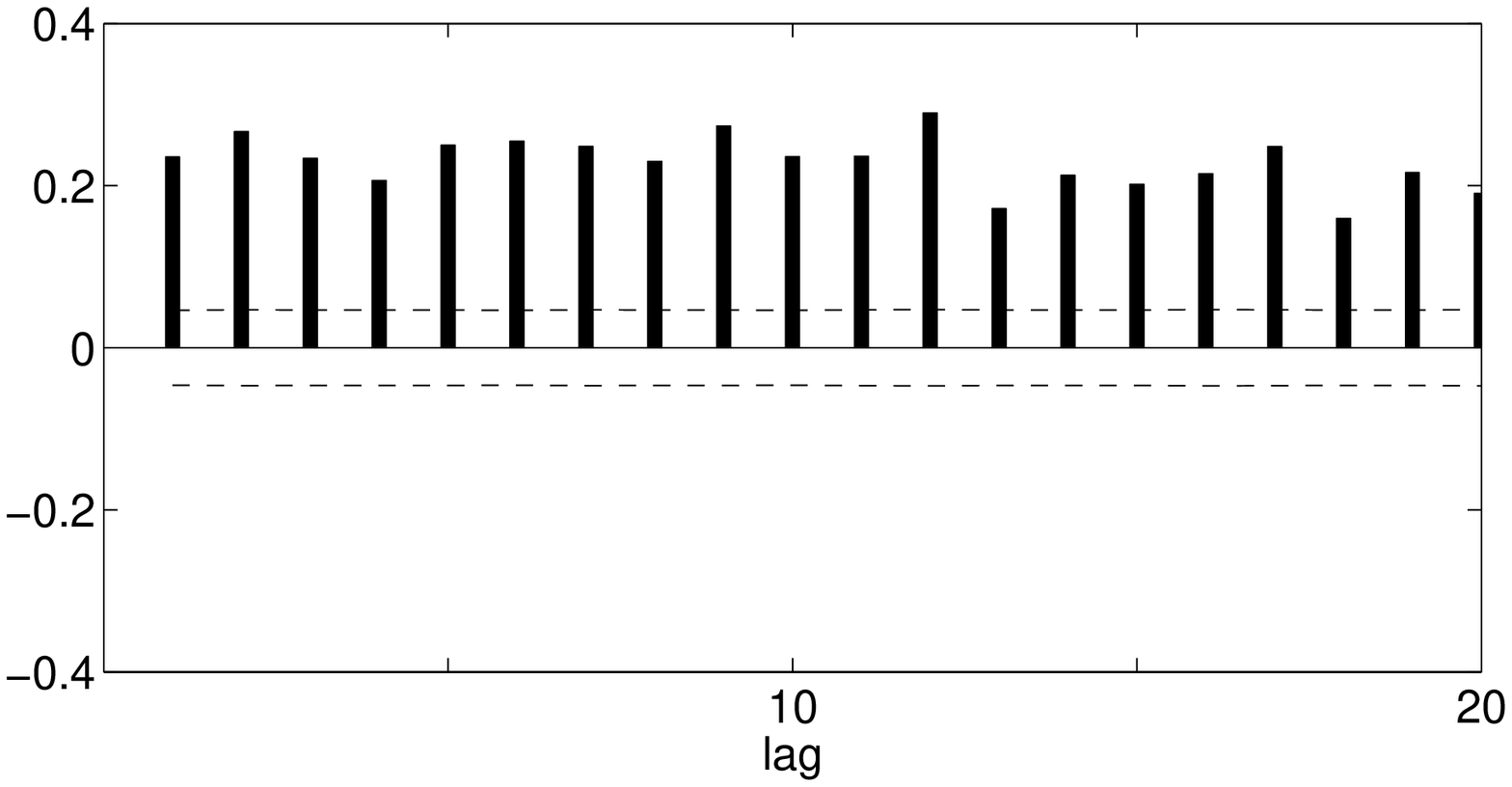}
}
\quad
\subfigure[$\Corr(X_t, X_{t-\ell} | X_t < 0, X_{t-\ell} > 0)$]{
\includegraphics[width=5cm]{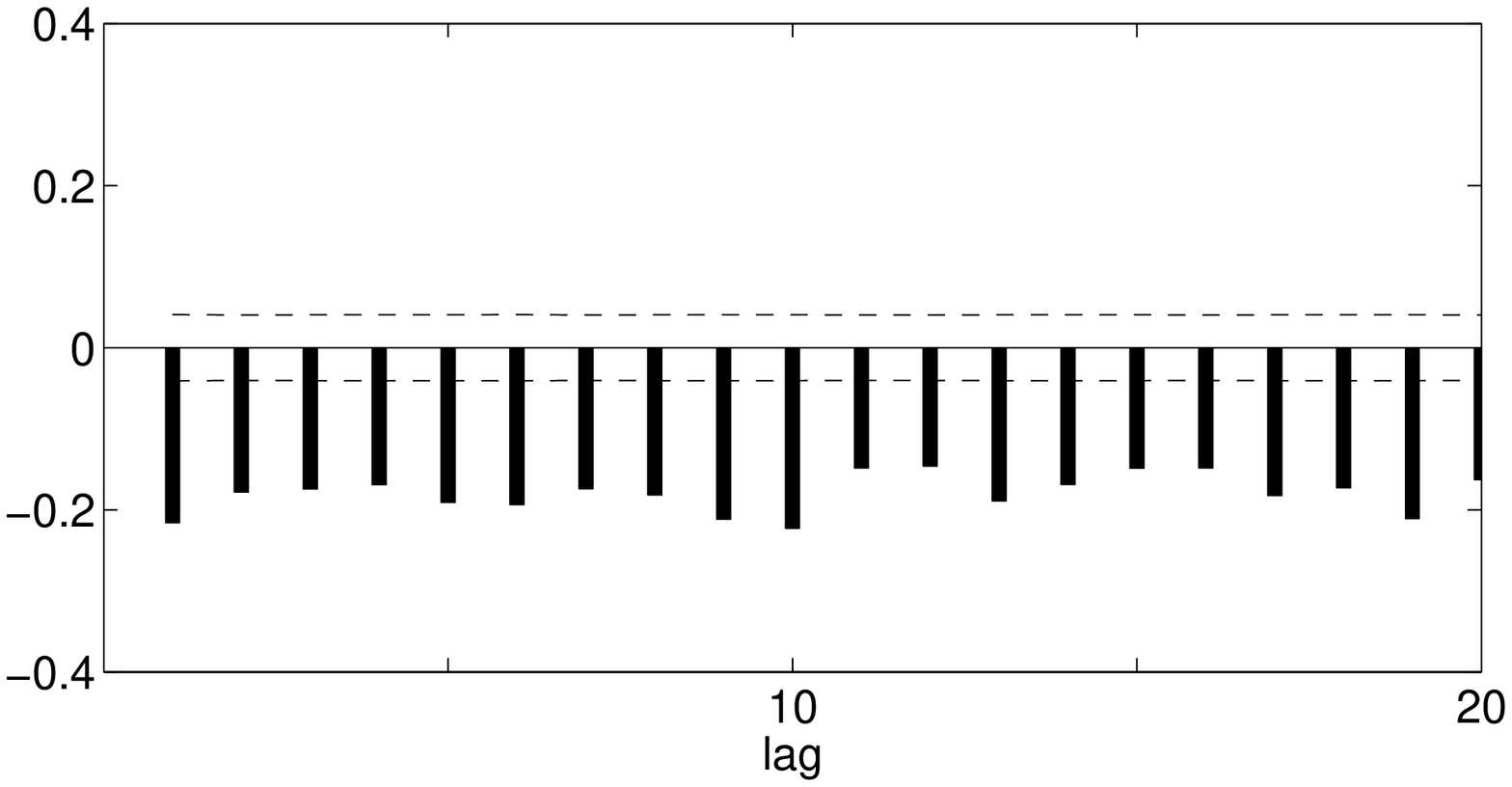}
}
\end{center}
\caption{Basic GARCH intensity: conditional correlations depending on current and past returns' signs}\label{Fig:GARCH_asymmetry}
\end{figure}

\begin{figure}
\begin{center}
\includegraphics[width=5cm]{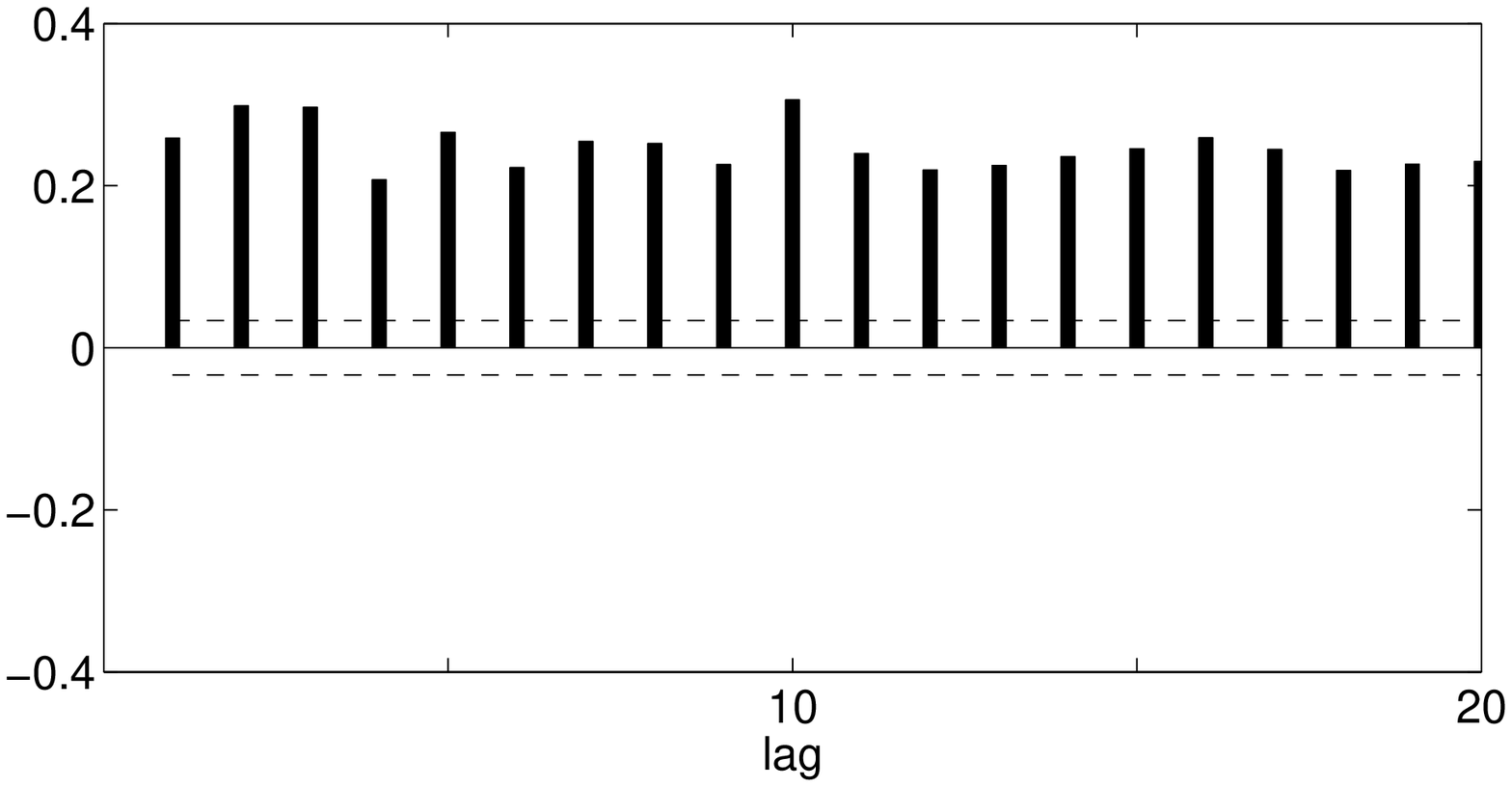}
\quad
\includegraphics[width=5cm]{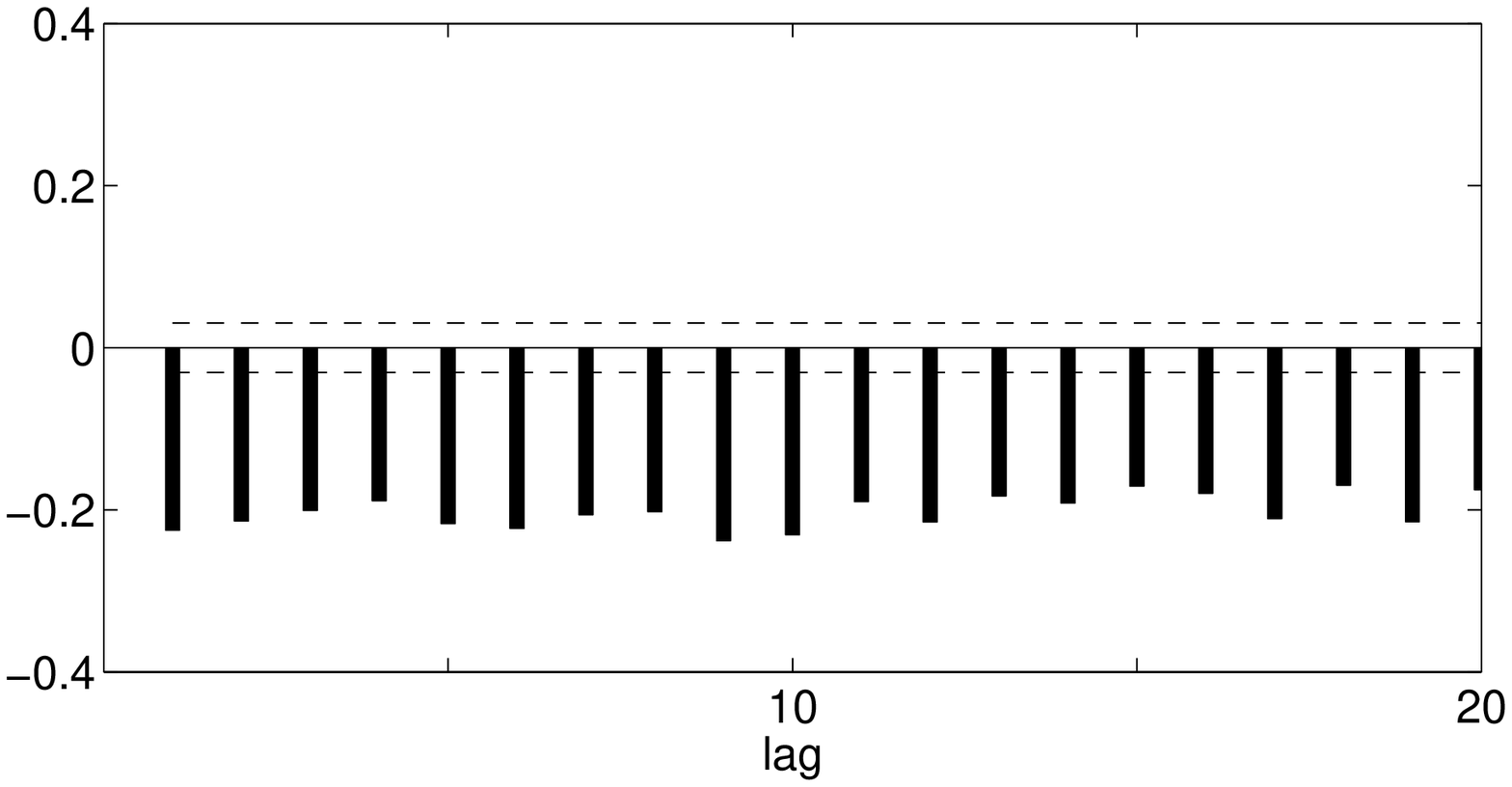}
\end{center}
\caption{Basic GARCH intensity: $\Corr(X_t, |X_{t-\ell}| \,| X_t>0)$
and $\Corr(X_t, |X_{t-\ell}| \,| X_t<0 )$ for $\ell\geq 1$ (from left to right)}
\label{Fig:GARCH_leverage22}
\end{figure}

With the obtained estimates, we simulate a return series and compute serial conditional correlations.
The conditional asymmetry is represented in Figures~\ref{Fig:GARCH_asymmetry} and \ref{Fig:GARCH_leverage22},
where the magnitudes of bars in the bottom of Figure~\ref{Fig:GARCH_asymmetry} are smaller than the bars in the top and where the bars on the right of Figure~\ref{Fig:GARCH_leverage22} are smaller than those on the left.
Compare the figure with Figures~\ref{Fig:asymmetry} and \ref{Fig:leverage22}.
Various kinds of interesting correlograms are plotted in Appendix~\ref{Append:GARCH}.
We also observe that the likelihoods slightly increase compared with the previous case where we have the constraint $\beta_+ = \beta_-$.
The p-values for the null hypothesis that $\alpha_{+}=\alpha_{-}$ are reported in Table~\ref{Table:alpha2}.
For all cases, we reject the null hypothesis and this implies the conditional symmetry at the $1\%$ level.
We also present normalized estimates of parameters in Table~\ref{Table:normalized_estimation1}.

\begin{table}
\caption{Parameter sets for basic GARCH type intensity model}\label{Table:estimation1}
$$
\begin{array}{ccccc}
\hline
\delta & 0.01 & 0.005 & 0.002 & 0.001\\
\hline
\omega_{+} & 0.0128 & 0.0168 & 0.0687 & 3.4907 \\
           & (0.0024) & (0.0035) & (0.015) & (0.7236) \\
\beta_{+}  & 0.9288 & 0.9342 & 0.9356 & 0.8492\\
           & (0.0079) & (0.0049) & (0.0091) & (0.0314)\\
\alpha_{+} & 296.2 & 1171.0  & 7492.4 & 28498\\
           & (35.6) & (76.9) & (1234.1) & (7158) \\
\omega_{-} & 0.0092 & 0.0105 & 0.0530 & 3.2870\\
           & (0.0026) & (0.0035) & (0.0118) & (0.6761)) \\
\beta_{-}  & 0.9381 & 0.9413 & 0.9388 & 0.8563\\
           & (0.0079) & (0.0057) & (0.0083) & (0.0294)\\
\alpha_{-} & 247.2 & 1024.2  & 7064.1 & 26701\\
           & (32.2) & (91.2)& (1116.0) & (6646) \\
\hline
-\textrm{loglikelihood}  & 7035 & 10258 & 14891 & 18990 \\
\hline
\end{array}
$$
\end{table}

\begin{table}
\caption{The p-values for the null hypothesis that $\alpha_{+}=\alpha_{-}$ without $\beta$ constraint}\label{Table:alpha2}
\centering
\begin{tabular}{ccccc}
\hline
$\delta$  & 0.01 & 0.005 & 0.002 & 0.001\\
\hline
p-value & 0.000 & 0.000 & 0.000 & 0.002 \\
\hline
\end{tabular}
\end{table}

\begin{table}
\caption{Normalized parameter sets for basic GARCH type intensity model}\label{Table:normalized_estimation1}
$$
\begin{array}{ccccc}
\hline
\delta & 0.01 & 0.005 & 0.002 & 0.001\\
\hline
\omega_{+}^{*} & 1.28\times10^{-6} & 4.20\times10^{-7} & 2.75\times10^{-7} & 3.49\times10^{-6}\\
\beta_{+} & 0.9040 & 0.9342 & 0.9356 & 0.8492\\
\alpha_{+}^{*} & 0.0296 & 0.0293 & 0.0300 & 0.0285\\
\omega_{-}^{*} & 9.19\times10^{-7} & 2.63\times10^{-7} & 2.12\times10^{-7} & 3.29\times10^{-6}\\
\beta_{-} & 0.9381 & 0.9413 & 0.9388 & 0.8563\\
\alpha_{-}^{*} & 0.0247 & 0.0256 & 0.0283 & 0.0267\\
\hline
\end{array}
$$
\end{table}

\subsection{Estimation for GJR GARCH-type intensity model}\label{Sect:Calibration2}

Now, we consider the GJR-type of intensity model to incorporate the leverage effect which cannot be captured in the basic GARCH model.
The original GJR model for volatility can capture the leverage effect as reported in \citet{EngleNg}.
By the parameters $\gamma_{\pm}$, we capture the leverage effect.
Suppose that $\gamma_{+}>0$ and $\gamma_{-}>0$.
If bad news arrives, i.e., $\varepsilon(t_{i-1})<0$, today's volatility increases more than in the case of good news arriving.
The estimates of parameters are presented in Table~\ref{Table:estimation2}.
Observe that $\omega_{+}, \alpha_{+}, \gamma_{+}$ are larger than $\omega_{-}, \alpha_{-}, \gamma_{-}$, respectively,
whereas $\beta_{+}$ is smaller than $\beta_{-}$.

\begin{table}
\caption{Parameter sets for GJR GARCH type intensity model}\label{Table:estimation2}
$$
\begin{array}{cccccc}
\hline
\delta & 0.01 & 0.005 & 0.002 & 0.001 \\
\hline
\omega_{+} & 0.0129 &  0.0210 & 0.0954 & 2.0169\\
           & (0.0023) &  (0.0031) & (0.0132) & (0.4861) \\
\beta_{+}  & 0.9345 &  0.9369 & 0.9382 & 0.9193\\
           & (0.0063) &  (0.0061) & (0.0054) & (0.0143)\\
\alpha_{+} & 1.4324 &  86.99 & 979.0 & 4288.9 \\
           & (13.87) &  (67.90) & (449.3) & (1504)\\
\gamma_{+} & 497.8 &  1899 & 11356 & 18044\\
           & (61.59) &  (175.9) & (1090) & (2755)\\
\omega_{-} & 0.0097 &  0.0167 & 0.0838 & 1.4537\\
           & (0.0022) &  (0.0033) & (0.0153) & (0.3852)\\
\beta_{-}  & 0.9430 &  0.9425 & 0.9405 & 0.9378\\
           & (0.0078) &  (0.0063) & (0.0058) & (0.0112) \\
\alpha_{-} & 0.8528 &  38.23 & 832.2 & 2913.8\\
           & (40.00) &  (72.05) & (499.2) & (1420)\\
\gamma_{-} & 428.8 &  1791.9 & 11114 & 16421\\
           & (87.50) &  (166.8) & (1117) & (2371)\\
\hline
-\textrm{loglikelihood}  & 6992 & 10198 & 14828 & 18733\\
\hline
\end{array}
$$
\end{table}

\begin{figure}
\begin{center}
\subfigure[$\Corr(X_t, X_{t-\ell} | X_t > 0, X_{t-\ell} < 0)$]{
\includegraphics[width=5cm]{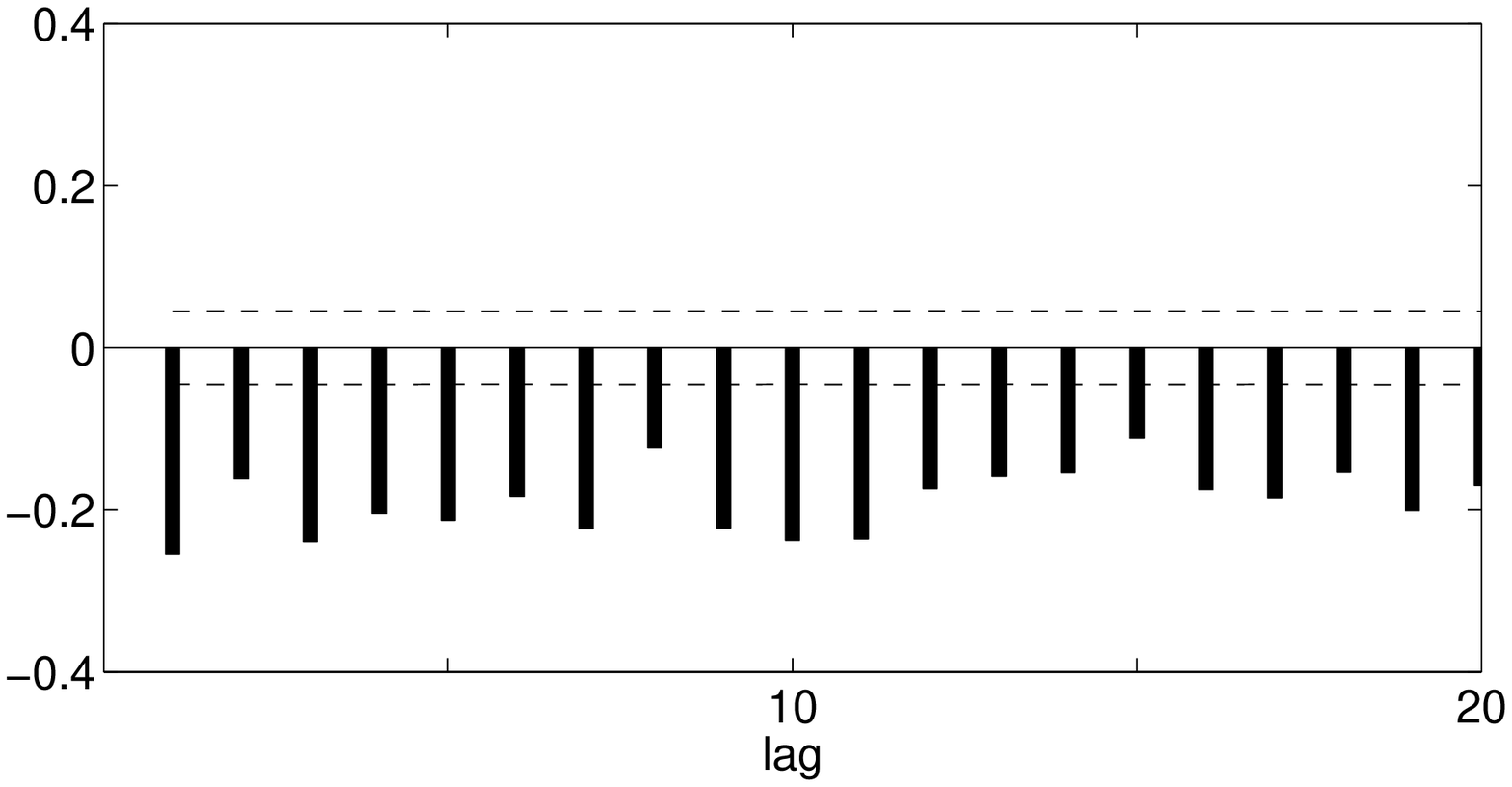}
}
\quad
\subfigure[$\Corr(X_t, X_{t-\ell} | X_t > 0, X_{t-\ell} > 0)$]{
\includegraphics[width=5cm]{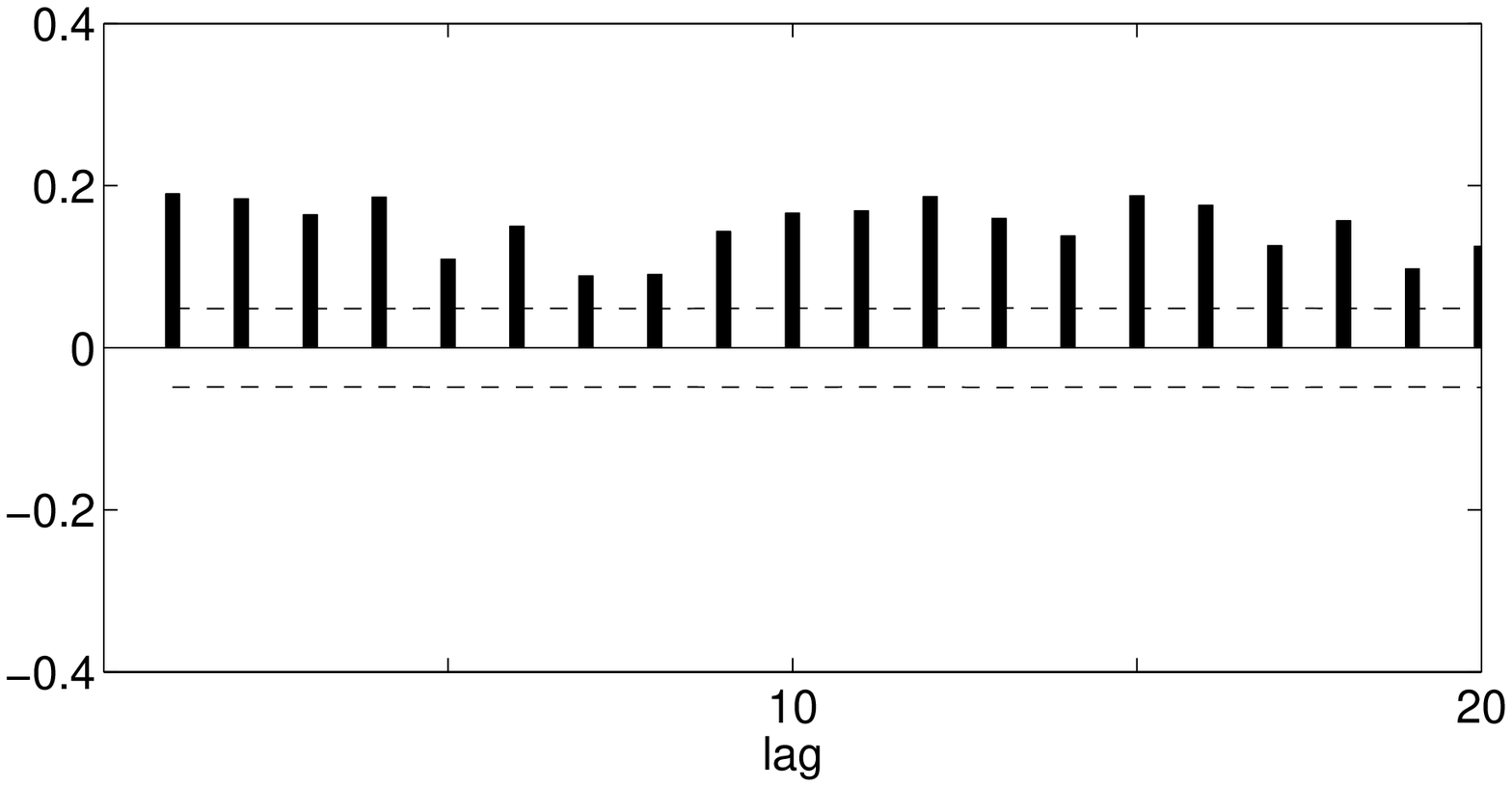}
}
\vspace{0.5cm}
\subfigure[$\Corr(X_t, X_{t-\ell} | X_t < 0, X_{t-\ell} < 0)$]{
\includegraphics[width=5cm]{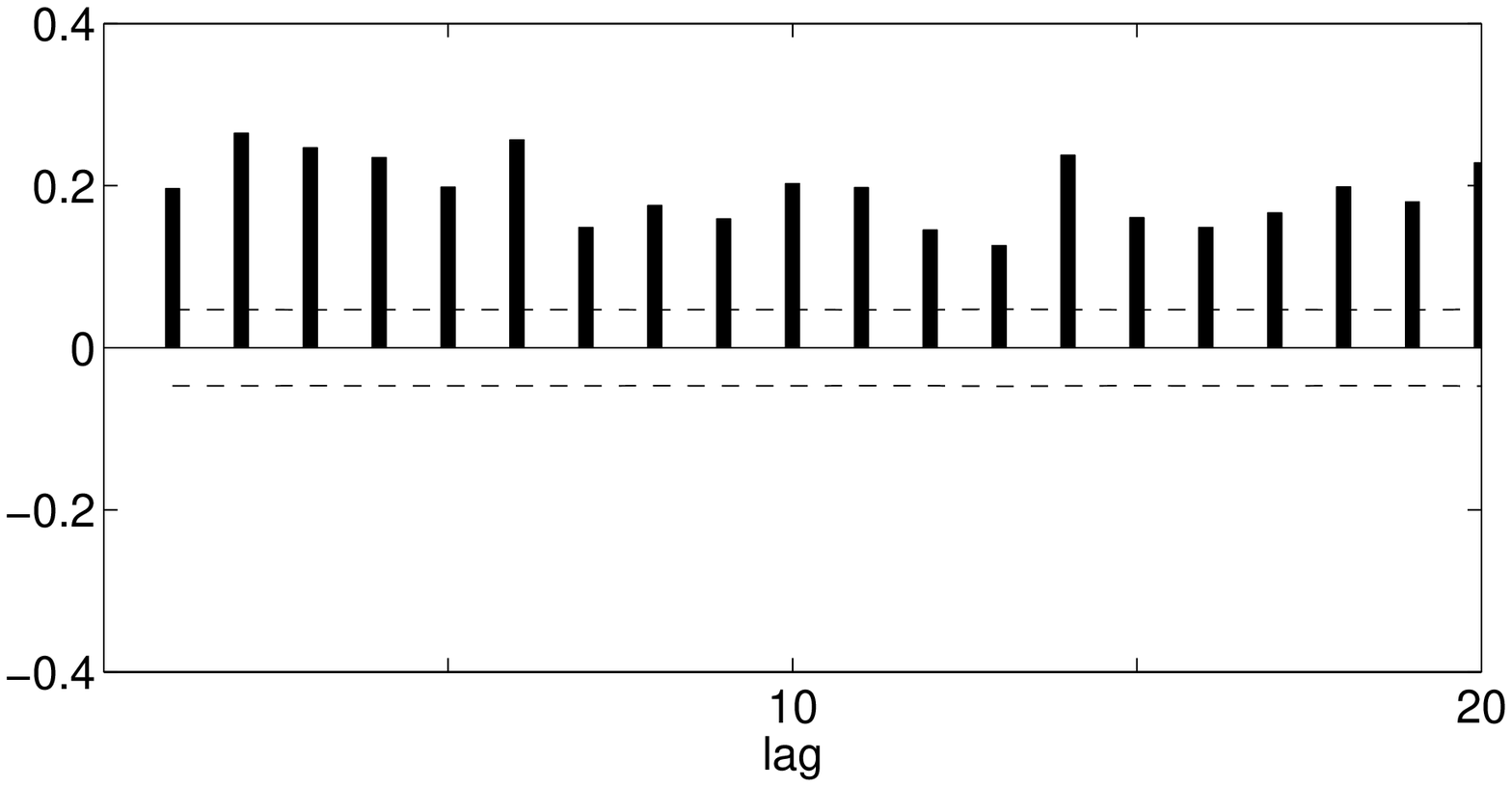}
}
\quad
\subfigure[$\Corr(X_t, X_{t-\ell} | X_t < 0, X_{t-\ell} > 0)$]{
\includegraphics[width=5cm]{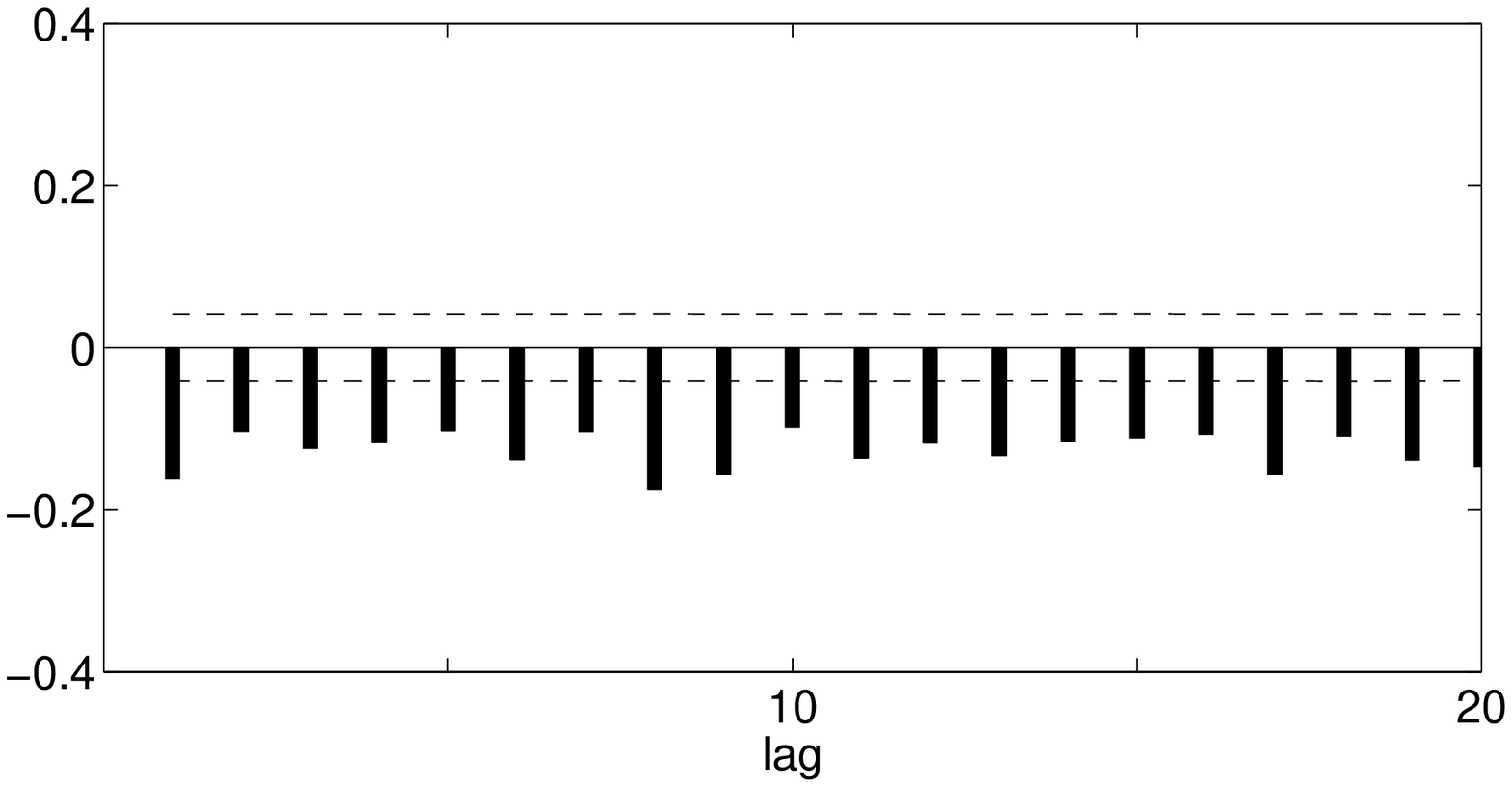}
}
\end{center}
\caption{GJR intensity: conditional correlations depending on current and past returns' signs}\label{Fig:GJR_asymmetry}
\end{figure}

\begin{figure}
\begin{center}
\includegraphics[width=5cm]{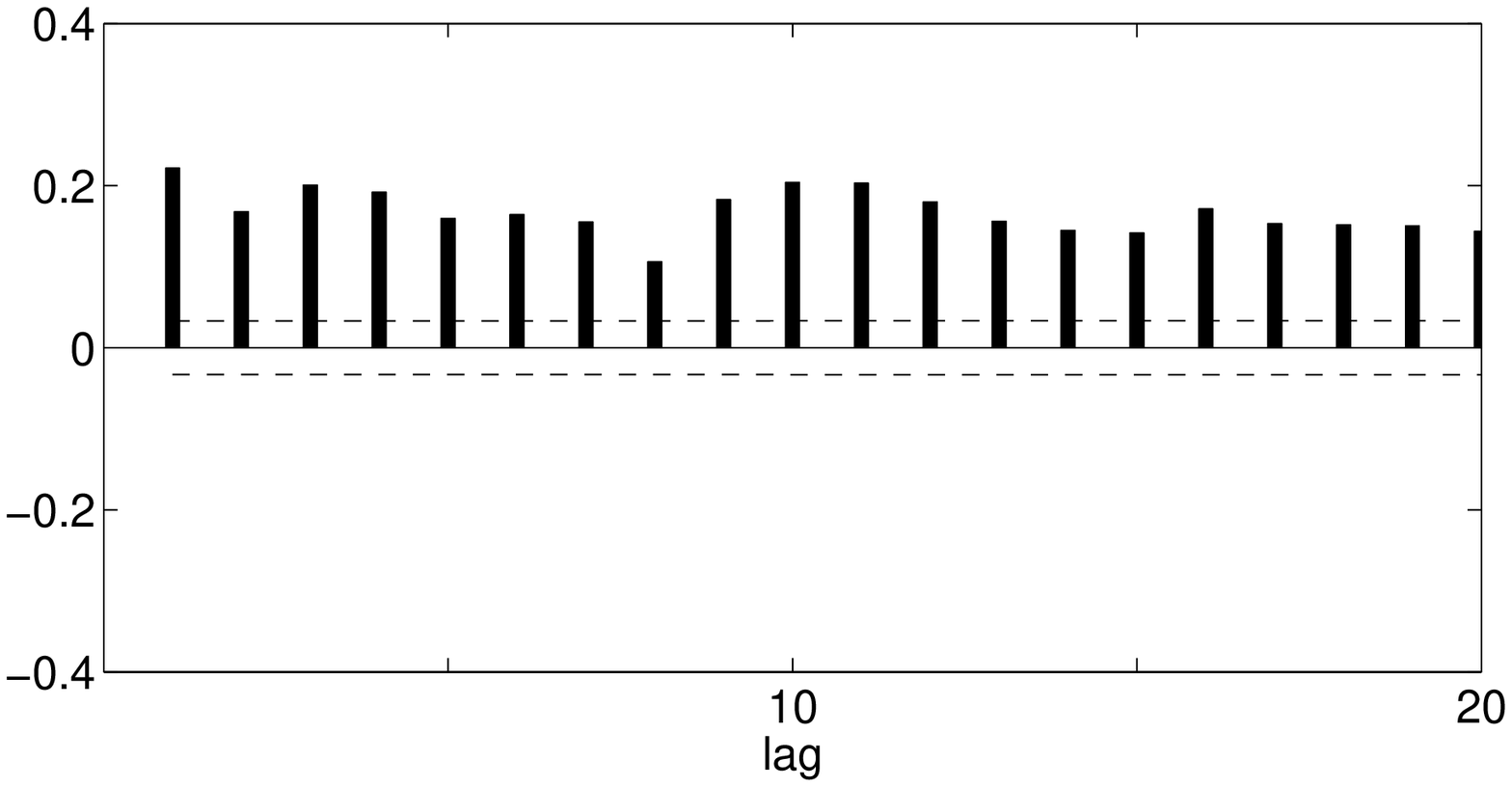}
\quad
\includegraphics[width=5cm]{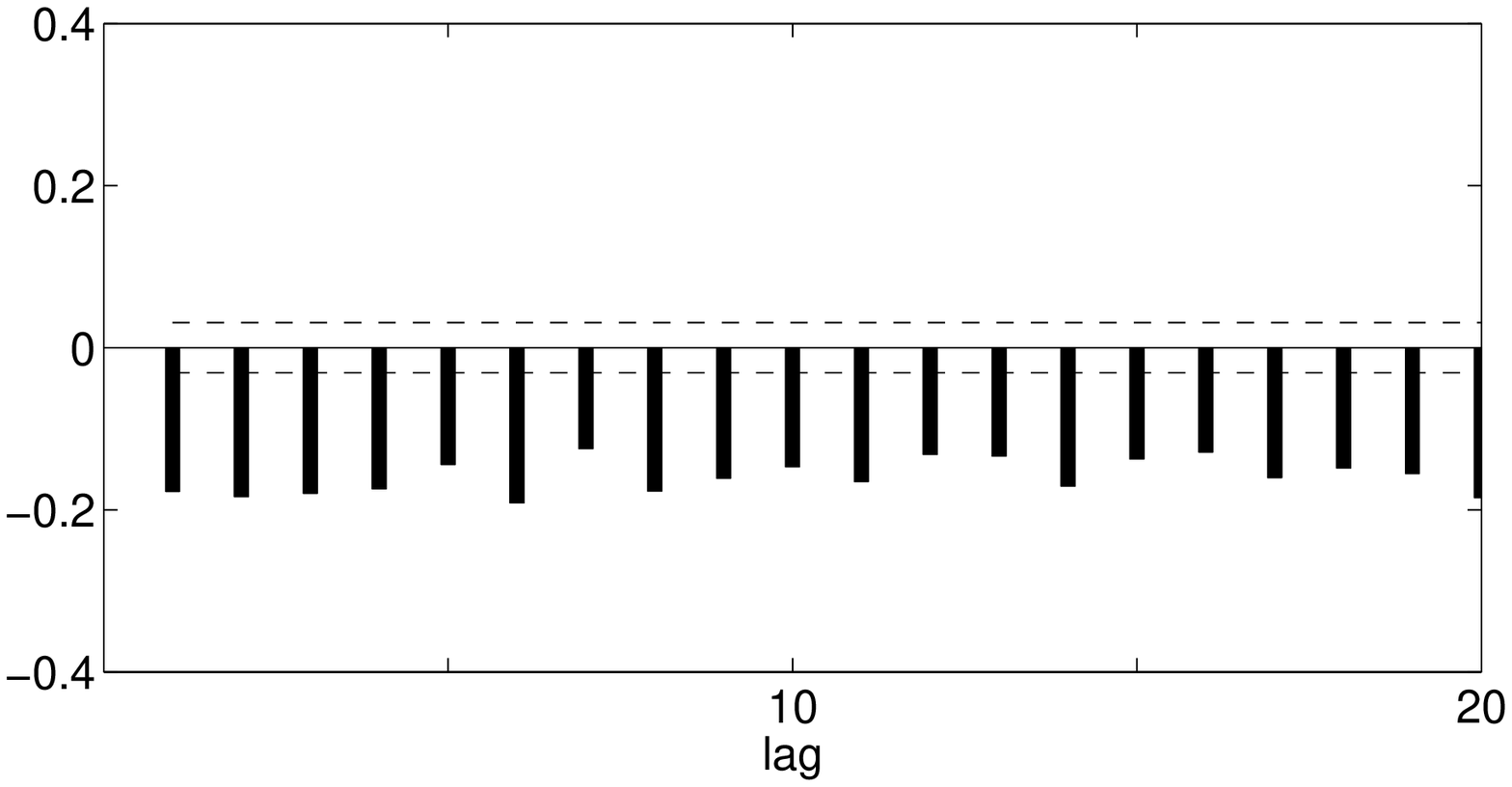}
\end{center}
\caption{GJR intensity:
$\Corr(X_t, |X_{t-\ell}| \,| X_t> 0 )$ and $\Corr(X_t, |X_{t-\ell}| \,| X_t< 0)$ for $\ell\geq 1$ (from left to right).}
\label{Fig:GJR_leverage22}
\end{figure}

Similarly with the previous cases, we observe $\alpha_{+}  > \alpha_{-}$ and $\gamma_{+} > \gamma_{-}$
which imply that down movement is less affected by the previous information of volatility than up movement.
We also plot conditional correlograms in Figures~\ref{Fig:GJR_asymmetry} and \ref{Fig:GJR_leverage22}.
Compare the figures with Figures~\ref{Fig:asymmetry} and \ref{Fig:leverage22}, respectively.
Autocorrelations for the volatility clustering and the leverage effect are plotted in Appendix~\ref{Append:GJR}.
We compare the the modified Ljung-Box test statistics of simulated GJR GARCH intensity model with the statistics of the S\&P 500 in Table~\ref{Table:LjungBox}.
When $X_t<0$, the statistics are smaller than when $X_t<0$, representing the conditional asymmetry.

\begin{table}
\caption{Normalized parameter sets for GJR GARCH type intensity model}\label{Table:normalized_estimation2}
$$
\begin{array}{cccccc}
\hline
\delta & 0.01 & 0.005 & 0.002 & 0.001 \\
\hline
\omega^*_{+} & 1.29\times10^{-6} & 5.24\times10^{-7}  & 3.82\times10^{-7} & 2.02\times10^{-6}  \\
\beta^*_{+}  & 0.9345 & 0.9369 & 0.9382 & 0.9193\\
\alpha^*_{+} & 0.0001 & 0.0022 & 0.0039 & 0.0043\\
\gamma^*_{+} & 0.0498 & 0.0475 & 0.0454 & 0.0180\\
\omega^*_{-} & 9.73\times10^{-7}  & 4.17\times10^{-7}  & 3.35\times10^{-7}  & 1.45\times10^{-6} \\
\beta^*_{-}  & 0.9430 & 0.9425 & 0.9405 & 0.9378\\
\alpha^*_{-} & 0.0001 & 0.0010 & 0.0033 & 0.0029\\
\gamma^*_{-} & 0.0429 & 0.0448 & 0.0445 & 0.0164\\
\hline
\end{array}
$$
\end{table}

\begin{table}
\centering
\caption{Comparison of modified Ljung-Box test}\label{Table:LjungBox}
\begin{tabular}{lll}
\toprule
Conditional correlation & S\&P 500 & GJR intensity\\
 & $Q_{20}$ & $Q_{20}$ \\
\midrule
$\Corr(X_t, |X_{t-\ell}| \; | X_t >0)$ & 4139.6 & 4236.9 \\
$\Corr(X_t, |X_{t-\ell}| \; | X_t <0)$ & 2329.7 & 3399.0 \\
\midrule
$\Corr(X_t, X_{t-\ell} | X_t>0, X_{t-\ell} >0)$ & 1993.0 & 1899.8 \\
$\Corr(X_t, X_{t-\ell} | X_t > 0, X_{t-\ell}<0)$ & 2239.3 & 2527.7 \\
$\Corr(X_t, X_{t-\ell} | X_t<0, X_{t-\ell} <0)$ & 1470.5 & 1904.2 \\
$\Corr(X_t, X_{t-\ell} | X_t<0, X_{t-\ell}>0)$ & 968.0 & 1537.4 \\
\bottomrule
\end{tabular}
\end{table}

Now we compare the likelihoods of various models:\\
(i) Model I : $\lambda_{\pm}(t_i) = \omega_{\pm}  + \beta\lambda_{\pm}(t_{i-1}) + \alpha\varepsilon^2(t_i) $\\
(ii) Model II : $\lambda_{\pm}(t_i) = \omega_{\pm} + \beta\lambda_{\pm}(t_{i-1}) + (\alpha + \gamma I(t_i))\varepsilon^2(t_i) $\\
(iii) Model III : $\lambda_{\pm}(t_i) = \omega_{\pm} + \beta_{\pm}\lambda_{\pm}(t_{i-1}) + \alpha_{\pm}\varepsilon^2(t_i)$\\
(iv) Model IV : $\lambda_{\pm}(t_i) = \omega_{\pm} + \beta_{\pm}\lambda_{\pm}(t_{i-1}) + (\alpha_{\pm} + \gamma_{\pm} I(t_i))\varepsilon^2(t_i) $\\
where $\alpha:=\alpha_+=\alpha_-$ and $\gamma:=\gamma_+=\gamma_-$.
Model I is the simplest one and similar to the original GARCH.
Model II is an improved version to incorporate the leverage effect.
Model III is for the conditional asymmetry.
By Model IV, we capture both the leverage effect and the conditional asymmetry.

In Table~\ref{Table:loglikelihood} we compare the minus loglikelihood of the models with various jump size.
In Model IV, we have the smallest minus loglikelihood for all jump sizes.
Note that, in terms of the likelihood ratio we have more improvement when we adopt Model II (to take into account the leverage effect) than
when we adopt Model III (for the conditional asymmetry).

\begin{table}
\centering
\caption{Comparison of minus loglikelihood}\label{Table:loglikelihood}
\begin{tabular}{ccccc}
\hline
delta & 0.01 & 0.005 & 0.002 & 0.001\\
\hline
Model I & 7038 & 10263 & 14897 & 19023\\
Model II & 6994 & 10201 & 14834 & 18788\\
Model III & 7035 & 10258 & 14891 & 18990\\
Model IV & 6992 & 10198 & 14828 & 18733 \\
\hline
\end{tabular}
\end{table}

\subsection{Comparison with usual GARCH}\label{Subsect:comparison}
In this subsection, we compare our model with the usual GJR GRACH model.
Using the same data of the S\&P 500 index in the previous subsections, the estimates of the autoregressive moving average (ARMA) (1,1)-GJR GARCH are
\begin{align*}
X(t_i) &= 2.42 \times 10^{-4} - 0.0183X(t_{i-1}) + \varepsilon(t_i) + 0.0142\varepsilon(t_{i-1})\\
\sigma^2(t_i) &= 8.82 \times 10^{-7} + 0.9412 \sigma^2(t_{i-1}) + (0.0014 + 0.0963I(t_{i-1})) \varepsilon^2(t_{i-1})
\end{align*}
where the innovations are assumed to follow the normal distribution and $\sigma(t_i)$ denotes the conditional volatility of $\varepsilon(t_i)$.
The estimates of the GJR GARCH variance model are similar with the normalized estimates of the intensity models in Table~\ref{Table:normalized_estimation2}.

\begin{figure}
\begin{center}
\subfigure[GJR intensity model with $\delta = 0.01$]{
\includegraphics[width=6cm]{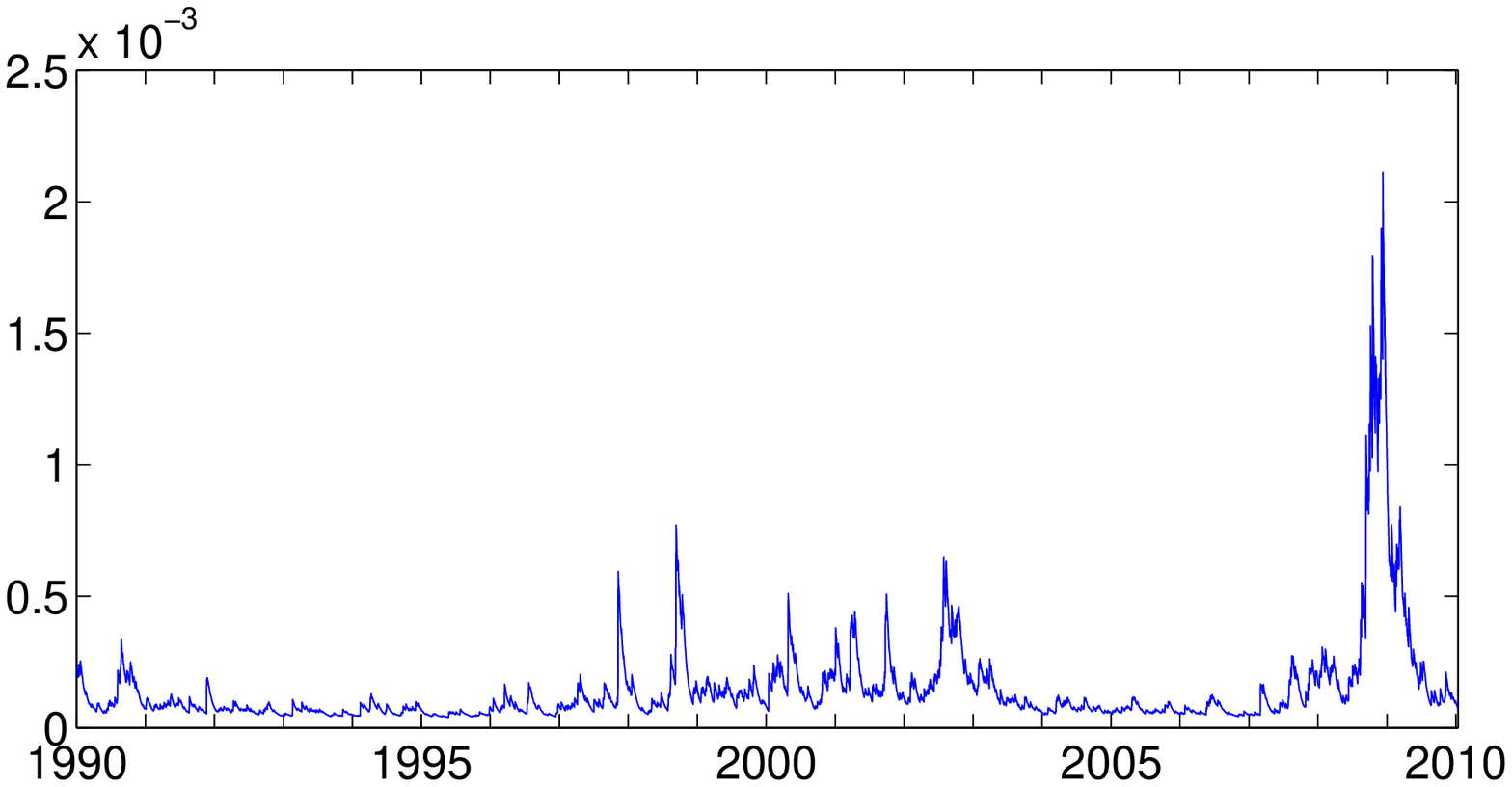}
}
\quad
\subfigure[GJR intensity model with $\delta = 0.005$]{
\includegraphics[width=6cm]{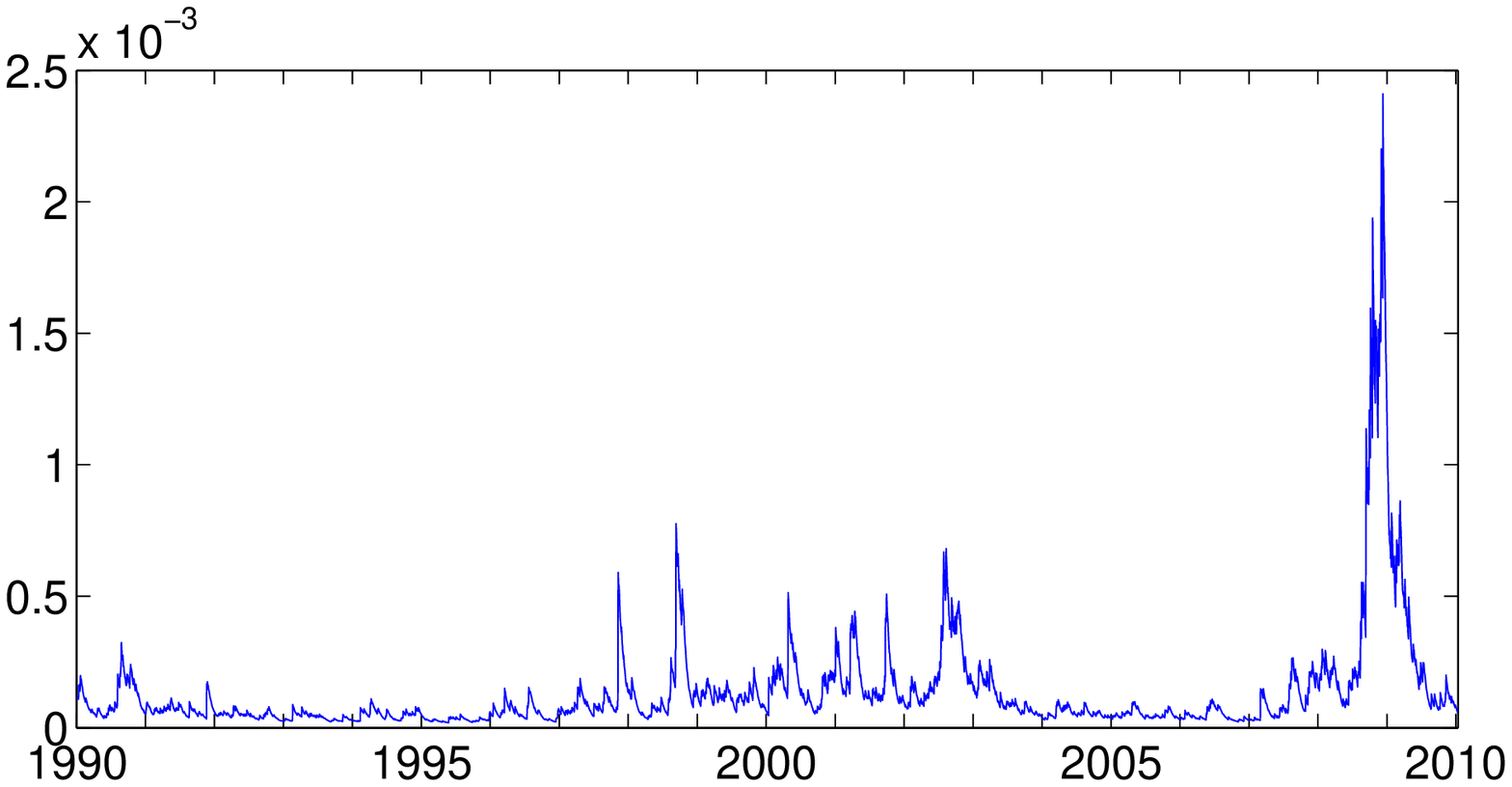}
}
\vspace{0.5cm}
\subfigure[GJR intensity model with $\delta = 0.002$]{
\includegraphics[width=6cm]{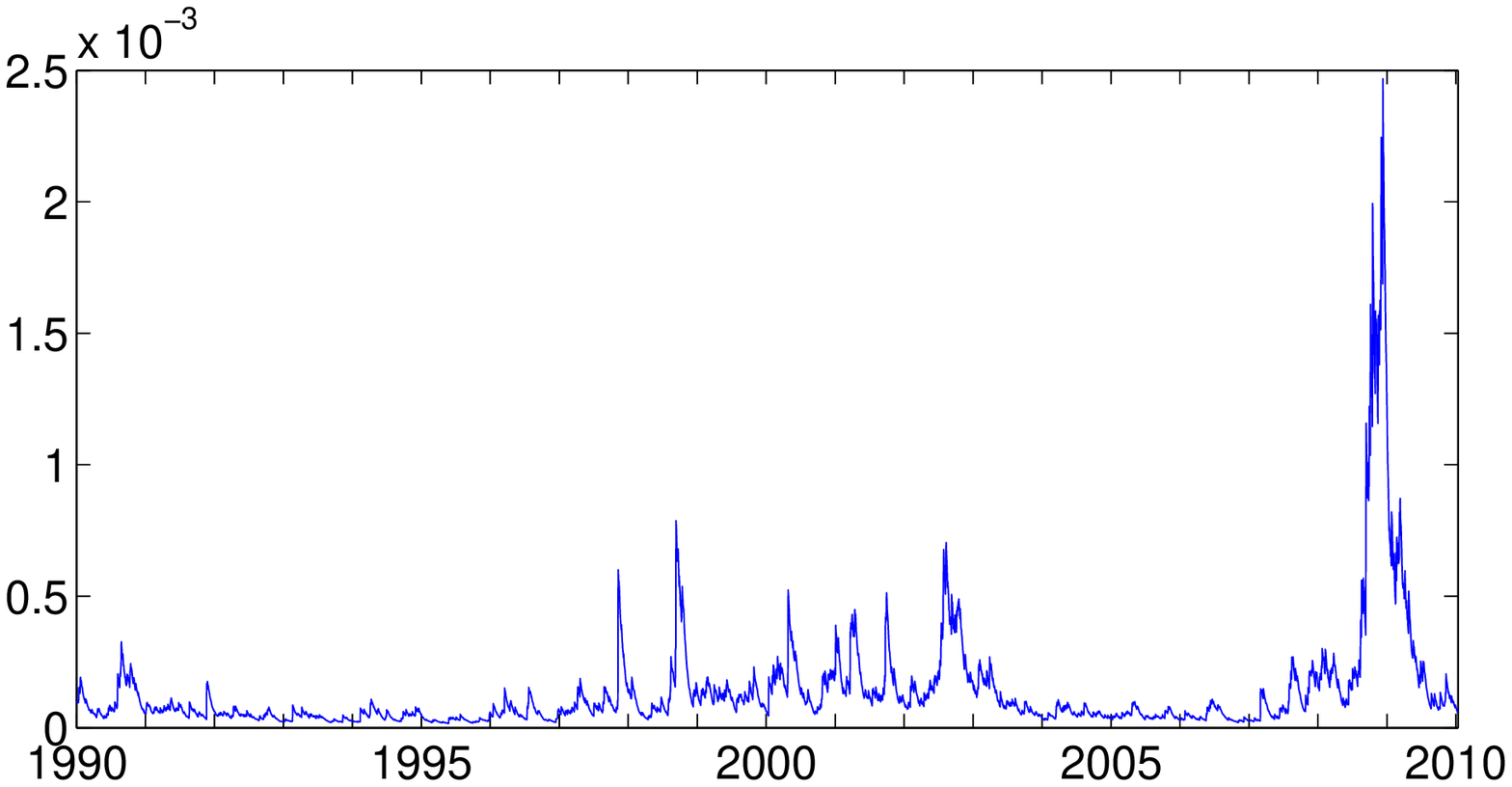}
}
\quad
\subfigure[Original GJR GARCH]{
\includegraphics[width=6cm]{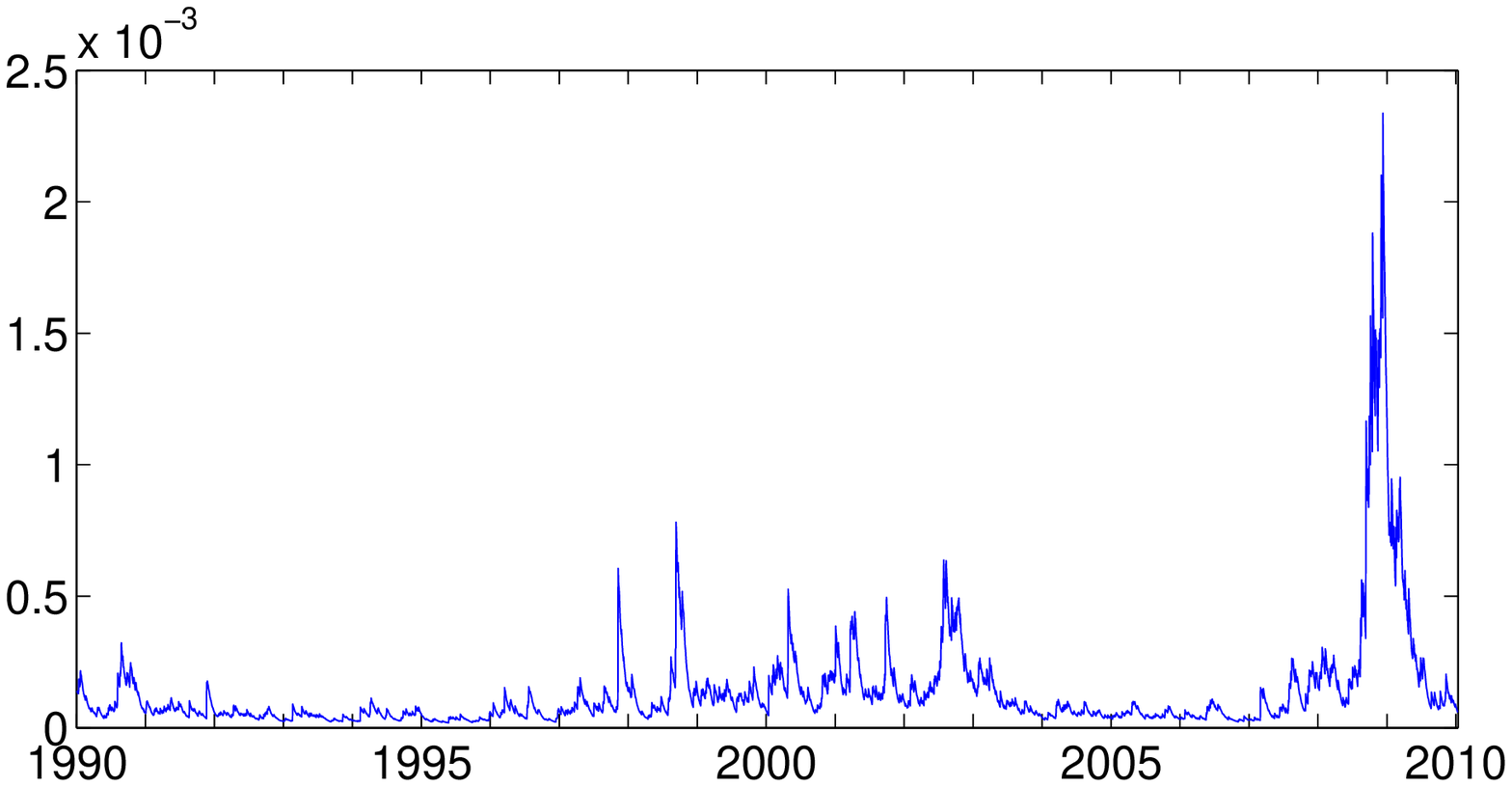}
}
\end{center}
\caption{Inferred conditional variance of S\&P 500 return series}\label{Fig:CV}
\end{figure}

The dynamics of the inferred conditional variances, $\Var[X(t_i) | \mathcal F(t_{i-1})]$, of the intensity models and the usual GARCH are plotted in Figure~\ref{Fig:CV}.
The jump sizes of the intensity model are $0.01$, $0.005$ and $0.002$.
Although the jump size $\delta$ varies, the behaviors of the conditional variances are similar.
During the periods of the dot-com bubble and after the financial crisis of 2008, we observe high volatilities.

\begin{figure}
\begin{center}
\subfigure[GJR intensity model with $\delta = 0.01$]{
\includegraphics[width=6cm]{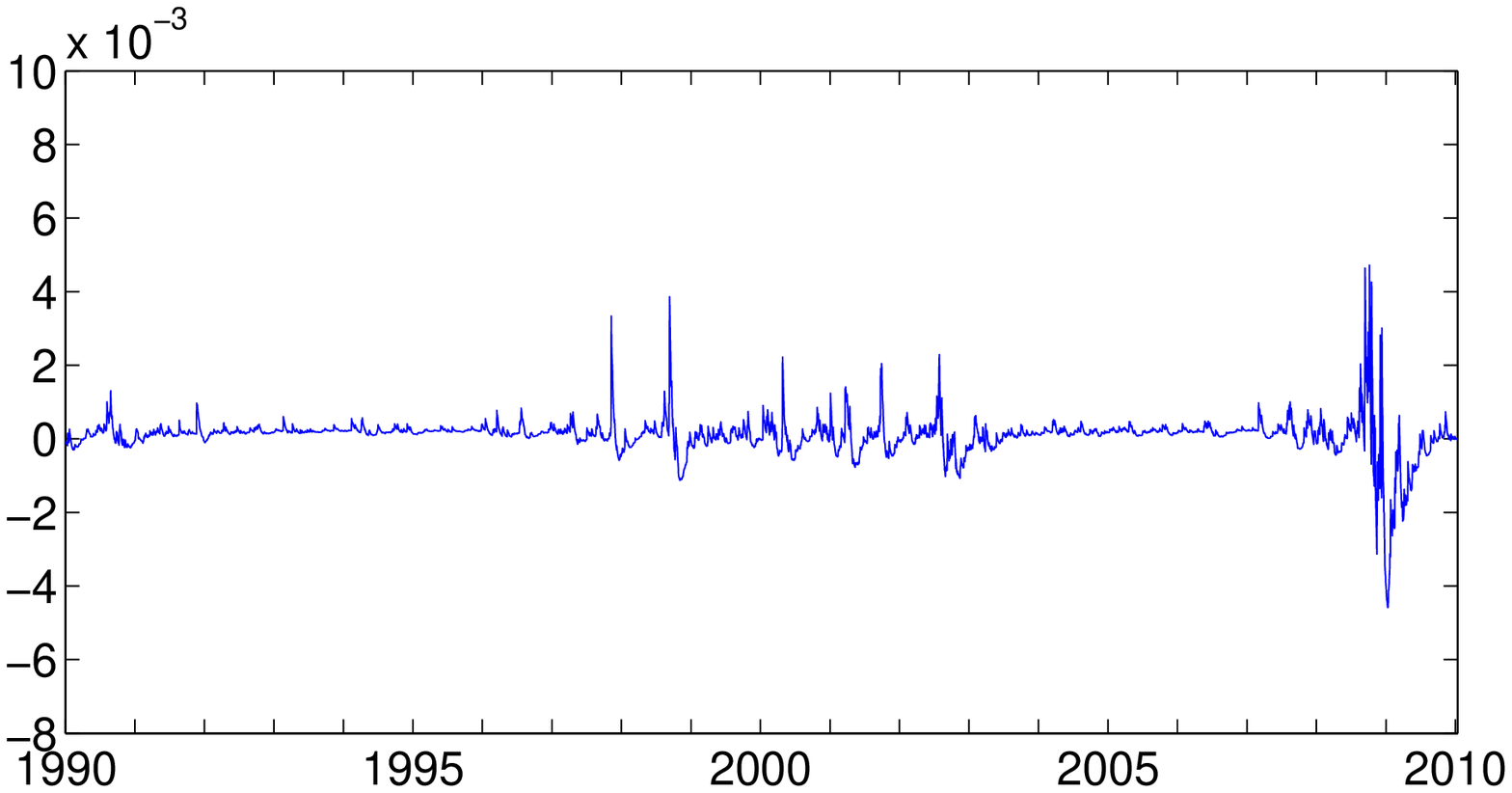}
}
\quad
\subfigure[GJR intensity model with $\delta = 0.005$]{
\includegraphics[width=6cm]{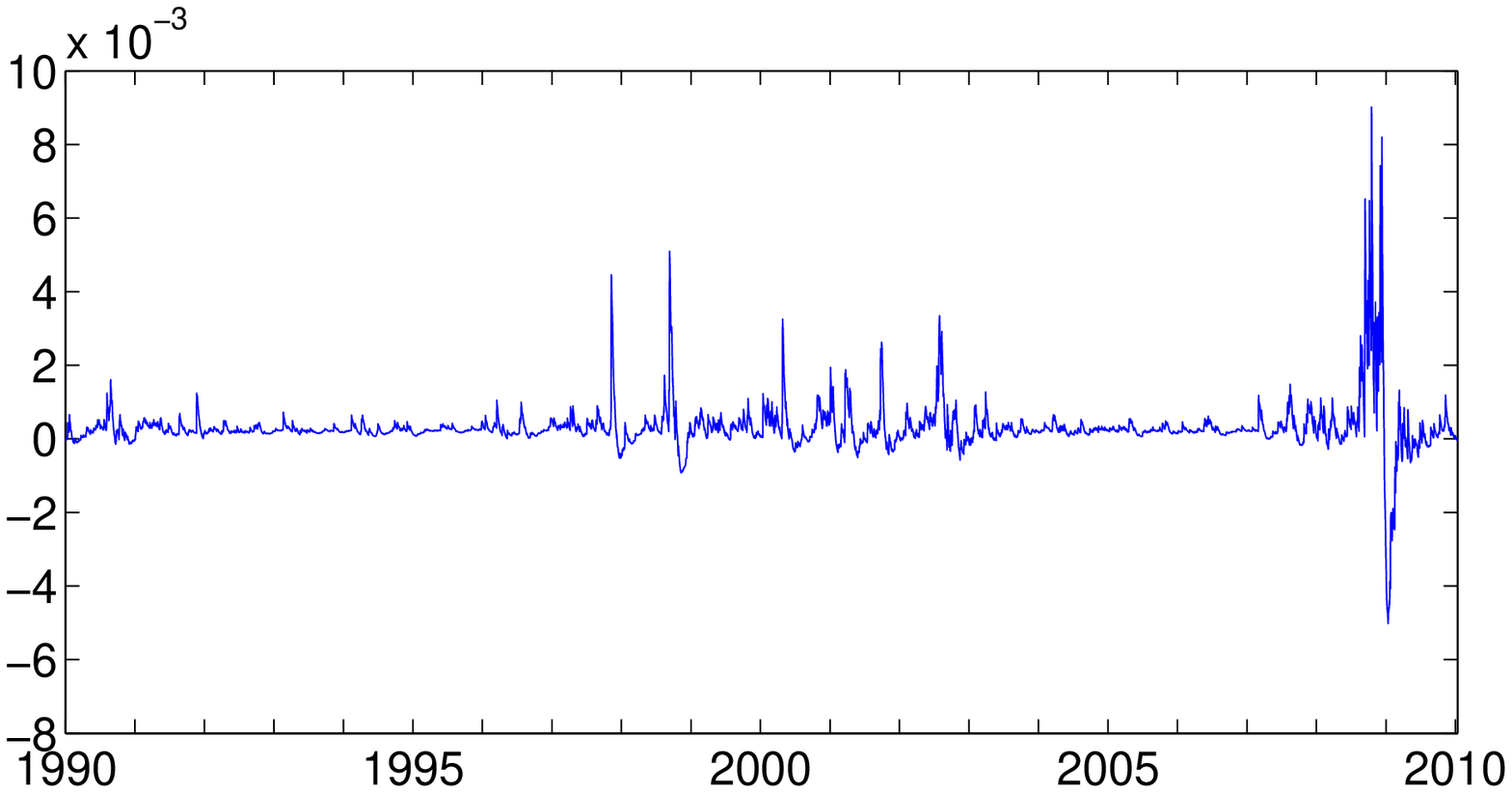}
}
\vspace{0.5cm}
\subfigure[GJR intensity model with $\delta = 0.002$]{
\includegraphics[width=6cm]{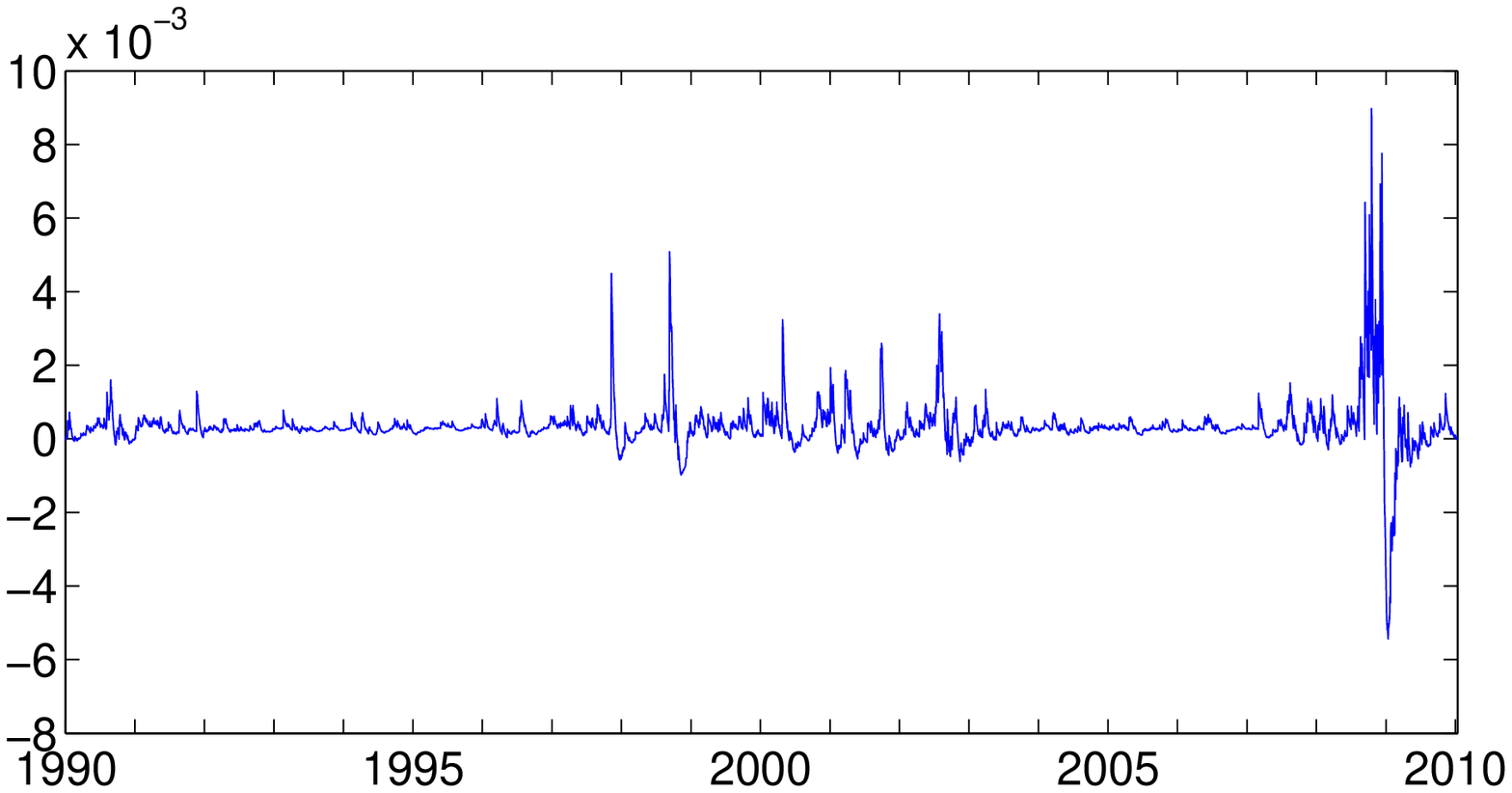}
}
\quad
\subfigure[Original GJR GARCH]{
\includegraphics[width=6cm]{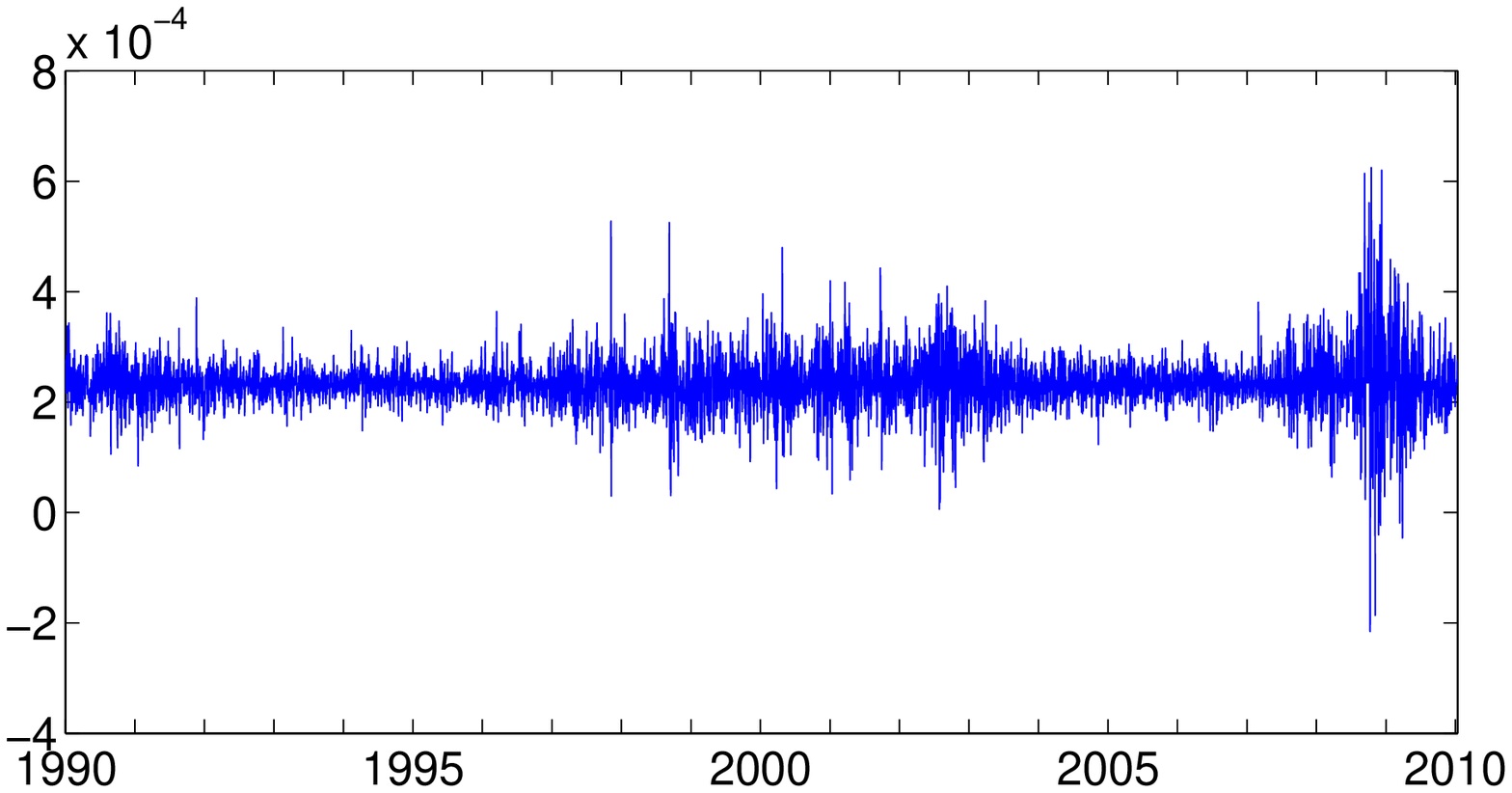}
}
\end{center}
\caption{Inferred conditional mean of S\&P 500 return series}\label{Fig:CM}
\end{figure}

The dynamics of the conditional means of the intensity models are different from the dynamics of the conditional mean of the usual ARMA(1,1)-GJR GARCH as plotted in Figure~\ref{Fig:CM}.
In the conditional mean structure of the usual ARMA(1,1)-GJR GARCH, the absolute value of the autoregressive parameter of AR(1)$=-0.0183$ is relatively close to zero and the conditional mean series is more like white noise.
Note that if the parameter of AR(1) approaches to 1, then the time series has a strong autoregressive property, since the current value is largely affected by the previous terms relative to the shocks.
In contrast with the usual GARCH, the conditional mean processes of the intensity models have stronger autoregressive structures,
as we derive the recursive formula for the conditional mean process in Eq.~\eqref{Eq:mean-intensity} when $\beta_+ = \beta_-$ and by the empirical studies with the fact that $\beta_{\pm}$ are close to 1.
In Figure~\ref{Fig:CM}, when $\delta=0.002$ and $0.005$, the dynamics of the conditional means are very close to each other and $\delta=0.01$ the values are relatively small in magnitude.

Using the inferred conditional mean in the intensity model, investors may get an improved idea about current risk by predicting the future conditional mean of return.
As shown in the figure, the intensity models have more predicting power than the usual GARCH model where the conditional mean is more like white noise.
In the intensity model, we model the up and down movements intensities separately, where each intensity process is predictable because of its clustering property.
In addition, by the conditional asymmetry, up and down intensities are different functions of past information from each other, and the conditional mean, the difference of the intensities has predictability, even though the predicting power is less than the volatility process.
More exact modeling for the intensities beyond the GARCH type modeling would improve the forecasting ability of the conditional variance as well as the conditional mean.

\subsection{Discussion about $\delta$}

We have explained that $\delta$ is rather exogenously chosen, since the estimates of the parameters and the behaviors of the inferred conditional variance and mean are consistent across various $\delta$.
This is based on the perspective that $\delta$ is a unit size of measurement of the changes in asset price to be counted.
One can estimate $\delta$ in the sense of the method of moments as explained in Eq.~\eqref{Eq:variace}, by matching the theoretical second moments and the sample moments.
However, the exact form of the theoretical second moments are only available in some specific kinds of intensity models, and the expected computational complexity is high in order to find the best estimates of $\delta$.
Our suggestion based on the heuristic methods is to assume that $\delta$ is around 0.005, implying that we count the economic events that change by $0.5\%$ in the asset return.
The reasons are that the analysis based on the second moment condition is satisfactory and the informal arguments of the behaviors of the conditional mean and variance are plausible compared with the benchmark GARCH model.

\section{Conclusion}\label{Sect:Conclusion}
The conditional asymmetry is the asymmetric relation of current return to past information depending on the current return's sign.
We introduced a new approach based on the Poisson intensity model for asset price movements
to incorporate the asymmetry as well as well-known properties
such as the leverage effect and the volatility clustering.
In our model the frequencies of up and down movements of returns are represented using two separate stochastic intensities.
To model the intensities we employ the GARCH-type models to capture the asymmetric effects of past shocks
to current return as well as time-varying volatility.
We tested various kinds of intensity models and had a consistent results in which there is the conditional asymmetry in the S\&P 500 return series.
By enabling the capture of conditional asymmetry, the intensity models are slightly improved in terms of maximum likelihood. 

\appendix\label{Appendix}

\section{Correlograms for simulation}
\subsection{GARCH intensity}\label{Append:GARCH}
\begin{figure}
\begin{center}
\includegraphics[width=5cm]{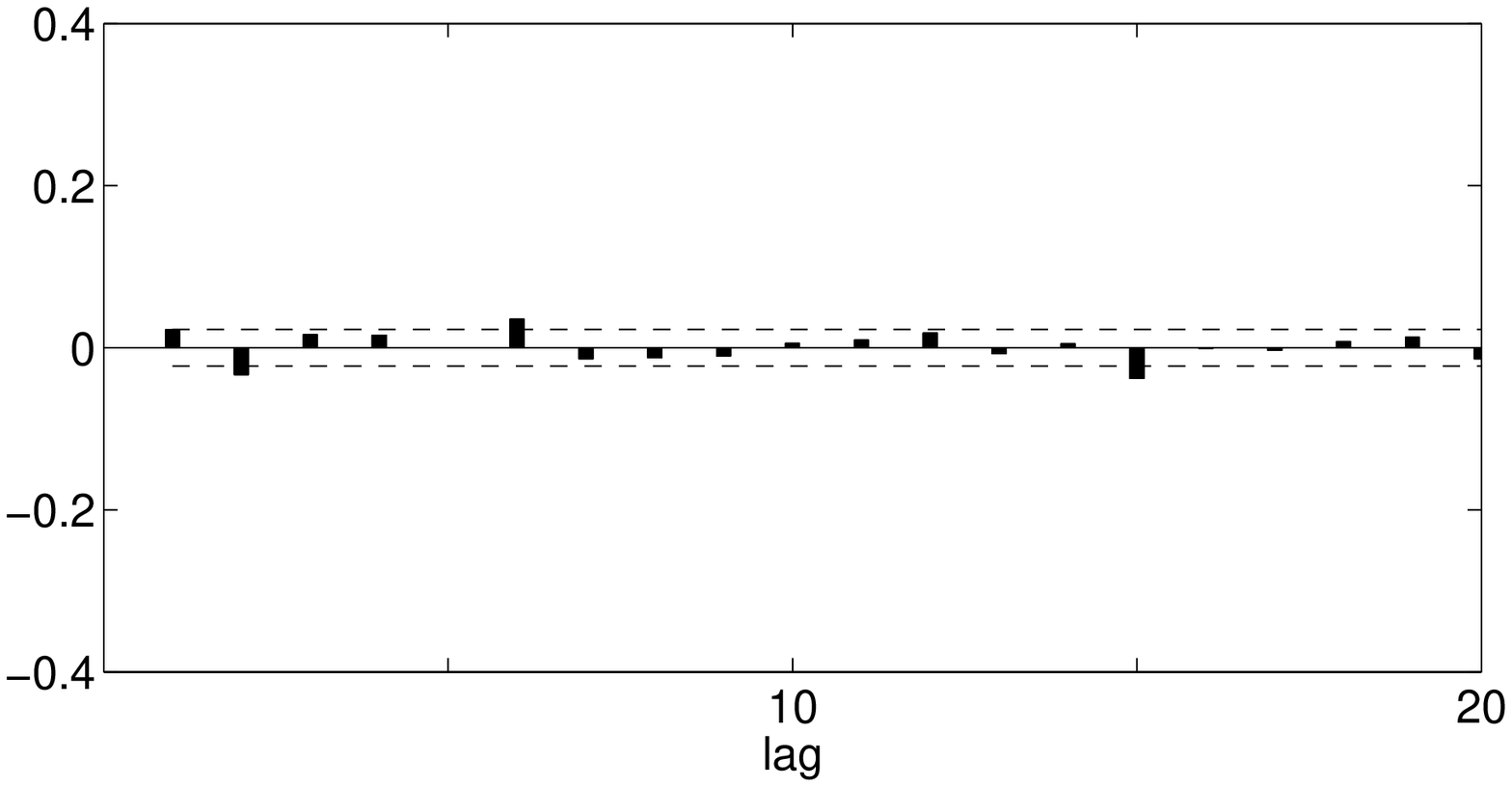}
\quad
\includegraphics[width=5cm]{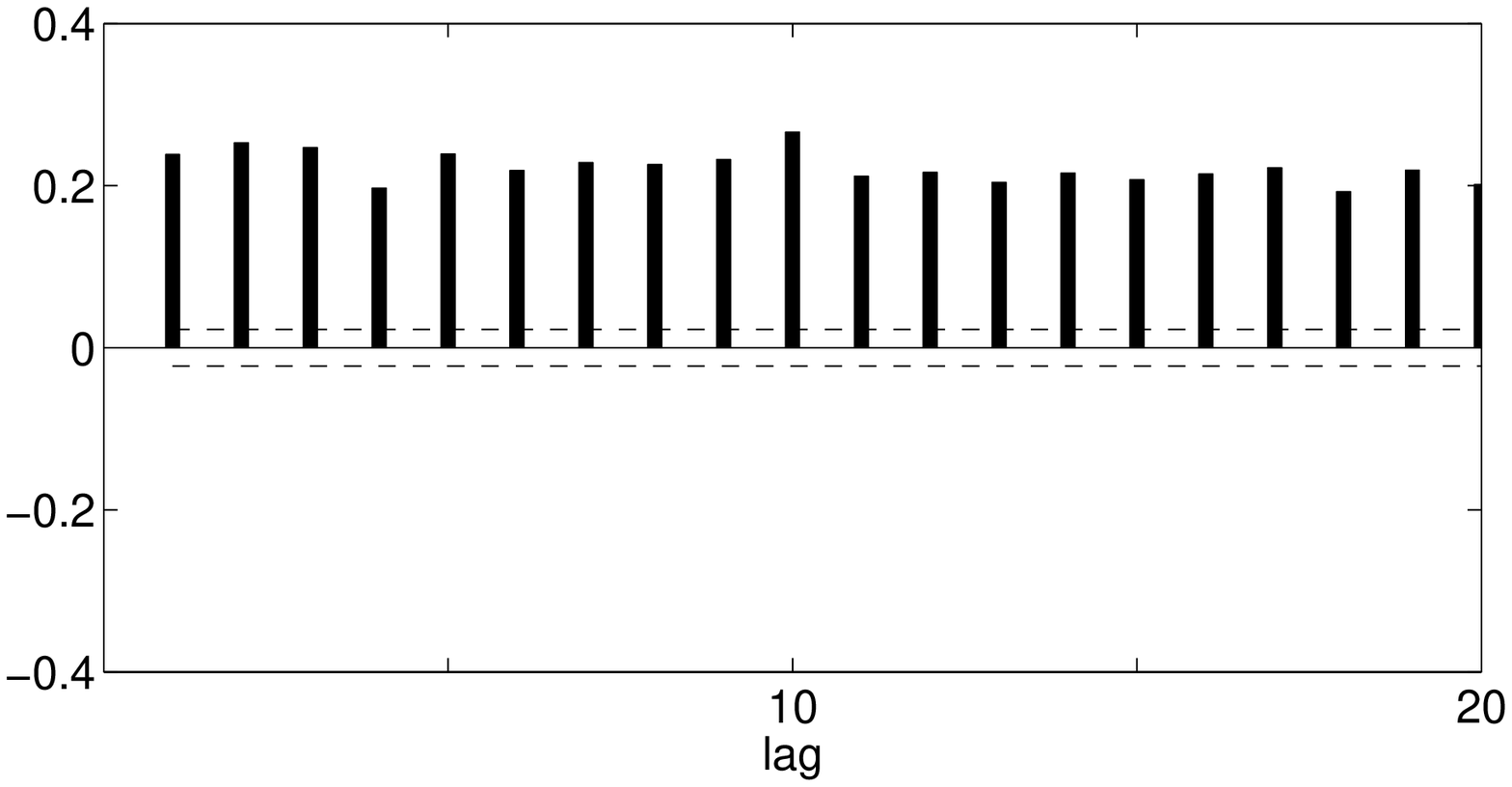}
\end{center}
\caption{Basic GARCH intensity: $\Corr(X_t, X_{t-\ell})$ and $\Corr(|X_t|, |X_{t-\ell}|)$ for $\ell \geq 1$ (from left to right).
Volatility clustering is observed in the right.}
\label{Fig:GARCH_ACF}
\end{figure}

\begin{figure}
\begin{center}
\includegraphics[width=5cm]{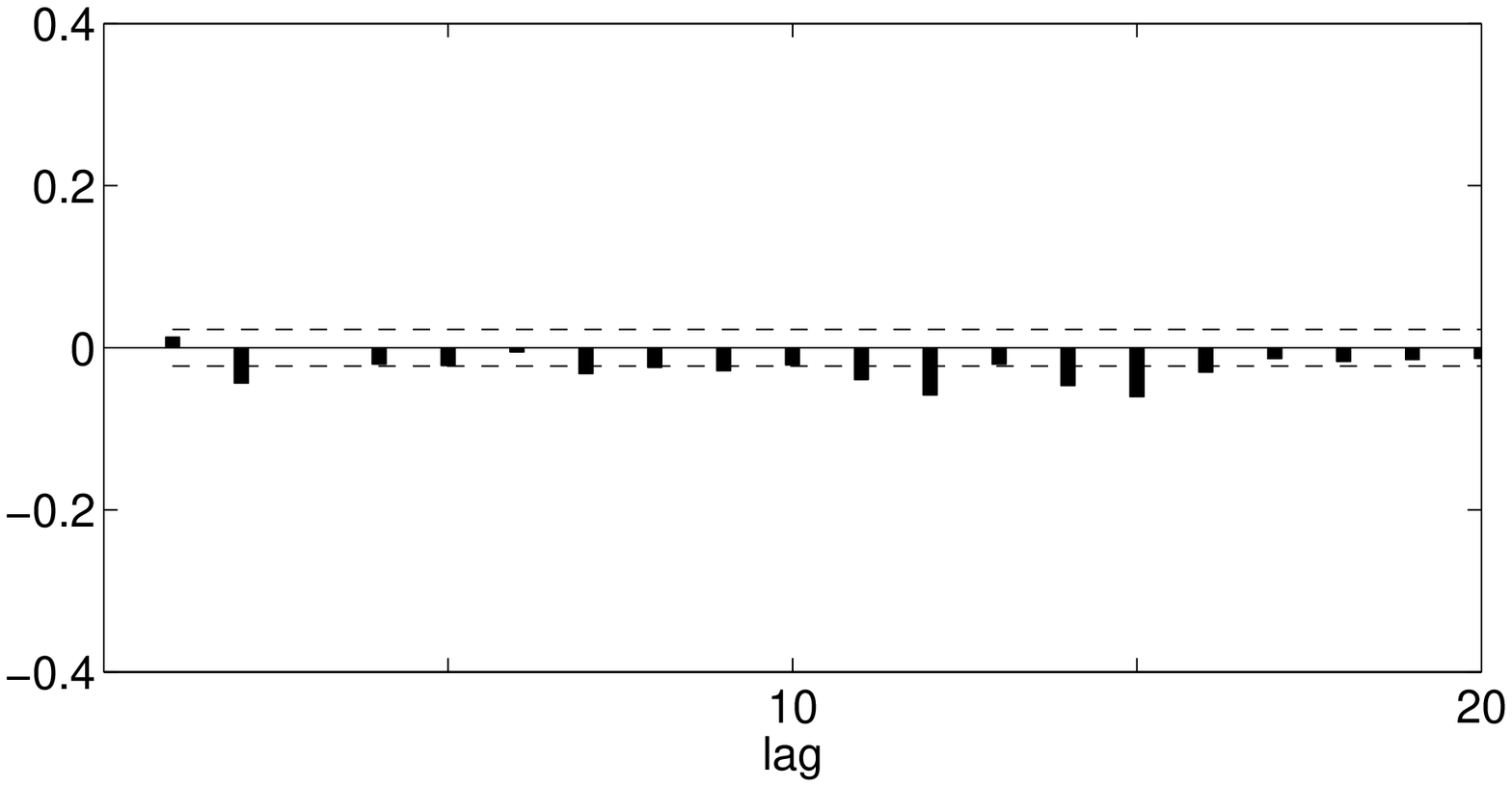}
\quad
\includegraphics[width=5cm]{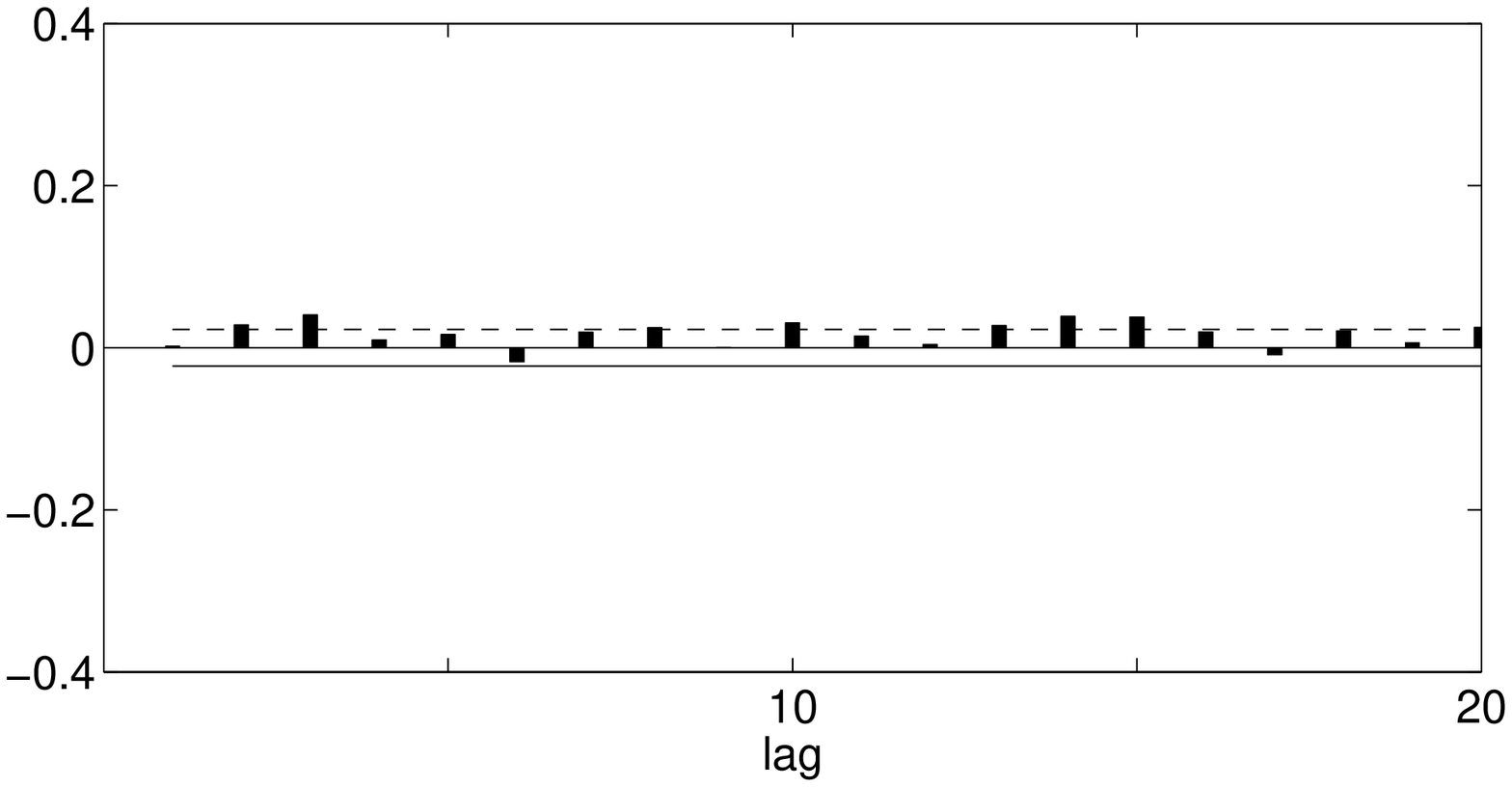}
\end{center}
\caption{Basic GARCH intensity: $\Corr(|X_t|, X_{t-\ell})$ and $\Corr(X_t, |X_{t-\ell}|)$  for $\ell \geq 1$ (from left to right).
The leverage effect is not observed in the left.}
\label{Fig:GARCH_leverage1}
\end{figure}

\begin{figure}
\begin{center}
\includegraphics[width=5cm]{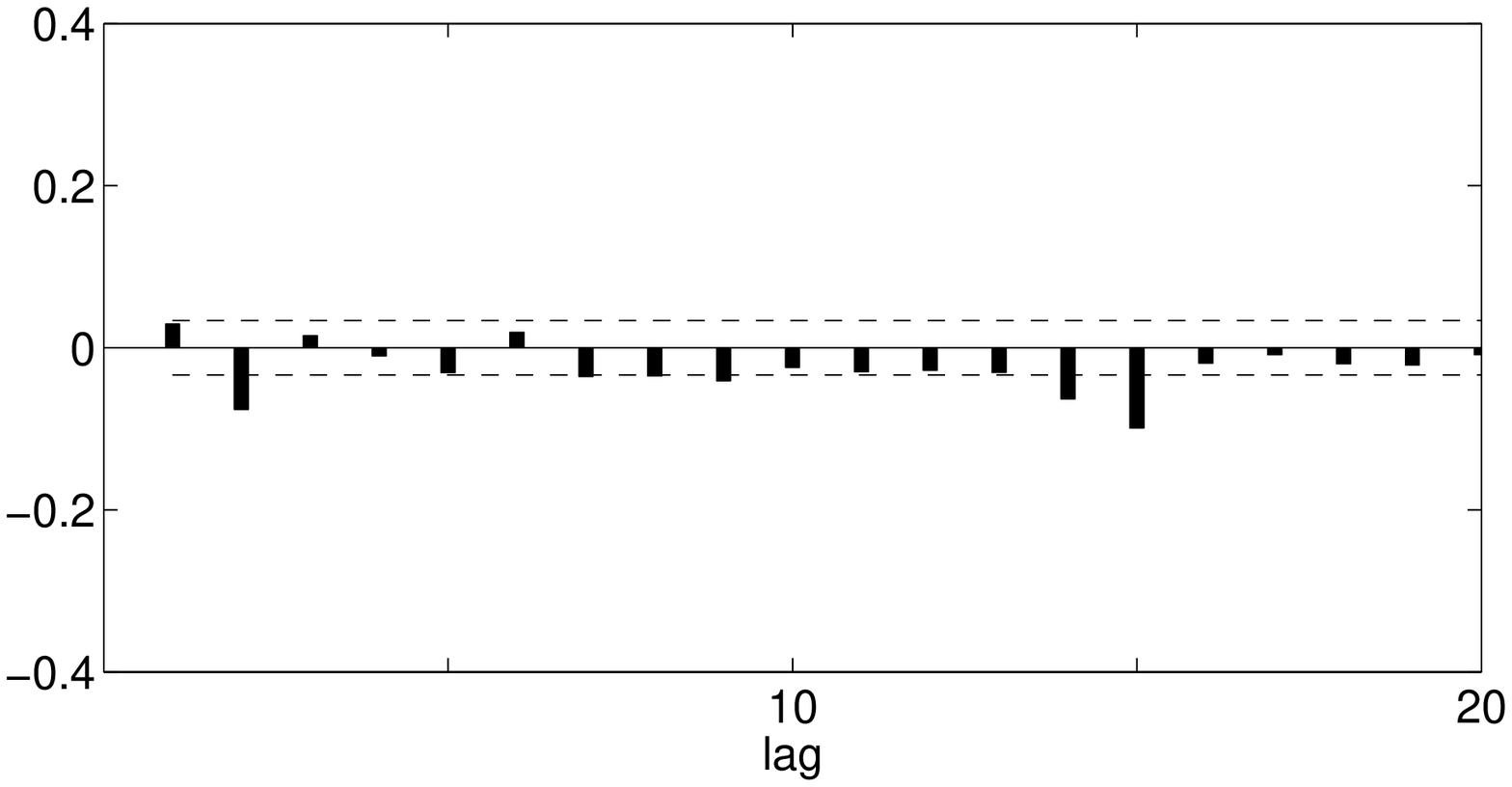}
\quad
\includegraphics[width=5cm]{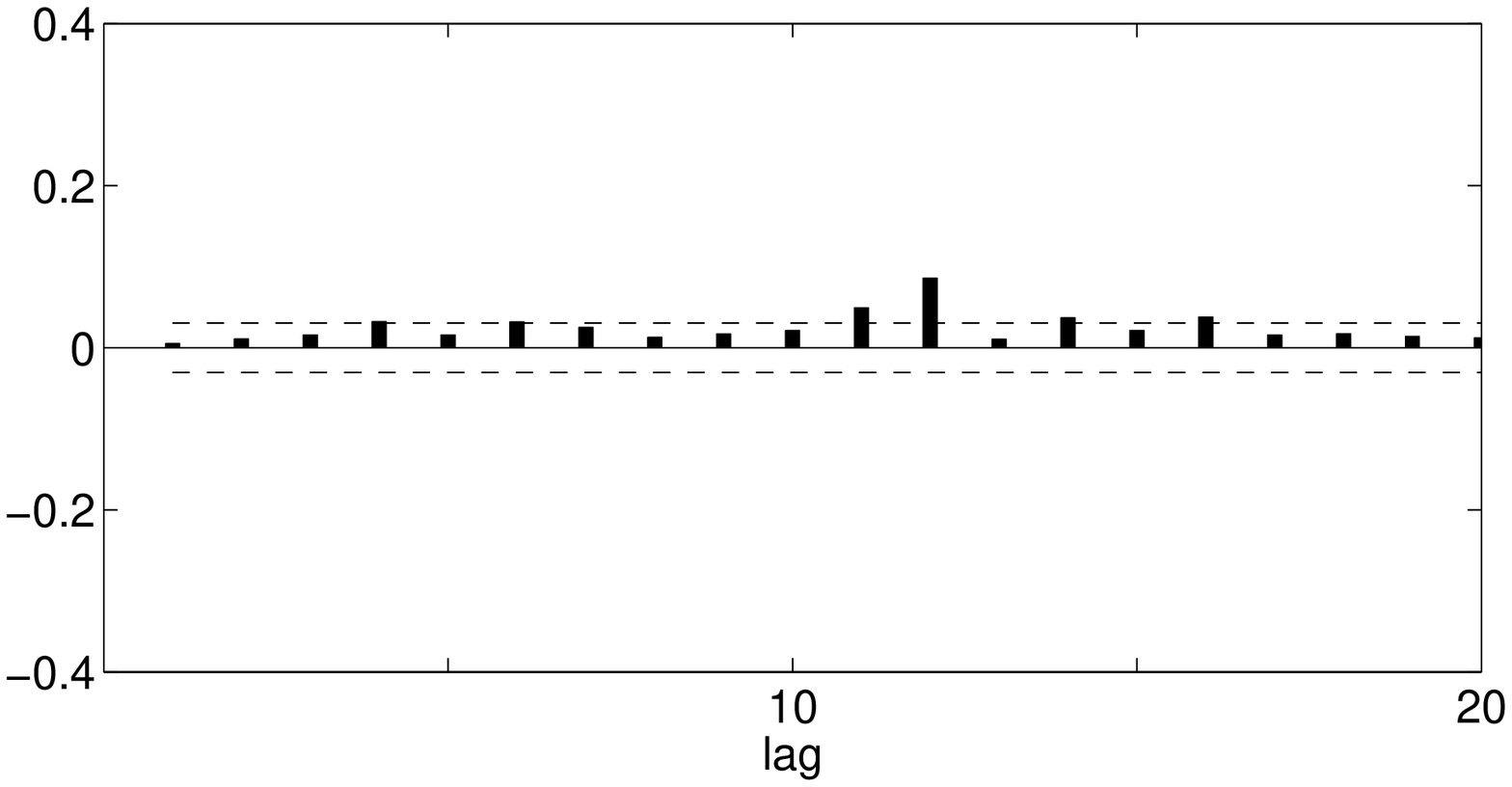}
\end{center}
\caption{Basic GARCH: $\Corr(X_t, X_{t-\ell} | X_t>0 )$ and $\Corr(X_t, X_{t-\ell} | X_t<0)$ for $\ell\geq 1$ (from left to right)}
\label{Fig:GARCH_leverage2}
\end{figure}

In Figure~\ref{Fig:GARCH_ACF}, the positive autocorrelation of absolute returns presented in the right panel indicates volatility clustering.
However, the basic GARCH intensity model is not rich enough to capture leverage effect
as demonstrated in Figures~\ref{Fig:GARCH_leverage1} and \ref{Fig:GARCH_leverage2}.
The leverage effect would imply that $\Corr(|X_t|, X_{t-\ell})$ is negative.
The absence of the leverage effect can also be seen in Figure~\ref{Fig:GARCH_leverage2}
where conditional correlations are negligible.

\subsection{GJR GARCH intensity}~\label{Append:GJR}
\begin{figure}
\begin{center}
\includegraphics[width=5cm]{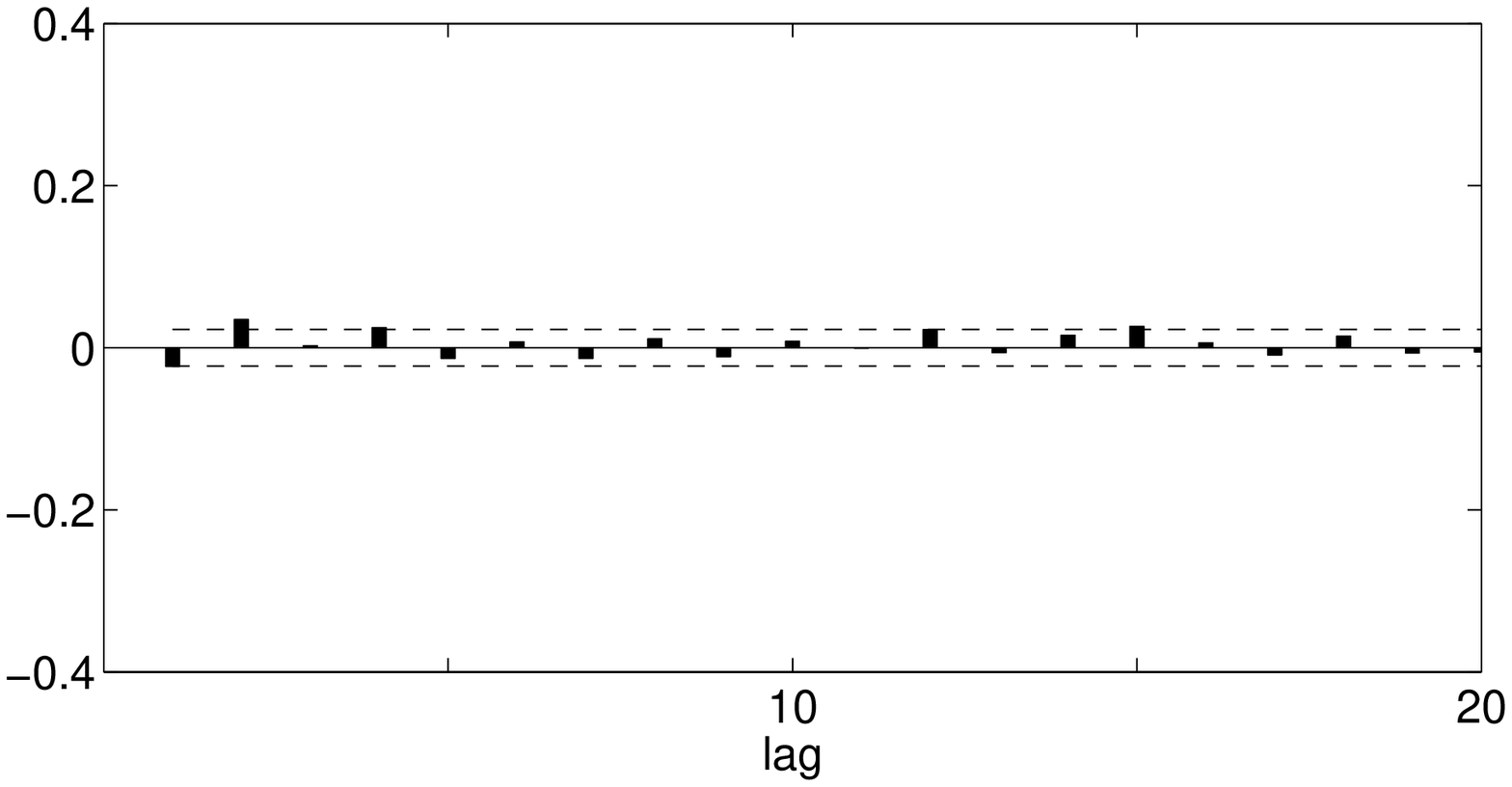}
\quad
\includegraphics[width=5cm]{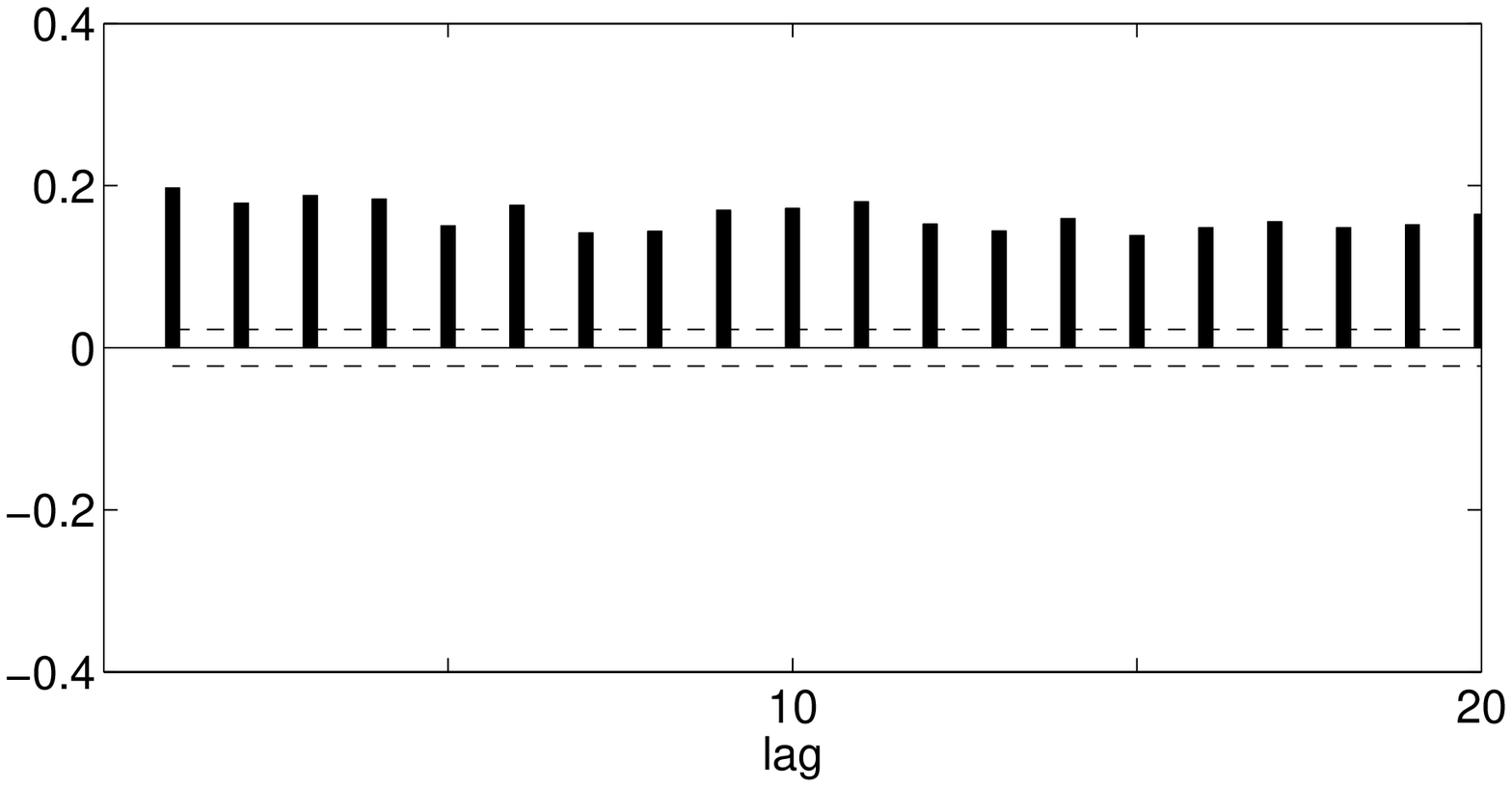}
\end{center}
\caption{GJR: $\Corr(X_t, X_{t-\ell})$ and $\Corr(|X_t|, |X_{t-\ell}|)$  for $\ell \geq 1$ (from left to right).
Volatility clustering is observed in the right.}
\label{Fig:GJR_ACF}
\end{figure}

\begin{figure}
\begin{center}
\includegraphics[width=5cm]{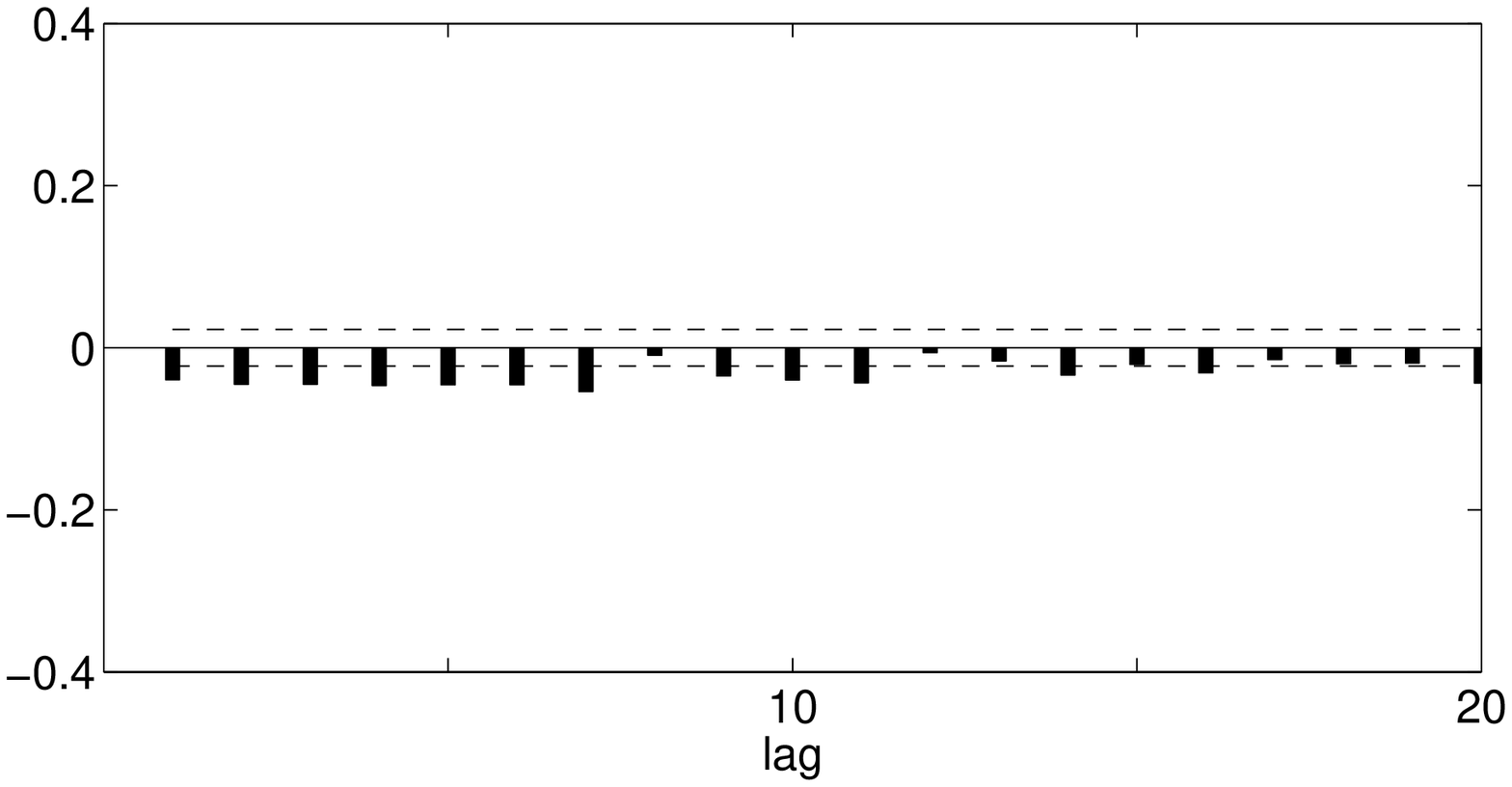}
\quad
\includegraphics[width=5cm]{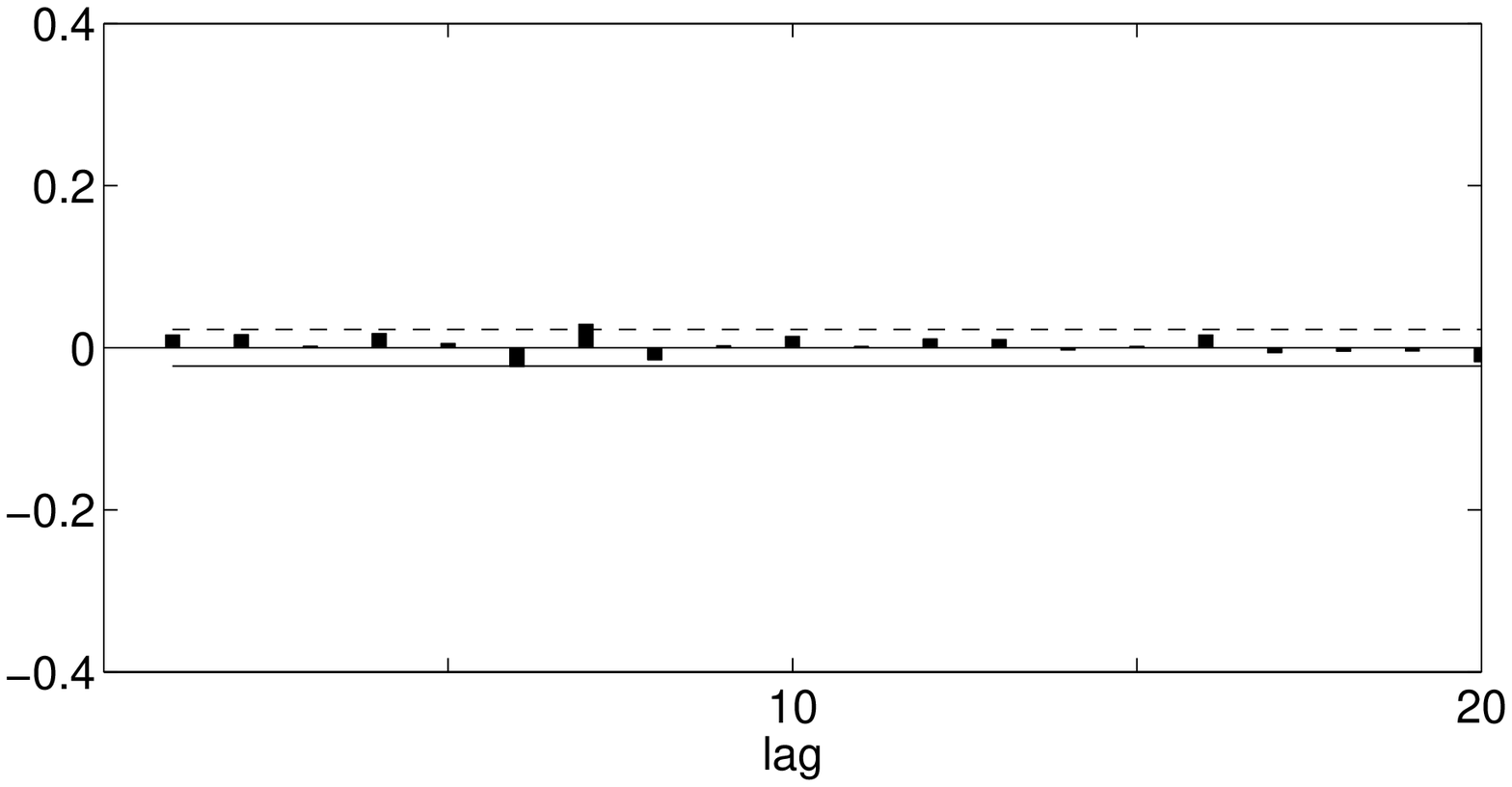}
\end{center}
\caption{GJR: $\Corr(|X_t|, X_{t-\ell})$ and $\Corr(X_t, |X_{t-\ell}|)$  for $\ell \geq 1$ (from left to right).
Leverage effect is observed in the left.}
\label{Fig:GJR_leverage1}
\end{figure}

\begin{figure}
\begin{center}
\includegraphics[width=5cm]{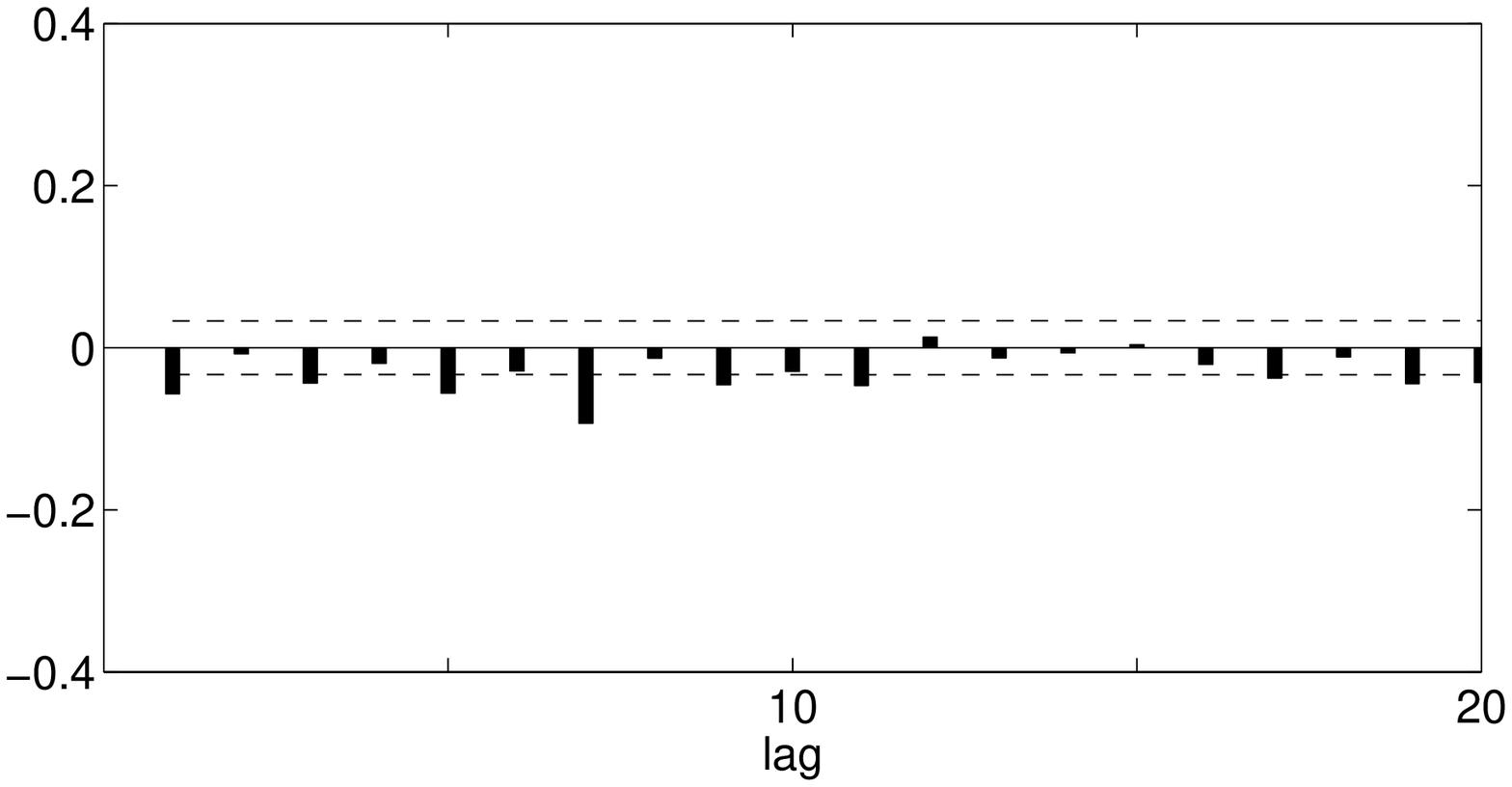}
\quad
\includegraphics[width=5cm]{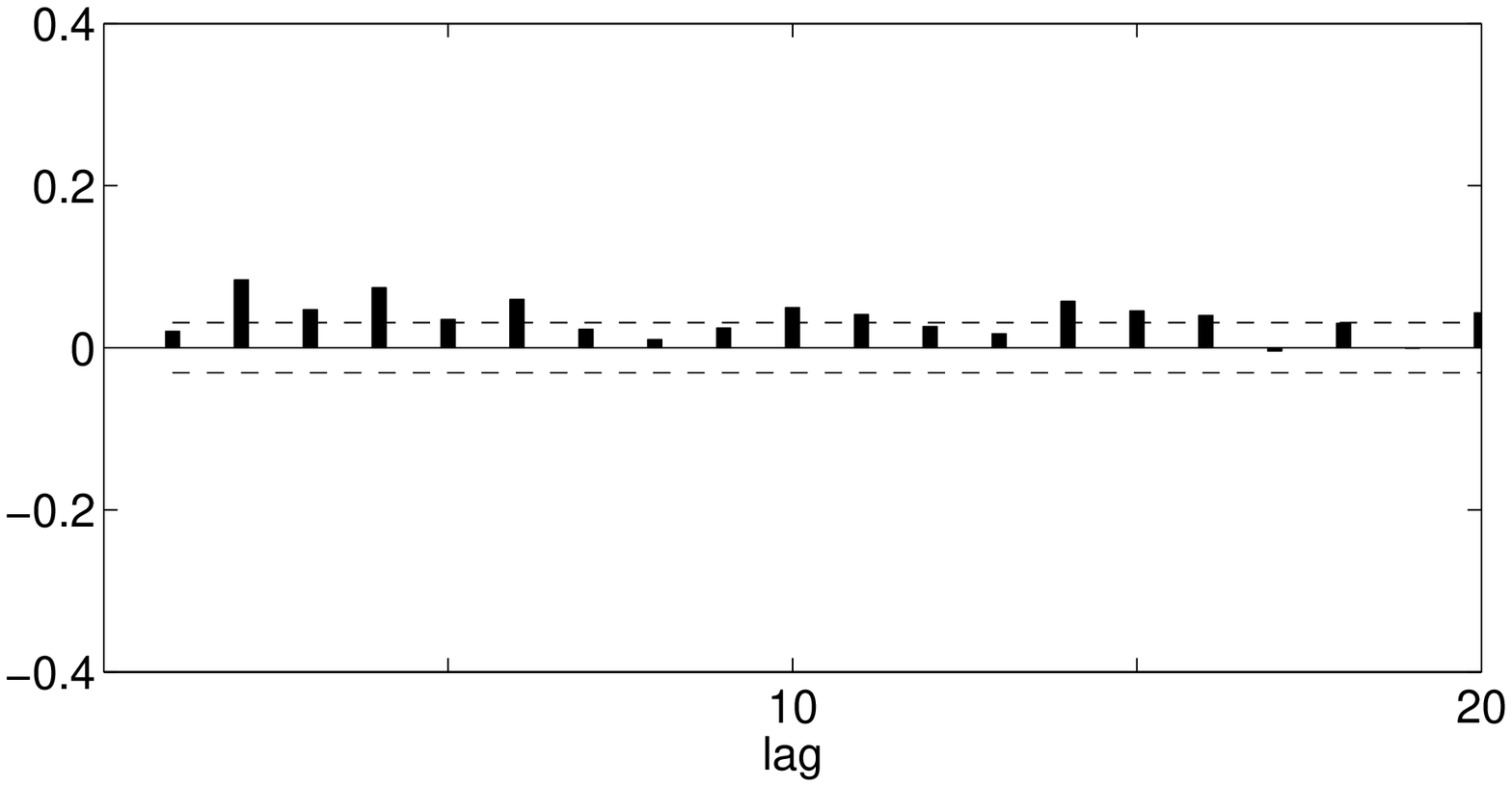}
\end{center}
\caption{GJR: $\Corr(X_t, X_{t-\ell} | X_t>0 )$ and $\Corr(X_t, X_{t-\ell} | X_t<0)$ for $\ell\geq 1$ (from left to right).}
\label{Fig:GJR_leverage2}
\end{figure}

\begin{figure}
\begin{center}
\includegraphics[width=4cm]{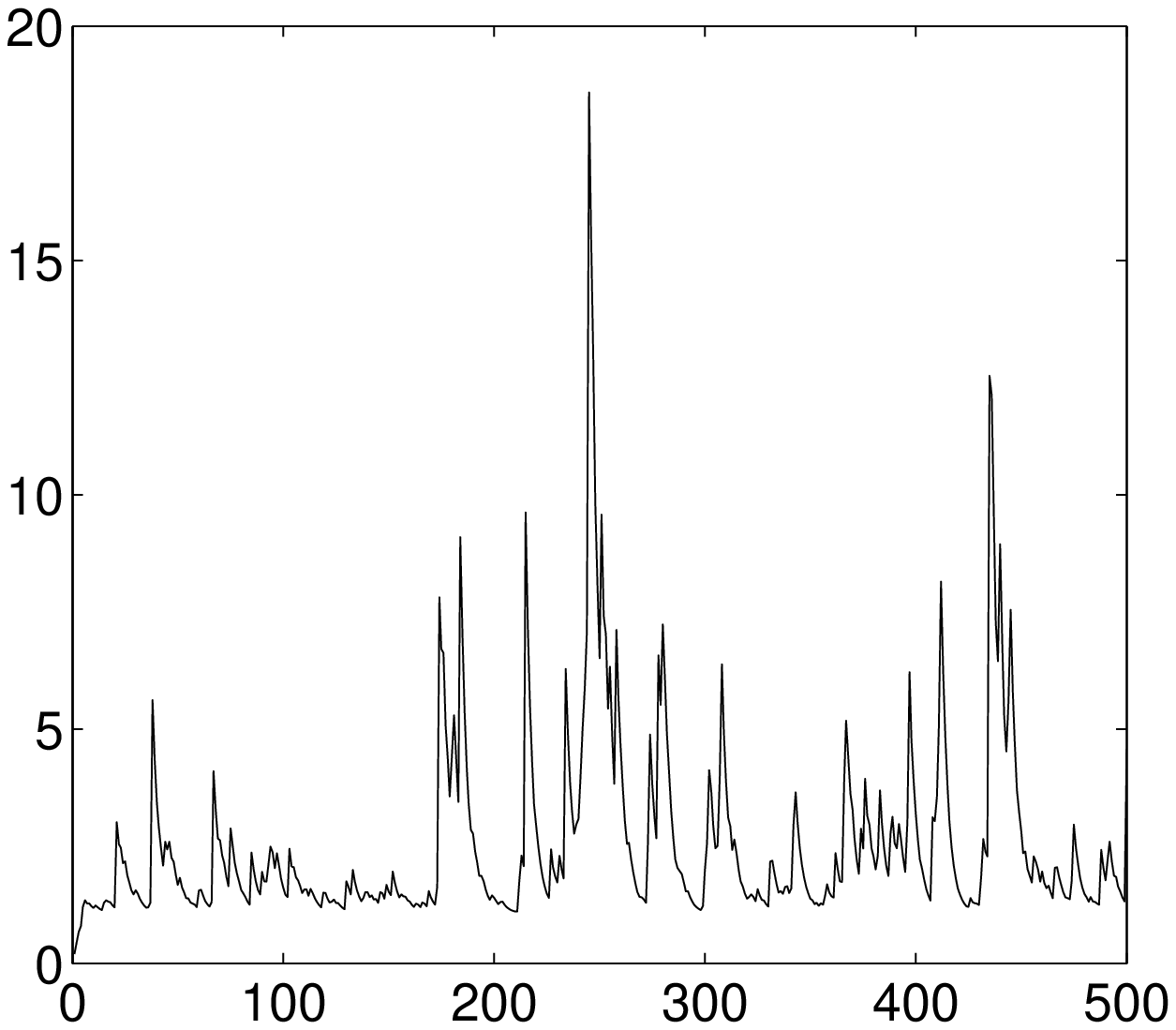}
\qquad
\includegraphics[width=4cm]{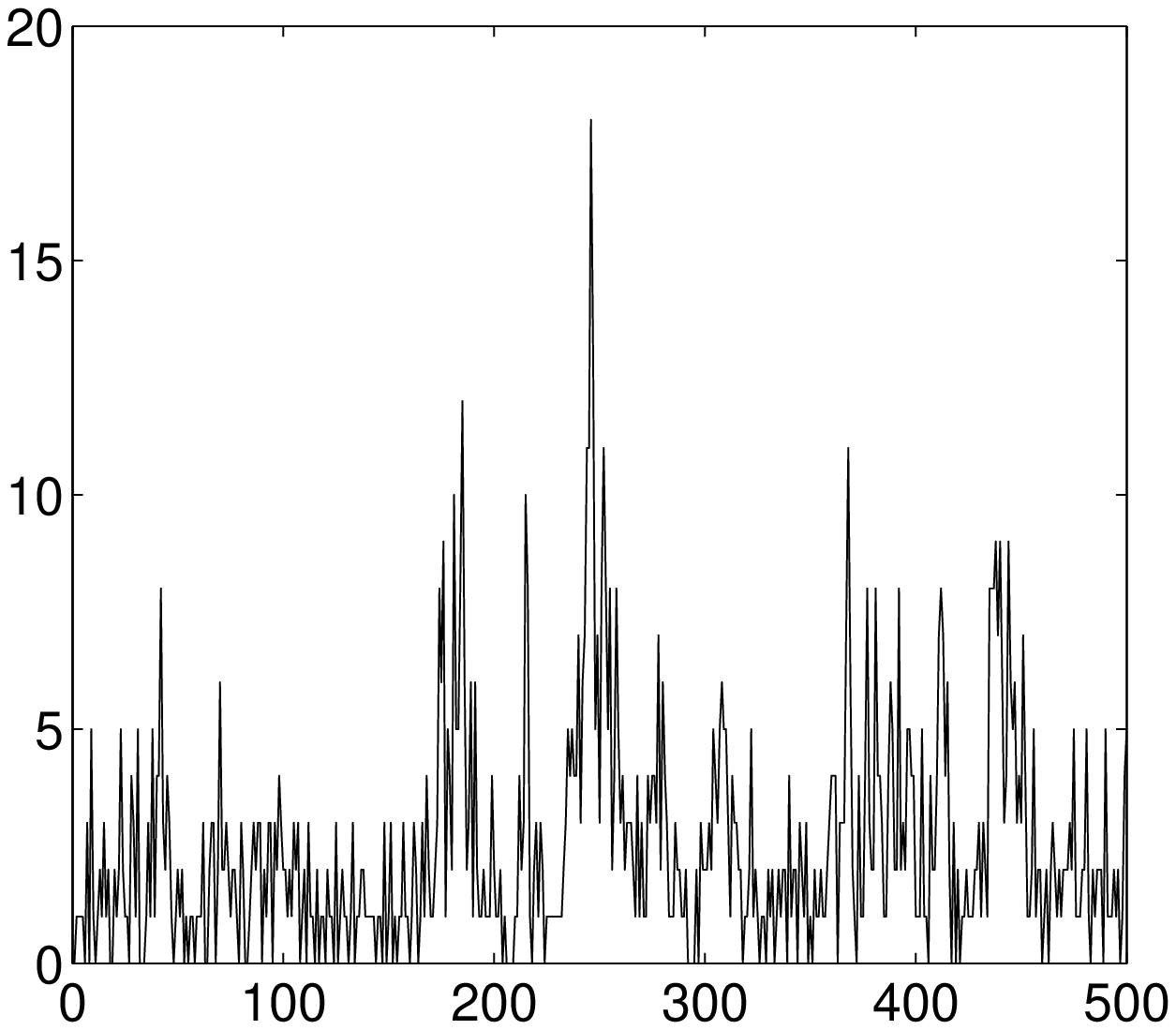}
\end{center}
\caption{GJR: $\lambda_{+}(t)$  and $N_{+}(t)$ (from left to right)}
\label{Fig:lambda+N+}
\end{figure}

\begin{figure}
\begin{center}
\includegraphics[width=4cm]{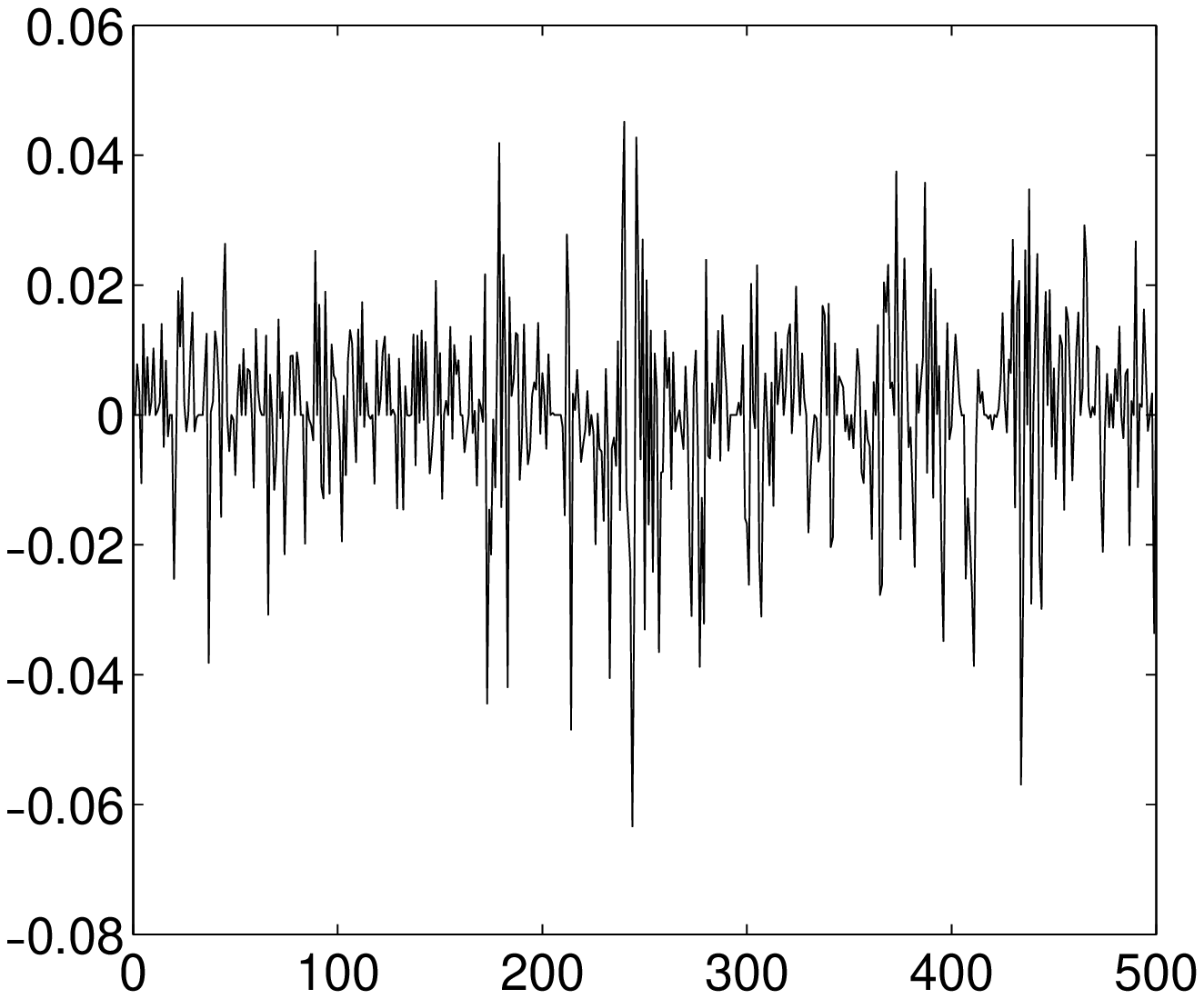}
\qquad
\includegraphics[width=4cm]{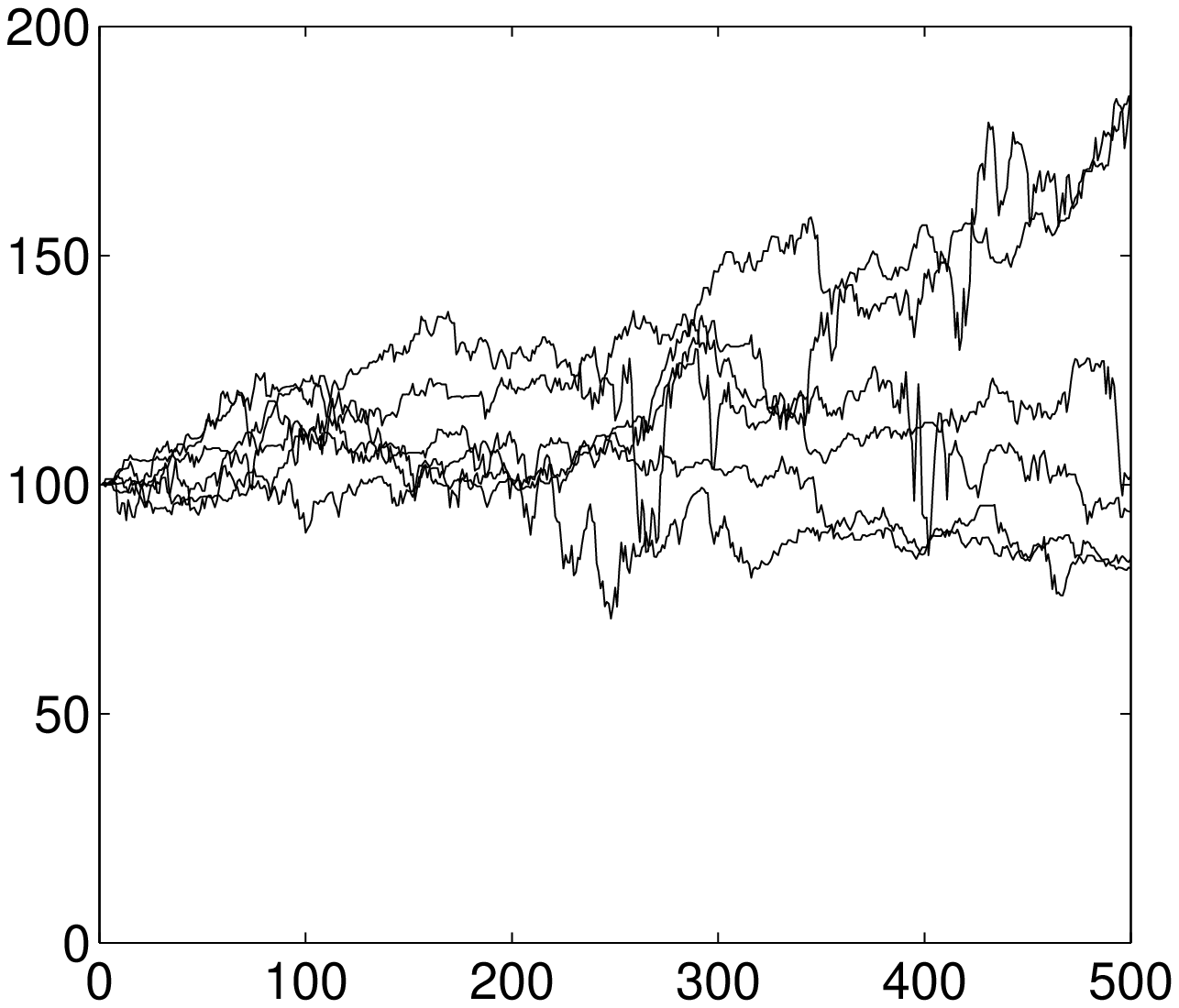}
\end{center}
\caption{GJR: $X(t)$  and $S(t)$ (from left to right)}
\label{Fig:XS}
\end{figure}

We plot correlations and conditional correlations of simulated returns obtained by GJR GARCH type intensity model.
In the right of Figure~\ref{Fig:GJR_ACF} we observe volatility clustering.
The leverage effect is shown in Figure~\ref{Fig:GJR_leverage1}
that the magnitude of today's return is negatively correlated with past return.
The correlation between today's return and past absolute return is negligible.
Figure~\ref{Fig:GJR_leverage2} shows that if today's return is positive then the correlation with past return is negative
while if today's return is negative then the correlation with past return is positive.
Figure~\ref{Fig:lambda+N+} presents simulations of $\lambda_+(t_{i})$ and $N_+(t_{i})$,
and Figure~\ref{Fig:XS} presents $X(t_{i})$ and $S(t_{i})$, $0\leq i \leq 500$.


\bibliography{Poisson_intensity_bib}
\bibliographystyle{apalike}
\end{document}